\documentclass[a4paper,11pt,final]{article}
\usepackage{import}
\usepackage{Preamble}
\title{
A Reverse Black Hole Information Problem
}

\author[a]{Jan de Boer,}
\author[a]{Jildou Hollander,}
\author[b]{and Andrew Rolph}

\affiliation[a]{
Institute for Theoretical Physics, University of Amsterdam, 1090 GL Amsterdam, The Netherlands}
\affiliation[b]{Vrije Universiteit Brussel (VUB) and The International Solvay Institutes, Pleinlaan 2, B-1050 Brussels, Belgium}

\emailAdd{j.deboer@uva.nl}
\emailAdd{jildouhollander@gmail.com}
\emailAdd{andrew.d.rolph@gmail.com}

\abstract{
We study the formation, detection and coarse-graining of black holes in AdS/CFT, with an emphasis on the tension between boundary unitarity and the production of mixed state Hawking radiation in the bulk.
We construct CFT states dual to black hole formation and evaporation by colliding bulk particle wavepackets at trans-Planckian energy. We propose boundary probes which are able to distinguish small AdS black holes from other states within the microcanonical ensemble. We investigate different coarse-graining prescriptions acting on the evolving CFT state, including averaging over CFT data, Hamiltonians and time windows, and compare their purities to those expected from the bulk semiclassical description. Our results clarify how semiclassical black hole behaviour can arise from an ensemble-averaging of the exact unitary dynamics, and take a step towards a better understanding of coarse-graining in the single-sided black hole information problem.
}

\makeatletter
\gdef\@fpheader{}
\makeatother
\begin{document}
\nolinenumbers
\maketitle
\flushbottom

\section{Introduction}

Euclidean wormholes play an important role in the semiclassical approach to quantum gravity, for example, in calculating Page curves~\cite{almheiriReplicaWormholesEntropy2020, Penington:2019kki} and in probing the chaotic properties of gravity~\cite{Saad:2018bqo}. But these wormholes also present challenges, such as the factorisation puzzle~\cite{Maldacena:2004rf}.
Multiboundary wormholes imply correlations between the boundary theories that are inconsistent with semiclassical gravity being dual to a single fixed boundary theory. Indeed, there are known and hypothesised holographic dualities between two and three-dimensional gravitational theories and ensemble-averaged theories~\cite{Saad:2019lba,belinApproximateCFTsRandom2024,chandraSemiclassical3DGravity2022}. But there are also the holographic dualities between \textit{single} CFTs and string theories in AdS. It is interesting to ask whether the two types of duality can be connected and whether semiclassical AdS gravity can arise as a coarse-graining of a single boundary theory. Recent developments have provided substantial evidence that the relevant coarse-graining is over all ``theories" that are semiclassically indistinguishable, a principle of maximum ignorance~\cite{deBoer:2023vsm}.

A plethora of other types of coarse-graining and averaging prescriptions have appeared in the literature, including, for example, diagonal projections, averages over approximate solutions of CFT data, and large $N$ averages. We will not give a systematic account of all of these, but instead point out that several open questions remain.
For semiclassical AdS gravity, is the maximal ignorance prescription the single ``correct" way of ensemble-averaging the boundary theory, or can multiple different types of averaging simultaneously match the same semiclassical gravitational prediction? 
Does the averaging prescription to use depend on what result from the semiclassical gravitational prediction one is trying to match? 
Previous work has primarily focused on Euclidean setups, often on matching semiclassical Euclidean wormhole amplitudes to boundary-averaged quantities, whereas our focus, on black hole formation and evaporation, is intrinsically dynamical and Lorentzian.

Understanding black hole evaporation is one of the central problems in quantum gravity. The information problem is a tension between the semi-classical evolution from pure to mixed states through the production of thermal Hawking radiation and putative unitarity in semiclassical quantum gravity~\cite{hawking1975particle, hawking1976breakdown}.
In holography, when the boundary theory is unitary, any mixedness in the state must arise from effective non-unitary dynamics emerging from the exact microscopics through some sort of coarse-graining. 

In this paper, we will revisit the idea that semiclassical dynamics in AdS/CFT can be understood as emerging from coarse-graining the exact unitary dynamics of the boundary CFT. AdS/CFT is an excellent framework in which to study these questions. 
There are manifestly unitary holographic CFTs, and we can construct exact CFT states dual to the formation and evaporation of small, thermodynamically unstable AdS black holes. There are many ways to construct such states~\cite{Giddings:2001ii,Danielsson:1999fa, Danielsson:1999zt,Heemskerk:2012mn,Bhattacharyya:2009uu,Arsiwalla:2010bt, Anous2016,banerjeeSignaturesBulkBlack2025,Bizon:2011gg,Dias:2011ss}. In section~\ref{sec:creating_small_BH}, we will do so by exciting the vacuum with a pair of smeared boundary operators to create bulk particle wavepackets that collide head-on at trans-Planckian energies~\cite{Polchinski:1999ry,Polchinski:1999yd}. 
Our construction of an exact CFT state allows us to control the kinematics -- the centre of mass energy, impact parameter and collision location -- of black hole formation.
Our setup does not couple the boundary theory to a bath, so the system is closed and the thermodynamic ensemble is microcanonical. 
For sufficiently high energy, the AdS black hole formed from the collision is thermodynamically stable within the microcanonical ensemble. 
At lower energies, there are small, thermodynamically unstable black holes that will eventually evaporate away, leaving behind a
mixed state gas of Hawking radiation (in the semiclassical description). This allows for a possible comparison to the black hole information problem for black holes in the flat space limit. 
We are thus naturally led to considering small black holes in AdS, because only these can fully evaporate.

Small black holes in AdS have various puzzling features. They want to localise on compact internal spaces if their horizon radius is less than the size of the internal space \cite{DiaSan16}.
Moreover, sufficiently small black holes do not dominate the microcanonical ensemble, and,
since these black holes evaporate, they are intrinsically time-dependent and cannot correspond to energy eigenstates. 
The endpoint of the evaporation will semiclassically be a mixed radiation state. On the boundary, the exact state is always pure, so even after the black hole has fully evaporated, some coarse-graining is needed in order to match the semiclassical description. This implies that most of the radiation states are chaotic, as the existence of chaos is intrinsically linked to the need for coarse-graining. Of course, not all radiation states will be chaotic, as there are boundary operators consisting of products of simple operators with small anomalous dimensions, and these boundary operators will be dual to non-chaotic weakly bound multi-particle states in the bulk. However, the majority of the radiation states are presumably chaotic. Therefore, the microcanonical Hilbert space will consist of an ``integrable" sector and a ``chaotic" sector, and the chaotic sector will contain both small black hole states and radiation states.
This situation is quite different from large black holes in AdS, which dominate their microcanonical ensemble, and where chaotic radiation states are too entropically suppressed to play a significant role. 

One may wonder whether we have seen evidence for the existence of chaotic radiation states before. They are reminiscent of the fact that generic states, even at low energies, are susceptible to turbulent instabilities~\cite{Bizon:2011gg, Dias:2011ss}. These turbulent instabilities are perhaps indicative of a lack of predictability and a precursor for thinking of these states as being chaotic. 

Another challenging aspect of small black holes is the decomposition of the chaotic sector into black hole and radiation states. This decomposition cannot be based on conserved charges, as time evolution mixes the two sectors. From the gravitational point of view, black hole states localise energy in the centre of AdS as much as possible. 
This suggests that a probe that is sensitive to radial location might be a suitable candidate to distinguish black hole and radiation states. 
Black hole formation can be detected using boundary correlators, for example, in~\cite{balasubramanianHolographicParticleDetection2000, balasubramanianTypicalityThermalityAnalytic2008,takayanagiMeasuringBlackHole2010,BalBer112,BalBer11,Anous2016}, and, in section \ref{sec:detect_BH}, we will propose that two-point functions of simple operators are suitable probes for distinguishing black hole and radiation states, and conjecture that black holes are the states which minimise two-point functions. 
This is particularly interesting for small, evaporating black holes in AdS, as they are atypical states and their existence is transient, so it is useful to have a diagnostic for when they are present.

The existence of chaotic sectors in CFTs for conformal dimensions in the small, unstable black hole range, well below the threshold for large black holes, is mysterious from the CFT point of view. In $\cN=4$ SYM, these are operators with dimensions scaling as $N^{\alpha}$ with some $\alpha<2$, and for 2d CFTS as $c^{\alpha}$ with $\alpha<1$. For gauge theories, it has been proposed that these theories admit partially deconfined phases where a subgroup deconfines, and that these phases are relevant for small black holes \cite{Ber06,AspBer09,HanMal17,Ber18,HanIsh19,HanJev19}, but we are not aware of many quantitative tests of this idea. We will discuss some aspects of partial deconfinement in section \ref{sec: deconfinement}, but we will not consider the subtleties of how to break the R-symmetry in the gauge theory that corresponds to the localisation of the dual black hole on the compact space \cite{HolKum07}. 

Given the lack of an explicit separation of the microcanonical Hilbert space (in the small black hole energy range) into integrable and chaotic black hole and radiation states, it is difficult to test any proposed coarse-graining prescription against the bulk semiclassical physics. We will therefore mostly consider coarse-grainings relevant for large black holes, in section \ref{sec:HowToCoarseGrain}, before discussing the differences with a putative coarse-graining for small black holes. Such a coarse-graining cannot involve microcanonical typicality, as that would mix small black hole and radiation states. Instead, it presumably involves a more refined version of typicality which applies separately to the black hole states, as well as to different subsectors of the radiation states. A first attempt to construct a toy model with this refined structure can be found in section \ref{sec: small black holes} and appendix \ref{ap: toy model}. 
The fact that bulk radial locality is important is one of the motivations for these toy models. 

Coarse-graining over all theories that are semiclassically indistinguishable leads to a natural coarse-graining of the exact CFT state coming from the uncertainty of the high-energy CFT data not being directly accessible to a low-energy semiclassical observer.
That said, some \textit{statistical} properties of the UV data are known from bootstrap constraints and bulk semiclassical wormhole amplitudes~\cite{Pappadopulo:2012jk, changBootstrapping2DCFTs2016, changSemiclassicalOPECoefficients2016, cardyNewHandleThreepoint2017, krausCardyFormulaThreePoint2017, dasModularCrossingsOPE2018, collierUniversalDynamicsHeavy2020, dasUniversalityAsymptoticBounds2021, belinRandomStatisticsOPE2021, anousOPEStatisticsHigherpoint2022, belinGeneralizedSpectralForm2022, belinNonGaussianitiesStatisticalDistribution2022, benjaminUniversalAsymptoticsHigh2024, boerMultiboundaryWormholesOPE2024}. 
The uncertainty in the UV data induces uncertainty in the time-evolved state, dynamically mixing it, and should result in a coarse-grained state whose entropy increases with time, qualitatively matching the behaviour of the semiclassical bulk state. 

We will analyse and explore several different types of state coarse-graining prescriptions. These include averaging over OPE coefficients, Hamiltonians, time windows, and density matrices.
Each type of coarse-graining prescription represents a mechanism by which a low-energy observer's 
uncertainty in the UV physics induces uncertainty in the time-evolved state.

To summarise briefly, this paper explores how semiclassical AdS gravity arises from coarse-graining the boundary theory, with an emphasis on black hole dynamics and the information problem. The novel aspects of our work include explicit constructions of exact CFT states dual to AdS black holes that can evaporate, a Lorentzian rather than Euclidean focus to ensemble-averaging holographic CFTs, and a comparison of multiple coarse-graining prescriptions.

The paper is structured as follows. In section~\ref{sec:creating_small_BH}, we analyse the thermodynamic stability of AdS black holes in the microcanonical ensemble and determine exact CFT states dual to the formation of AdS black holes by colliding two particles. 
In section~\ref{sec:detect_BH}, we explore how boundary probes can distinguish AdS black holes from radiation states with the same total mass and connect this to (partial) deconfinement. 
In section~\ref{sec:HowToCoarseGrain}, we discuss coarse-graining prescriptions, apply them to time-evolving pure states, including our constructed black hole-producing CFT state, and compare the purities of the resulting coarse-grained states with bulk semiclassical expectations.
Lastly, in appendix \ref{ap: toy model} we construct an explicit toy model for the coarse-graining of a small black hole as it is evaporating. 

\section{How to create an AdS black hole} \label{sec:creating_small_BH}

In this section, we will construct exact CFT states dual to AdS black holes. We will be particularly interested in states dual to the formation and evaporation of a small, unstable AdS black hole, because then we can explore the tension between boundary unitarity and the production of mixed state Hawking radiation. We begin with an analysis and review of the AdS black hole thermodynamic stability threshold.

\subsection{Microcanonical stability of AdS black holes
} \label{sec:HowSmall}

\paragraph{Mass and stability.} All small AdS black holes are thermodynamically unstable in the canonical ensemble, but not necessariliy in the microcanonical ensemble. Large AdS black holes have positive heat capacity while small ones have negative heat capacity. 
The AdS spacetime acts as a confining box for the emitted radiation, not letting it escape, and, in the microcanonical ensemble, the temperature of the black hole's surroundings increases as energy is added to it, and the small black hole can equilibrate. 

The black hole is unstable because black hole states do not dominate the microcanonical ensemble. The threshold of stability is derived by comparing the entropy of an AdS black hole to the entropy of a gaseous ball of radiation with the same total mass~\cite{Horowitz:1999uv}. The black hole is thermodynamically stable in the microcanonical ensemble when $S_{\mathrm{BH}} \gg S_{\mathrm{rad.}}$. 

AdS-Schwarzschild black holes with small horizon radii 
($r_h \ll \ell_{\text{AdS}}$) are locally indistinguishable from asymptotically flat Schwarzschild black holes, which gives us an approximation for their thermodynamic properties. The entropy of a small AdS$_{d+1}$-Schwarzschild black hole scales with its ADM mass $M$
as%
\footnote{We use $\approx$ in this subsection to denote an $r_h \ll \ell_{\rm AdS}$ approximation for which we have dropped prefactors that are independent of the parameters and quantities of interest $(S,M, r_h, \ell_p, \ell_{\rm AdS}\dots)$.}
\bne \label{eq:BHentropy} S_{\tsc{BH}} \approx (\ell_p M)^{\frac{d-1}{d-2}} .\ene
The thermal entropy density of radiation scales as $T^d$, so the entropy of an AdS-scale ball of radiation gas scales with its total mass as
\bne S_{\mathrm{rad.}} \propto (\ell_{\text{AdS}} T)^d = (\ell_{\text{AdS}} M)^{d/d+1}. \label{eq:gasEntropy} \ene

So, the mass range of AdS$_{d+1}$-Schwarzschild black holes that are both thermodynamically \textit{unstable} in the microcanonical ensemble and have a horizon radius much larger than the Planck length is (we want $r_h \gg \ell_p$ so that we can trust the gravitational EFT)
\bne \label{eq:mass_bound} \frac{\ell_{\text{AdS}}}{\ell_p} \ll M \ell_{\text{AdS}} \ll \left( \frac{\ell_{\text{AdS}}}{\ell_p}\right)^{\frac{d^2 - 1}{2d-1}}.\ene
 We used that the horizon radius of a small AdS$_{d+1}$-Schwarzschild black hole with $d>2$ scales with its mass as
\bne r_h^{d-2} \approx \ell_p^{d-1} M . \label{eq:smallBHr}\ene 

\paragraph{Size and stability.}

The stability bound can be rewritten in terms of the horizon radius:
\bne \frac{r_h}{\ell_{\text{AdS}}} \ll \left( \frac{\ell_p}{\ell_{\text{AdS}}}\right)^{\frac{d-1}{2d-1}} ,\ene
which shows that all unstable AdS black holes are small, but also that not all small black holes are unstable. This justifies, after the fact, our use of the small AdS black hole entropy formula~\eqref{eq:BHentropy} to derive the stability bound.

\paragraph{Lifetime and stability.} 
The lifetime of an asymptotically flat Schwarzschild black hole is, in any dimension,\footnote{The lifetime can be estimated from the Stefan-Boltzmann law $\frac{dM}{dt} \approx - A T^{d+1} \approx -r_h^{-2}$. Calculating the precise prefactor requires accounting for the greybody factors.
}
\bne t_{\mathrm{evap.}}\approx \, r_h\, S_{\mathrm{BH}}. \ene
So, the lifetime of a small, unstable AdS black hole scales with its mass as
\bne \frac{t_{\mathrm{evap.}}}{\ell_{\text{AdS}}} \approx \left( \frac{\ell_p}{\ell_{\text{AdS}}}\right)^{\frac{2(d-1)}{d-2}}(M \ell_{\text{AdS}})^{\frac{d}{d-2}}. \label{eq:BHlift} \ene 
Substituting this into our mass range of unstable AdS black holes~\eqref{eq:mass_bound} gives the following range of allowed lifetimes:
\bne \frac{\ell_p}{\ell_{\text{AdS}}} \ll \frac{t_{\mathrm{evap.}}}{\ell_{\text{AdS}}} \ll \left( \frac{\ell_{\text{AdS}}}{\ell_p} \right)^{\frac{(d-1)^2}{2d-1}} .\ene
Therefore, small, unstable AdS black holes can be both short and long-lived in AdS units. From~\eqref{eq:BHlift}, the ADM mass of a black hole whose lifetime is AdS-scale, is
\bne M \ell_{\text{AdS}} \approx \left(\frac{\ell_{\text{AdS}}}{\ell_p} \right) ^{\frac{2(d-1)}{d}}. \ene

\paragraph{BTZ black holes.}
For BTZ black holes ($d=2$), the horizon radius formula~\eqref{eq:smallBHr} is no longer applicable. Instead, the BTZ horizon radius is
\bne r_h^2 = 8 \ell_{\text{AdS}} G_N \left(M \ell_{\text{AdS}}-\frac{c}{12}\right). \ene
Here, $M$ denotes the difference in ADM masses between the BTZ and vacuum geometries.\footnote{We use this convention so that $\Delta = M \ell_{\text{AdS}}$ holds in any dimension.}

There are no unstable BTZ black holes in the microcanonical ensemble.
The stability bound for BTZ black holes is (setting $d=2$ in~\eqref{eq:mass_bound}) $M \ell_{\text{AdS}} \ll \ell_{\text{AdS}}/\ell_p$, which cannot be satisfied given that the lower bound on BTZ black hole mass is $M \ell_{\text{AdS}} > c/12$. 
Moreover, since adding charge or angular momentum to a black hole while keeping the mass fixed can only decrease its entropy, by the first law of black hole mechanics, all BTZ black holes are stable in the microcanonical ensemble. The takeaway is that one must work in more than three spacetime dimensions to have an evaporating AdS black hole.  

\paragraph{Compact internal spaces.} 
We have so far neglected the possibility that the bulk has compact internal dimensions. Suppose instead that the bulk spacetime contains a compact internal space $\mathcal{M}^{D-d}$ of AdS scale, where $D+1$ is the total spacetime dimension, such as AdS$_{d+1}\times S^{D-d}$. 
In this spacetime, the entropy of a small AdS black hole localised on the internal space scales with mass as
\bne S_{\tsc{BH}} \approx (\ell_{p}^{(D+1)} M)^{\frac{D-1}{D-2}} .\ene
We compare this to the entropy~\eqref{eq:gasEntropy} of a volume of radiation which is AdS-scale on the AdS factor and delocalised on the internal space. When comparing entropies, we need not consider the $D$-dim localised ball of radiation because it has a smaller entropy than the one spread over the internal space.%
\footnote{Maximising the volume occupied by the radiation (subject to confinement by the AdS potential) maximises the entropy, because
\bne S_{\rm rad.} \approx V^{\frac{1}{D+1}} E^{\frac{D}{D+1}}. \ene
}
The entropy of the delocalised radiation is the same as~\eqref{eq:gasEntropy} with $d \mapsto D$.
The new mass range of unstable AdS black holes is
\bne \label{eq:compactbound} \frac{\ell_{\text{AdS}}}{\ell_p^{(D+1)}} \ll M \ell_{\text{AdS}} \ll \left( \frac{\ell_{\text{AdS}}}{\ell_p^{(D+1)}}\right)^{\frac{D^2 - 1}{2D-1}}.\ene
Note that the exponent in~\eqref{eq:compactbound} increases monotonically with $D$, so having extra compact internal dimensions always increases the mass of the heaviest unstable black hole.
This allows us to have unstable black holes in AdS$_3$ whenever there are large compact internal dimensions.
For the example of AdS$_3 \times S^3$ with $R_{S^3} = \ell_{\rm AdS}$, the range~\eqref{eq:compactbound} gives us that the mass of the largest unstable black hole is of order%
\footnote{The notation $f(n) = \Theta (g(n))$ means that $f$ is asymptotically of the same order as $g$: there exists an $n_0$ and positive constants $k_1$ and $k_2$ such that $k_1 g(n) \leq f(n) \leq k_2 g(n)$ for all $n \geq n_0$.
}
\bne M \ell_{\text{AdS}} = \Theta(c^{2/3}). \ene
This mass is both smaller than the BTZ black hole threshold $M\ell_{\text{AdS}} = c/12$ and the 3d Planck mass, but larger than the 6d Planck mass.
It is not an issue that this black hole is smaller than the 3d Planck length; it is outside the regime of validity of the 3d gravitational EFT, but not the 6d EFT. 

The unstable black hole evaporation time is given by~\eqref{eq:BHlift} with $d$ replaced by $D$, and $\ell_p$ by $\ell_p^{(D+1)}$, the $(D+1)$-dimensional Planck length. Since AdS-scale black holes in AdS$_{d+1}\times \mathcal M^{D-d}$ are stable, all unstable black holes are localised on the internal space\footnote{This assumes the absence of scale separation between the compact space and AdS.} and $D$-dimensional.
The heaviest unstable 6d black hole in AdS$_3 \times$ S$^3$ has a long lifetime in AdS units:
\bne \frac{t_{\mathrm{evap.}}}{\ell_{\text{AdS}}} = \Theta(c^{1/6}). \ene

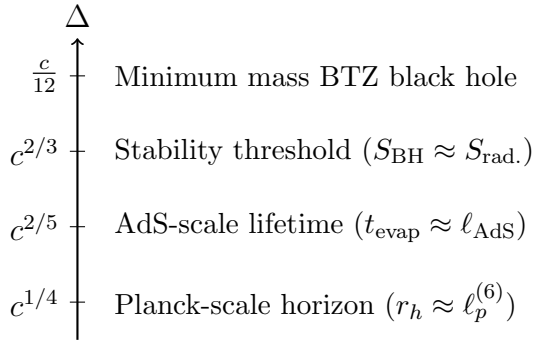
\begin{figure}
  \centering
\begin{tikzpicture}[scale=1]

% Vertical axis
\draw[->, thick] (0,0.5) -- (0,4.5) node[above] {\large $\Delta$};

% Tick positions (bottom to top)
\def\yplanck{1}
\def\yevapo{2}
\def\ystab{3}
\def\ybtz{4}

% Ticks and left labels
\draw (-0.1,\yplanck) -- (0.1,\yplanck)
  node[left=6pt] {\large $c^{1/4}$};

\draw (-0.1,\yevapo) -- (0.1,\yevapo)
  node[left=6pt] {\large$c^{2/5}$};

\draw (-0.1,\ystab) -- (0.1,\ystab)
  node[left=6pt] {\large$c^{2/3}$};

\draw (-0.1,\ybtz) -- (0.1,\ybtz)
  node[left=6pt] {\large$\frac{c}{12}$};

% Right-hand annotations
\node[right=10pt, align=left] at (0,\yplanck)
  {Planck-scale horizon $(r_h \approx \ell_p^{(6)})$};

\node[right=10pt, align=left] at (0,\yevapo)
  {AdS-scale lifetime $(t_{\rm evap} \approx \ell_{\rm AdS})$};

\node[right=10pt, align=left] at (0,\ystab)
  {Stability threshold $(S_{\rm BH} \approx S_{\rm rad.})$};

\node[right=10pt, align=left] at (0,\ybtz)
  {Minimum mass BTZ black hole};

\end{tikzpicture}
\caption{The mass hierarchy of black holes in AdS$_3 \times S^3$ and the corresponding operator dimensions
    in the 2d D1-D5 CFT. $\Delta = M \ell_{\rm AdS}$, where $M$ is the difference in ADM masses between the black hole and vacuum solutions. The black holes with mass $M \ell_{\rm AdS} \ll c$ are 6d (localised on the $S^3$) with an $S^4$ horizon topology.
    }
    \label{fig:hierarchy0}
\end{figure}

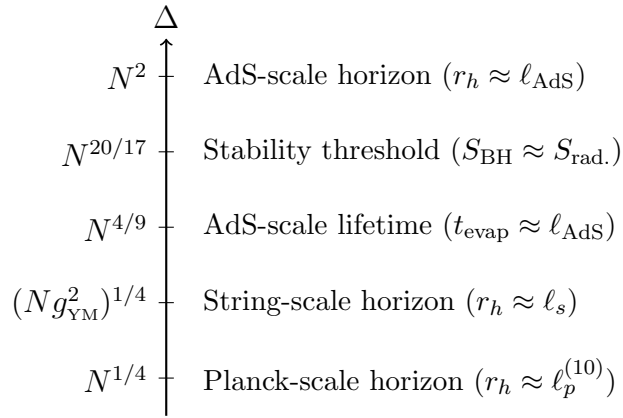
\begin{figure}
  \centering
\begin{tikzpicture}[scale=1]

% Vertical axis
\draw[->, thick] (0,0.5) -- (0,5.5) node[above] {\large $\Delta$};

% Tick positions (bottom to top)
\def\yplanck{1}
\def\ystring{2}
\def\yevapo{3}
\def\ystab{4}
\def\yads{5}

% Ticks and left labels
\draw (-0.1,\yplanck) -- (0.1,\yplanck)
  node[left=6pt] {\large $N^{1/4}$};

\draw (-0.1,\ystring) -- (0.1,\ystring)
  node[left=6pt] 
  {\large$(N g_{\tsc{YM}}^2)^{1/4}$};

\draw (-0.1,\yevapo) -- (0.1,\yevapo)
  node[left=6pt] {\large$N^{4/9}$};

\draw (-0.1,\ystab) -- (0.1,\ystab)
  node[left=6pt] {\large$N^{20/17}$};

\draw (-0.1,\yads) -- (0.1,\yads)
  node[left=6pt] {\large$N^{2}$};

% Right-hand annotations
\node[right=10pt, align=left] at (0,\yplanck)
  {Planck-scale horizon $(r_h \approx \ell_p^{(10)})$};

\node[right=10pt, align=left] at (0,\ystring)
  {String-scale horizon $(r_h \approx \ell_s)$};

\node[right=10pt, align=left] at (0,\yevapo)
  {AdS-scale lifetime $(t_{\rm evap} \approx \ell_{\rm AdS})$};

\node[right=10pt, align=left] at (0,\ystab)
  {Stability threshold $(S_{\rm BH} \approx S_{\rm rad.})$};

\node[right=10pt, align=left] at (0,\yads)
  {AdS-scale horizon $(r_h \approx \ell_{\rm AdS})$};

\end{tikzpicture}
\caption{The mass hierarchy of black holes in AdS$_5 \times S^5$ and the corresponding operator dimensions
    in the 4d $\cN=4$ SYM CFT. $\Delta = M \ell_{\rm AdS}$, where $M$ is the ADM mass. The black holes with mass $M \ell_{\rm AdS} \ll N^2$ are 10d with an $S^8$ horizon topology.
    }
    \label{fig:hierarchy}
\end{figure}

The results of this subsection are brought together and illustrated in Fig.~\ref{fig:hierarchy0}, which shows the hierarchy of black hole masses for AdS$_3 \times S^3$ and the D1-D5 CFT, and Fig.~\ref{fig:hierarchy}, which shows the hierarchy for AdS$_5 \times S^5$ and 4d $\cN =4$ super-Yang Mills (SYM).%
\footnote{For reference: $g_s = g_{\tsc{YM}}^2$, $\ell_{\text{AdS}} = \ell_s (N g_{\tsc{YM}}^2)^{1/4}$, $G_{N}^{(10)}= g_s^2 \ell_s^8$.}
In the figures, besides the BTZ mass threshold, we only give the scaling (e.g. with $c$) of the black hole masses.

Which position the string scale horizon comes in the hierarchy depends on the coupling $\lambda = N g_{\tsc{YM}}^2$ for the 4d $\cN =4$ SYM case. We wish to have unstable black holes which are larger than the string scale, for which the very strong coupling limit ($N\to \infty$ with $g_{\tsc{YM}}$ fixed) suffices.
In this regime, the masses of the string and Planck-scale horizons have the same $N$ scaling, and they match the dimension of the lightest massive string states $\Delta \approx \lambda^{1/4}$ such as the Konishi operator.
The $r_h \approx \ell_s$ regime is the Horowitz-Polchinski transition between string and black hole states~\cite{Horowitz:1996nw}.

Two important points are that not all small black holes are unstable, and not all unstable black holes are short-lived. 

\subsection{Colliding two particles to form an AdS black hole} \label{sec:colling_particles}
We wish to construct an exact CFT state whose time evolution describes the formation and eventual evaporation of a small AdS black hole, to study the tension between the bulk semiclassical description and boundary unitarity. 

\begin{figure}
    \centering
    \includegraphics[width=0.4\linewidth]{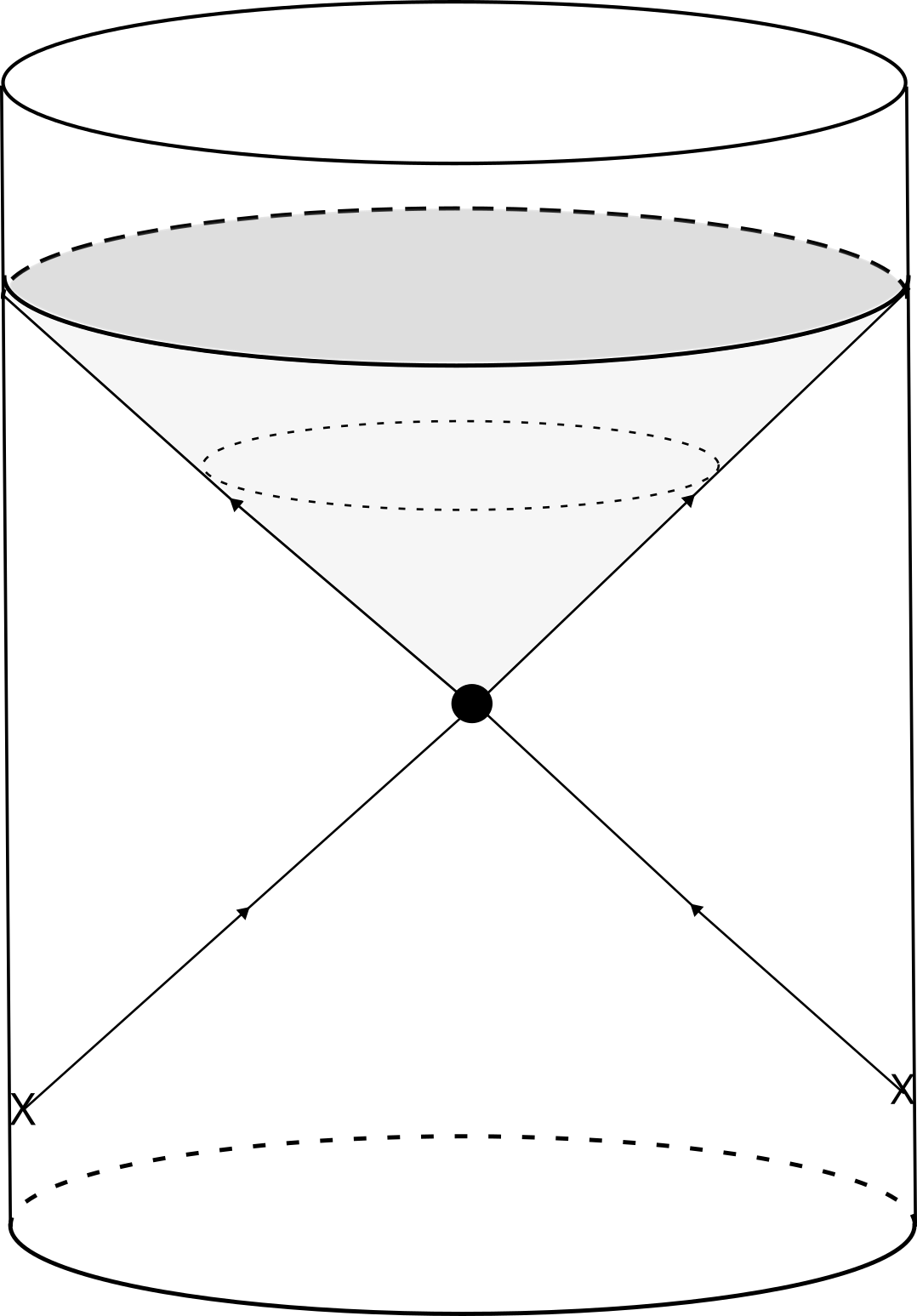}
    \caption{The creation of a small AdS black hole by colliding two particles. The operators inserted on the boundary have a finely-tuned smearing so that the bulk particle wavepackets are beams that are localised and stay localised until they collide. Black holes that are thermodynamically unstable can be short or long-lived in AdS units; the black hole depicted is short-lived, which is why the Hawking radiation is a spherical shell (the grey cone).}
    \label{fig:WavepacketCollision}
\end{figure}

There are many ways to form a small AdS black hole -- for completeness, we will review a number of these in Sec.~\ref{sec:OtherWaysToSmash} -- but the method that we will focus on is 
colliding a pair of bulk particles at trans-Planckian energies.  
This is depicted in Fig.~\ref{fig:WavepacketCollision}. The essence of creating a black hole is to localise energy in as small a region as possible.

We will create these particles at opposite sides of the AdS boundary by exciting the vacuum with a pair of smeared operators inserted at antipodal points on the boundary sphere. The kinematics of the bulk particles are controlled by how the boundary operators are smeared, as we will derive.

Black hole production via particle scattering has a long history in the literature. Some relevant seminal papers include~\cite{Eardley:2002re}, who calculated the production cross section at non-zero impact parameter using colliding shockwaves, and~\cite{Polchinski:1999yd}, one of the first papers to explore black hole production in AdS/CFT, including a discussion of boundary operator insertions. Several subtleties have been raised and resolved in the literature, both in the classical and semiclassical regimes; see, for example,~\cite{Giddings:2004xy} and references therein. We will discuss some of these subtleties along the way, as we describe our bulk particle wavepackets and their boundary duals. For more recent work on black hole production via particle collision in AdS/CFT, see~\cite{Chowdhury:2025bud}. 

In global AdS, if the two particles both have (dimensionless) frequency $p_t = \omega$ and collide at $r=0$, then the centre-of-mass collision energy $\sqrt{s}$ is
\bne s=-g^{tt}(2p_t)^2 = 4\omega^2 / \ell_{\text{AdS}}^2 \label{eq:CoMen} .\ene 
So, by increasing $\omega$, the centre-of-mass energy can be made arbitrarily large, including above the Planck mass scale for black hole formation. 

We treat the bulk particles as freely propagating, localised Gaussian wavepackets while they are sufficiently separated and redshifted by the AdS potential to neglect gravitational backreaction. 
For our purposes, it suffices that the 
free approximation holds for the propagation of the wavepackets from the AdS boundary to a separation of order the Schwarzschild radius associated with their CoM energy. Around this separation, the relative boost is large, and the gravitational interactions become strong. The particles backreact to form Aichelburg-Sexl-type gravitational shockwaves, and closed trapped surfaces form, which signals black hole formation~\cite{Eardley:2002re}.

In the rest frame of either of the particles, the gravitational backreaction from that (light) particle is negligible, but when a pair of particles' relative velocities are sufficiently high, there is no inertial frame in which both particles have sub-Planckian energy, so backreaction cannot be avoided with a frame shift; this is the frame-independence of black hole formation and centre-of-mass (CoM) energy.

\subsubsection{Bulk wavepacket and boundary smeared operator}

Here we will determine the exact CFT state that is dual to the black hole-producing pair of bulk particles. We will detail our bulk particle wavepackets and specify how to create them by exciting the vacuum with smeared boundary operators. This analysis is based on~\cite{Polchinski:1999ry, Hernandez:2025seq}. 

We take the frequencies of the incoming wavepackets to be large, $\omega \gg 1$. This is necessary to have sufficient energy to create a black hole (see Eq.~\eqref{eq:CoMen}), and it allows us to use a WKB approximation.
A Gaussian wavepacket solution to the free massless Klein-Gordon (KG) equation in the WKB approximation is~\cite{Polchinski:1999ry}
\bne \phi_{\omega \vec{e}} (t,\vec{x}) \approx F_{\omega \vec{e}} (t,\vec{x}) e^{-i\omega(t-\vec{e}\cdot \vec{x})} .\ene
Here $\vec{e}$ is the unit vector direction of travel and $F$ is the Gaussian envelope function
\bne \label{eq:wavepacket} F_{\omega, \vec{e}}(t,\vec{x}) = \exp \left ( -\frac{\omega}{2}(x_{\bot}^2+(t-\vec e \cdot \vec x)^2 ) \right ). \ene
The transverse spatial spread of this wavepacket is sub-AdS scale: $\sigma_\perp \approx 1/\sqrt{\omega}$ in AdS units.

The bulk wavepacket approaches the AdS boundary at the intersection points of the classical trajectory, at $t= \pm \pi/2$ and opposite poles of the boundary sphere. In other words, as $r\to \infty$,
\bne \phi_{\omega \vec{e}} (t,\vec{x}) \approx \left [G_- (t+\pi/2,|\vec{\theta}+\vec{e}|)+G_+ (t-\pi/2,|\vec{\theta}-\vec{e}|)\right] e^{-i\omega t}, \label{eq:bulk_field} \ene 
where $\vec x = r \, \vec{\theta}$ and $\vec \theta$ is a point on the boundary sphere.
$G$ is, up to a constant prefactor,%
\footnote{Note that the Gaussian factor in this wavepacket does not respect the angular periodicities. The exact wavepacket does, and one can create such a wavepacket through a sum of images, but equation~\eqref{eq:kernel} suffices as the narrow-angle, large-$\omega$ approximation of the exact wavepacket.}
\bne \label{eq:kernel} G_\pm (t,\vec{\theta}) = e^{-\frac{\omega}{2}(\vec{\theta}^2+t^2)} .\ene
$G$ is derived by matching the general large $r$ asymptotic solution to the transverse profile specified in~\eqref{eq:wavepacket}~\cite{Polchinski:1999ry}. This envelope function is sharply localised in both time and angular position on the boundary sphere: $\sigma_\theta = \sigma_t \approx \omega^{-1/2}$. But the proper transverse width of the wavepacket is much larger near the AdS boundary than in the deep interior, so this KG solution describes a wavepacket that focuses as it propagates to the collision point. 

There are two terms in~\eqref{eq:bulk_field}, the incoming wavepacket $G_-$ and the outgoing one $G_+$.
For us, the relevant one is the incoming $G_-$. The $G_+$ wavepacket is an artefact of the free approximation, which ceases to exist when one turns on gravitational interactions and collides these wavepackets to form black holes.

From~\eqref{eq:bulk_field}, through the extrapolate dictionary, the boundary operator that creates each wavepacket is the smeared operator
\bne \label{eq:wavepacket2}  \hat{\Phi}_{\omega \vec{e}} := \int dt d^d x \, G_- (t+\pi/2 , |\vec{x}+\vec{e}|)e^{-i\omega t}\hat \phi (t,\vec{x}) \ene
where $\hat \phi$ is the boundary operator dual to the bulk field we are creating the wavepacket from. We emphasise that $G_-$ is a carefully chosen smearing kernel. It has a tunable parameter $\omega$ controlling the CoM energy of the colliding particle wavepackets. 

Using~\eqref{eq:wavepacket2}, the initial CFT state dual to two bulk particles with equal frequencies $\omega$ at opposite sides of the AdS bulk is
\bne \label{eq:initial_state}
\ket{\psi(0)}= \cN^{-1} :\!\hat{\Phi}_{\omega \vec{e}_S}\hat{\Phi}_{\omega \vec{e}_N} \!:\ket{0}. \ene
Here $\omega$ was the frequency of the wavepacket, $\vec{e}_{S/N}$ denote antipodal points on the boundary sphere, and $\cN$ is a normalisation prefactor (which would not be needed if $\hat{\Phi}$ were unitary rather than Hermitian, as it is in some constructions). This equation for the initial state, with $\hat{\Phi}$ defined in~\eqref{eq:wavepacket2}, is an important result for this subsection. 

The time evolution of~\eqref{eq:initial_state} is the CFT state dual to two bulk particles colliding with zero impact parameter. The particles collide at half the light-crossing time. Then, with a choice of $\omega$ in the correct range (see Eqs.~\eqref{eq:mass_bound} and~\eqref{eq:CoMen}), a small unstable AdS black hole is formed and eventually evaporates, and what we have in the time evolution of~\eqref{eq:initial_state} is an exact constructed CFT state dual to that process. 

\paragraph{Requirements for black hole formation.} For general impact parameter, we assume that a black hole will form if the minimal separation for the classical trajectory of the two bulk particles is parametrically smaller than the horizon radius $r_h$ of the CoM energy $\sqrt{s} = 2\omega/\ell_{\text{AdS}}$.\footnote{For AdS$_3$, the precise classical condition for black hole formation from particle scattering is known~\cite{gott1991closed}. This is the Gott condition:
\bne 4\pi G_N \sqrt{s} > e^{b/2}\ene
where $b$ is the impact parameter. This condition is simple because there are no propagating gravitons in AdS$_3$. For higher-dimensional AdS, the condition is less simple, as there is some energy loss to gravitational wave emission before the particles collide, so the black hole mass is slightly less than the total incoming energy.
}
We can control this minimal separation easily through a suitable choice of the particles' initial positions and momenta.

A second localisation-type requirement is that the particle wavepacket's transverse spatial spread be much smaller than the AdS-Schwarzschild radius at the collision time $t_c = \pi/2$: 
\bne \sigma_\perp (t_c) \ll r_h(\omega). \label{eq:locre} \ene
Otherwise, the wavepackets will have large tails outside of the AdS-Schwarzschild radius ball centred on the collision point, so the probability of producing a black hole will be significantly less than one.

This means that we want to check the diffusive spreading of the transverse width of the wavepacket as it propagates.
In the free approximation, for a Gaussian wavepacket centred around momentum $p_0$, with momentum bandwidth $\sigma_p$ and dispersion relation $E(p)$, the transverse spatial spread is
\bne \sigma_\perp (t)^2 = \frac{1}{(2\sigma_p)^2} + \hbar^2 \sigma_p^2 E'' (p_0)^2 t^2 . \label{eq:tdwidth} \ene
For large frequency $\omega$ and a bulk field with mass $m \ell_{\rm AdS} \ll \omega$, we have $p_0 \ell_{\rm AdS} \approx \omega$ and $E''(p_0) = m^2/p_0^3 + O(p_0^{-4})$.
By varying the momentum bandwidth, the smallest we can make the wavepacket at the collision point is
\bne \min_{\sigma_p} \left( \sigma_\perp (t_c)^2 \right)= \hbar E''(p_0) t_c \label{eq:min_sig} \ene
This limits how small we can make a black hole, because we must increase $\omega$ to make the wavepacket narrower at the collision point, but this also makes the resulting black hole larger.
To give an example, for an AdS$_5$ Schwarzschild black hole of mass $M=2\omega/\ell_{\text{AdS}}$, whose radius is (dropping constant prefactors) 
$r_h^2 \approx G_N \omega/\ell_{\text{AdS}}$ when it is small ($r_h \ll \ell_{\text{AdS}}$), and $t_c \approx \ell_{\text{AdS}}$, the size of the black hole is lower-bounded: $r_h^8 \gtrsim m^2 \ell_{\text{AdS}} \ell_p^{9}$, which is a smaller radius than the Planck length.\footnote{As a reminder, $m$ is the mass of the bulk field. Note that we cannot take the RHS to zero with $m\to 0$ while keeping the wavepacket localised in the momentum domain, $\Delta p \ll p_0$, because the minimising $\Delta p$ in~\eqref{eq:min_sig} scales as $\omega^{3/2}/m\ell_{\text{AdS}}^2$.} 

\paragraph{Other comments.} In~\eqref{eq:wavepacket}, we took the bulk field to be massless, but this can be generalised to a light massive field, and the resulting boundary operator smearing kernel is the same as~\eqref{eq:kernel}~\cite{Polchinski:1999ry}. We will always take $\hat \phi$ to be a \textit{light} scalar operator, because the CFT duals of light bulk fields are better understood than those of heavy bulk fields, though the creation of BTZ black holes by colliding conical defects has also been studied~\cite{banerjeeSignaturesBulkBlack2025}. For us, the energy to create the black hole comes from the large relative velocities of the colliding particles, not the particles' rest mass. 

We chose to create the wavepacket out of a scalar field, but we can also create a black hole by colliding \textit{graviton} wavepackets at trans-Planckian energies, or with a sufficiently large number of low-energy graviton excitations.
This is an interesting setup because then the bulk EFT need not have any matter fields; it can be pure gravity. In the boundary CFT, the stress tensor would be the only single trace operator, and~\eqref{eq:wavepacket2} would be a smeared stress tensor. But, to avoid the expense of complicating calculations with tensor indices, we will restrict ourselves to scalar fields.

Not all Gaussian wavepacket solutions have a spatial width related to their frequency $\omega$ like in~\eqref{eq:wavepacket}. Eq.~\eqref{eq:wavepacket} is a special solution whose cross-sectional, transverse spatial profile at a given moment in time is constant in the longitudinal direction. This property isn't necessary for our analysis, and we only restrict to the one-parameter $\omega$-family of solutions for simplicity. Note also that, in contrast to~\eqref{eq:tdwidth}, the spatial width of the wavepacket~\eqref{eq:wavepacket} is constant in time because we neglect subleading $\hbar$ corrections in the WKB approximation.

In the literature, it has been shown that bulk wavepackets created by briefly turning on boundary-compact sources have tails which prevent one from determining flat-space S-matrix elements, but also that these tails are not an obstacle to creating a black hole~\cite{giddingsGravitationalSmatrixErice2013, Gary:2011kk}.
 
If the bulk spacetime has an internal compact space, then we can make a small black hole localised on that space, i.e. a 10d black hole for AdS$_5 \times S^5$, by also localising the particle wavepacket on the space, dual to giving the boundary operator insertion the appropriate R-charge. 
Alternatively, we can continue to use an $s$-wave bulk wavepacket. Then the black hole formed will initially be spread over the internal space but will subsequently dynamically localise via the Gregory-Laflamme instability~\cite{BanDou98, Gregory:1993vy}. The black hole localises on a horizon radius timescale, which is parametrically faster than it evaporates.

\subsubsection{Initial state in the spin-dimension eigenbasis} \label{sec:initial_state_hh_basis}

We will write our initial state~\eqref{eq:initial_state} as a sum over the primary states and their descendants:
\bne \ket{\psi(0)} = \cN^{-1} \sum_{h,\bar{h} \in \text{primaries}} c_{\phi\phi\cO_{h,\bar{h}}} \sum_{m, \bar{m}} \psi_{h,\bar{h},m,\bar m}  \ket{h+m,\bar{h} + \bar{m}} \label{eq:primd}.\ene 
Here $h$ and $\bar h$ are the conformal weights of the primary operators, $m$ and $\bar{m}$ are the global descendant levels, and $\ket{h+m, \bar h + \bar m}$ are the normalised primary and descendant states. 

There are two reasons to write the state in the spin-dimension eigenbasis. First, the time evolution is trivial, allowing us to track the dynamics in the changes of phases.
Secondly, because one of our coarse-graining maps in Sec.~\ref{sec:HowToCoarseGrain} will be averaging over an ensemble of OPE coefficients, which reflects the uncertainty in what~\eqref{eq:primd} is, induced from the uncertainty of the OPE coefficient values.

Black hole production depends sensitively on the kinematics of the bulk particles, which are controlled by the insertion positions and smearing of the boundary operators. For example, by increasing the impact parameter while keeping the CoM energy fixed, we transition from black hole production to two particles orbiting each other indefinitely.
The kinematics are fully determined by and encoded in the $\psi$ wavefunction coefficients%
\footnote{It is a slight misnomer to call $\psi$ the ``wavefunction coefficients'', because the full wavefunction coefficients include the normalisation and OPE coefficient factors in~\eqref{eq:primd}.}
in~\eqref{eq:primd},
so the only difference between these two very different sets of states is in these coefficients.

We have specialised to $d=2$ for simplification. In higher dimensions, the calculations are structurally similar to those we will do for two dimensions, with additional tensor structure and labels for the spherical harmonics. 
On the bulk side, restricting to $d=2$ is not a serious reduction in generality either, because one can still create black holes that evaporate away if one allows for compact internal dimensions, as we showed in Sec.~\ref{sec:HowSmall}.
We see no obstacle to generalising to higher dimensions, and we expect that the results would be qualitatively the same. 

\paragraph{Operator product expansion.}
The initial state~\eqref{eq:initial_state} involves two smeared local operator insertions on the Lorentzian cylinder at the two points
\bne w_1 = \theta_1 - t_1, \quad \bar{w}_1 = \theta_1 + t_1 \ene
and 
\bne \quad w_2 = \theta_2+ \pi -t_2 , \quad  \bar{w}_2 = \theta_2 + \pi + t_2.\ene

To write~\eqref{eq:initial_state} in the conformal multiplet basis, we want to do an operator product expansion of
\bne \phi (\theta_1, t_1) \, \phi(\theta_2 + \pi , t_2). \ene
To that end, we map from the Euclidean cylinder to the Euclidean plane with%
\footnote{
The cylinder $S^{d-1} \times \mathds{R}$ is conformally flat in any dimension, so this step works in higher dimensions too.}
\bne z = e^{- i w} , \quad\bar{z}= e^{ i \bar{w}}. \ene
with $w = \theta+ i\tau$. We will perform the expansion on the plane and analytically continue to complex times $\tau_1 = \epsilon + i t_1$ and $\tau_2 = it_2$. 

Since $\phi$ is a primary scalar operator, it transforms as
\bne \phi (w,\bar{w}) = \left (\frac{\del z}{\del w} \right)^{\Delta_\phi/2}\left (\frac{\del \bar{z}}{\del \bar{w}} \right)^{\Delta_\phi /2} \phi (z,\bar{z}). \ene
Once on the plane, we can do an operator product expansion:
\bne  \phi (z_1,\bar{z}_1) \phi (z_2,\bar{z}_2) \ket{0} = \sum_{h,\bar{h}} c_{\phi\phi\cO_{h,\bar{h}}} z_{12}^{h- \Delta_\phi}\bar{z}_{12}^{\bar{h} - \Delta_\phi} \cO_{h,\bar{h}} (0) \ket{0}. \label{eq:OPEex} \ene
Note that $c_{\phi\phi\cO_{h,\bar{h}}} \in \mathbbm{R}$ since $\phi = \phi^\dagger$. 

The sum~\eqref{eq:OPEex} is over all primary and descendant operators.
To make use of the conformal symmetry, we split the OPE~\eqref{eq:OPEex} into separate sums over primaries and descendant levels:
\bne  \phi (z_1,\bar{z}_1) \phi (z_2,\bar{z}_2) \ket{0} = \sum_{h,\bar{h}
} c_{\phi\phi\cO_{h,\bar{h}}} \sum_{m,\bar m} \frac{c_{\phi\phi L_{-1}^m \bar{L}_{-1}^{\bar m} \cO_{h,\bar{h}} }}{c_{\phi\phi \cO_{h,\bar{h}} }}  z_{12}^{h+m- \Delta_\phi}\bar{z}_{12}^{\bar{h}+\bar m - \Delta_\phi} L_{-1}^m L_{-1}^{\bar m} \cO_{h,\bar{h}} (0) \ket{0}. \label{eq:OPEex2} \ene
This sum over $h,\bar{h}$ is only over the primaries.
We have dropped the Virasoro descendants and kept only the global descendants because the Virasoro descendants are suppressed by $1/c$~\cite{Sachdeva:2024hvi} and the smearing (also, in higher dimensions, there are only global descendants). 

To write the initial state in the spin-dimension eigenbasis, i.e. to get~\eqref{eq:primd} from~\eqref{eq:OPEex2}, we will: (1) normalise $L_{-1}^m \bar{L}_{-1}^{\bar{m}} \cO_{h,\bar h} \ket{0}$, (2) use that conformal symmetry fixes the OPE coefficients of descendant operators in terms of the OPE coefficients of the primary operators, and (3) integrate the kinematical factors in~\eqref{eq:OPEex2} against the smearing kernels in~\eqref{eq:wavepacket2}. 

\paragraph{Normalisation of primary and descendant states.} We take the primary operators to be normalised such that $\braket{\cO_i | \cO_j} = \delta_{ij}$, with
\bne \cO (0) \ket{0} = \ket{\cO}, \ene
\bne \bra{\cO} = \bra{0} \cO (0)^\dagger = \bra{0} \cO(\infty) = \bra{0} \lim_{z \to \infty} z^{2\Delta}\cO(z). \ene
The \textit{unnormalised} descendant state 
$L_{-1}^m \bar{L}_{-1}^{\bar{m}} \cO_{h,\bar{h}}(0)\ket{0}$ 
is related to the normalised descendant state $\ket{h+m, \bar{h}+ \bar{m}}$ by~\cite{guoNoteETHDescendant2019}%
\footnote{In this context, $|\dots |^2$ denotes (holomorphic)$\times$(antiholomorphic).}
\bne 
L_{-1}^m \bar{L}_{-1}^{\bar{m}} \cO_{h,\bar h} (0) \ket{0} = \left| \sqrt{\frac{\Gamma(2h)}{m!\, \Gamma(2h+m)}}\,\right|^2 \ket{h+m,\bar{h}+\bar{m}} \label{eq:denor} \ene
with the asymptotics%
\footnote{We use the symbol $\sim$ in a precise fashion, to denote asymptotic equivalence, meaning $f(x) \sim g(x)$ as $x \to \infty$ iff $\lim_{x\to\infty} \frac{f(x)}{g(x)} = 1$. 
}
\bne \begin{split} 
\bra{h+m,\bar{h}+\bar m} L_{-1}^m \bar{L}_{-1}^{\bar{m}} \cO_{h,\bar h} (0) \ket{0} 
&\sim 
\begin{cases} \frac{1}{2\pi} \left|e^{m}m^{-(h +m)}\sqrt{\Gamma(2h)}\right|^2 
\qquad 
&m,\bar{m} \to \infty, \\
\quad \left| (m! (2h)^m)^{-1/2} \right|^2 \qquad\qquad\qquad &h,\bar{h} \to \infty .
\end{cases}
\end{split}\ene
Deriving~\eqref{eq:denor} is an exercise in combinatorics of commuting the conformal generators $L_{-1}$ and $L_{1}$ past each other. Note that the normalisation is singular when $h$ or $\bar{h}$ is zero, and this is because the vacuum state has no descendants: $L_{-1}\ket{h} = 0$ if $h = 0$, as follows from $\bra{h}L_1 L_{-1}\ket{h} = 2h$.

\paragraph{OPE coefficients of descendant states.} 
The OPE coefficients of descendants are fixed by conformal symmetry in terms of the OPE coefficients of the primary operator: 
\bne 
\frac{c_{\phi\phi L_{-1}^m \bar{L}_{-1}^{\bar m} \cO_{h,\bar{h}} }}{c_{\phi\phi \cO_{h,\bar{h}} }} 
= \left| (-1)^m\frac{\Gamma(h+m)}{\Gamma(h)}\right|^2 ,\label{eq:descOPE}
\ene
with the asymptotics
\bne 
\frac{c_{\phi\phi L_{-1}^m \bar{L}_{-1}^{\bar m} \cO_{h,\bar{h}} }}{c_{\phi\phi \cO_{h,\bar{h}} }} 
\sim 
\begin{cases}
2\pi \left| \frac{(-1)^m e^{-m} m^{-\frac{1}{2}+h+m}}{\Gamma(h)} \right|^2 \qquad &m,\bar{m} \to \infty, \\
\quad \left| (-h)^m \right |^2\qquad\qquad\qquad\qquad &h,\bar{h} \to \infty . 
\end{cases}
 \ene

\paragraph{Kinematical coefficients.}
We pull out the normalisation factors from the $\psi$ coefficients to define new coefficients $\varphi$:
\bne \psi_{h,\bar h, m,\bar{m}} = \frac{c_{\phi\phi L_{-1}^m \bar{L}_{-1}^{\bar m} \cO_{h,\bar{h}} }}{c_{\phi\phi \cO_{h,\bar{h}} }} \braket{h+m,\bar h+\bar m|L_{-1}^m L_{-1}^{\bar m} \cO_{h,\bar{h}} (0) | 0} \varphi_{h+m-\Delta_\phi} \varphi_{\bar h+ \bar m-\Delta_\phi}\, . \ene
Then we calculate $\varphi$ by integrating the $z_i$-dependent factor in~\eqref{eq:OPEex2} against the smearing kernels.
Namely, we take our state~\eqref{eq:initial_state}, perform the OPE expansion~\eqref{eq:OPEex}, and insert in the integration kernels from~\eqref{eq:wavepacket2} and~\eqref{eq:kernel}, absorbing~\eqref{eq:kernel}'s prefactor into the normalisation $\cN$. This gives 
\bne \begin{split}
\varphi_{h+m-\Delta_\phi} \varphi_{\bar h+ \bar m-\Delta_\phi} 
&= \int dt_1 \int d\theta_1 \int dt_2 \int d\theta_2 \; e^{-\frac{\omega}{2} (t_1^2 + \theta_1^2)} e^{-i\omega t_1} e^{-\frac{\omega}{2} (t_2^2 + \theta_2^2)} e^{-i\omega t_2}  \\
&\times \left(e^{-i(\theta_1 -t_1)} + e^{-i(\theta_2 -t_2)} \right )^{h+m-\Delta_\phi} \left(e^{i(\theta_1 +t_1)} + e^{i(\theta_2 +t_2)} \right )^{\bar h+ \bar m-\Delta_\phi} . \label{eq:bintegral} \end{split}\ene 
This can be manipulated into%
\footnote{The integrations over $\theta_i$ are approximated with an extension of the integration ranges to the whole real line, which is a large $\omega$, narrow Gaussian approximation.} 
\bne \begin{split} \varphi_{h+m-\Delta_\phi} \varphi_{\bar h+ \bar m-\Delta_\phi} &= \frac{2\pi}{\omega}  \left| 2^{h+m-\Delta_\phi} 
\int_{-\infty}^\infty dx e^{-\frac{\omega}{2} x^2} \cos^{h+m-\Delta_\phi} x 
e^{-\frac{(\omega-h-m+\Delta_\phi)^2}{2\omega}} \right|^2 
\\
& \sim \frac{4\pi^2}{\omega} \left |\frac{2^{h+m-\Delta_\phi}}{ \sqrt{\omega+h+m-\Delta_\phi}}  e^{-\frac{(\omega-h-m+\Delta_\phi)^2}{2\omega}} \right|, \quad \text{ as } \omega \to \infty. \label{eq:primc}
\end{split} \ene
\paragraph{Wavefunction coefficients.} Bringing things together, we get the following coefficients for the initial state in the spin-dimension eigenbasis
\bne \psi_{h,\bar h, m,\bar m} =    \left| (-1)^m \frac{2^{h+m} e^{-\frac{(\omega-(h+m-\Delta_\phi))^2}{2\omega}}}{\sqrt{\omega+h+m -\Delta_\phi}} \frac{\Gamma(h+m)}{\Gamma(h)} \sqrt{\frac{\Gamma(2h)}{m!\, \Gamma(2h + m )}} \,\right|^2 \label{eq:psico} .\ene
This is a large $\omega$ approximation, and we have absorbed constant prefactors (like $\frac{4\pi^2}{\omega}$ and $2^{-\Delta_\phi}$ in~\eqref{eq:primc}) 
into the overall normalisation prefactor $\cN$ in~\eqref{eq:primd}. 

\paragraph{Comments on~\eqref{eq:psico}.}With this result, we can determine which $\ket{\Delta,s}$ eigenstates our black hole-producing state has the largest overlap with.
The $\psi$ coefficients
in~\eqref{eq:psico} are peaked around conformal weight $h+m = \omega + \Delta_\phi \approx \omega$, because of the Gaussian factor. This is consistent with our CFT state being dual to two bulk particle wavepackets colliding with the CoM energy $2\omega$.
Also, the weights of the wavefunction coefficients are well-localised, as the spread is $\sigma = \sqrt{\omega} \ll \omega$. 

The peak of~\eqref{eq:psico} is the same for both the holomorphic and antiholomorphic factors, showing that our initial state~\eqref{eq:initial_state} is dominated by those states $\ket{h+m,\bar h+ \bar m}$ with spin $h+m - \bar h - \bar m$ centred around zero.
This is because the impact parameter $b$ of our collision is zero. Increasing the impact parameter introduces angular momentum into the bulk state, which increases the overlap of the boundary state with non-zero spin eigenstates. The angular momentum can be increased by changing the smearings of the boundary operators as described in~\cite{Hernandez:2025seq}, though we will not explore that in detail here. 

The descendant level that maximises~\eqref{eq:psico} for fixed weight $h+m$ is $m \sim \frac{2}{5} (h+m)$ as $(h+m) \to \infty$. This shows that the spin-dimension eigenstates that our initial state has the largest overlap with are \textit{not} heavy descendants of light primaries, which agrees with what we expect of a black hole microstate.

The $\psi$-coefficients are all real. This is non-trivial and a symptom of the atypicality of our initial pre-collision state. As time progresses, the state will dephase and approach typicality, with decorrelated phases. Typicality can be diagnosed by showing that the time-evolved state looks thermal for simple observables, e.g.
exponential decay of two-point correlators, 
$\bra{\psi(0)} \! O(t+t') O(t) \!\ket{\psi(0)} \sim e^{-t'/\tau_d}$ as $t' \to \infty$,
when $t$ is larger than the time it takes for the horizon to form.

\paragraph{Changing the kinematics.} We can prevent the creation of a black hole by changing the kinematics of the bulk particles, such as by delaying the release of one of the particles by $\delta t$. In that case, the particles will not collide in the centre of the AdS geometry, and this decreases the centre of mass energy. By changing the $t_1$ Gaussian factor in~\eqref{eq:bintegral} to $e^{-\frac{\omega}{2} (t_1 -\delta t)^2}$, the 
$\varphi$ coefficients change to
\bne \varphi_{h+m-\Delta_\phi} = \sqrt{\frac{2\pi}{\omega}} 2^{\alpha}  e^{-\frac{(\omega-\alpha)^2}{2\omega}} I(\delta t, \alpha), \label{eq:newba} \ene
where $\alpha =h+m-\Delta_\phi$ and
\bne \begin{split} I(\delta t, \alpha) 
&= \int dx \, e^{-\frac{\omega}{2} x^2} \cos^\alpha (x - \delta t) \\
&\approx \sqrt{\frac{2\pi}{\alpha  \sec
   ^2\left(\delta t+\frac{\alpha  \tan (\delta t )}{\omega }\right)+\omega}}e^{-\frac{\alpha ^2 \tan ^2(\delta t)}{2 \omega }} \cos ^{\alpha }\left(\delta t +\frac{\alpha  \tan (\delta t)}{\omega }\right) .
   \label{eq:Iinexp}
\end{split} \ene
This has a $\delta t \sim \delta t + 2 \pi$ periodicity, matching the bulk particle periodicity.
When $\tan (\delta t) \neq 0$, the large $\alpha, \bar{\alpha}$ coefficients in~\eqref{eq:newba} are exponentially suppressed, consistent with the delay lowering the CoM energy. This suppression can be seen in the integral representation of $I$ in~\eqref{eq:Iinexp}: the Gaussian factor is peaked around zero, and the $\cos^\alpha(\dots)$ factor around $\delta t + n\pi$, and the larger $\alpha$ is, the smaller the overlap between the terms when the peaks do not coincide. 

\subsection{Uncertainty in the initial state and dynamics} \label{sec:Uncertainty}

We have determined the initial state in the spin-dimension basis, and the wavefunction coefficients depend on the kinematics and conformal data.
Specifically, our state describing the formation and evaporation of a small AdS black hole, Eq.~\eqref{eq:primd}, in the spin-dimension eigenbasis is, when time-evolved,
\bne \ket{\psi(t)} = \cN^{-1} \sum_{h,\bar{h} \in \text{primaries}} c_{\phi\phi\cO_{h,\bar{h}}} \sum_{m, \bar{m}} \psi_{h,\bar{h},m,\bar m}  e^{-i(h+m+\bar h  + \bar m)t}\ket{h+m,\bar{h} + \bar{m}} \label{eq:primd2}.\ene 
In this subsection, we will see just how precisely we can determine the wavefunction coefficients of our black hole-producing state in this eigenbasis. This is of both intrinsic interest and motivates Sec.~\ref{sec:HowToCoarseGrain}'s coarse-graining prescriptions.

The $\psi$ coefficients in~\eqref{eq:primd2} are known exactly, they are given by~\eqref{eq:psico}. If we also know the \textit{complete} set of conformal data perfectly, then there is zero uncertainty in what the density matrix $\rho(t) = \ket{\psi(t)}\bra{\psi(t)}$ is. Conversely, any uncertainty in what the conformal data that go into~\eqref{eq:primd2} are induces an uncertainty in what the state is.

Here is the question: if we are given all the conformal data of all the \textit{light} primaries, and only the light primaries,
how precisely do we know what $\ket{\psi(t)}$ in~\eqref{eq:primd2} is? This is the data a low-energy observer in the bulk can measure; they can probe low-energy excitations of \textit{light} fields in low curvature regions to determine their masses and couplings. Light operators have finite dimension $\Delta = O(1)$ as $c\to\infty$, and heavy operators do not.

Evidently, we still know the overlaps 
$\braket{h+m,\bar h + \bar m|\psi(t)}$ 
between $\ket{\psi(t)}$ and the light primary states and their descendants. 
Naively, we know nothing about the conformal data of the heavy primaries given the light data, so we do not know the overlaps of $\ket{\psi(t)}$ with the heavy states, but this disregards UV/IR mixing. We know some of the UV data from CFT bootstrap constraints.

Furthermore, some heavy primary states are approximate multi-trace operator states, dual to high-energy but weakly interacting multi-particle states. For example, the double-trace operator $\phi^2_{n,l}$
with $n=0$ and $l = c$ is a heavy and approximately stationary state dual to a pair of particles with Planckian energy whose angular momentum keeps them sufficiently far apart to be weakly interacting. 
So, while some heavy primary states are dual to black holes, some are approximate Fock states.

We will investigate this further now and determine for which heavy primaries the weights and OPE coefficients are known, at least approximately, and for which primaries we only know the statistical properties of the conformal data. 

\paragraph{The free approximation.} \label{sec:GFF}
We start with the $c\to \infty$ limit, the generalised free field (GFF) limit. All boundary correlators factorise exactly, and all $k$-particle Fock states $:\!\cO_1 \dots \cO_k \!:\ket{0}$ are also energy eigenstates. In this limit, the OPE of two identical scalar primary operators is
\bne \phi(z,\bar{z})\phi (0) = \frac{1}{|z|^{2\Delta}} + \sum_{n,l}C_{n,l}(|z|^{2n+l}\phi^2_{n,l} (0) + \text{descendants}), 
\label{eq:GFFOPE}
\ene
where $\phi^2_{n,l}$ are the double-trace operators 
\bne \phi^2_{n,l} = \;:\!\phi \hat{\del}_{[\mu_1 \dots } \hat{\del}_{\mu_l]}\phi \!: \, .\ene
The square brackets denote the symmetric traceless combination, and $\hat{\del} \!:=\, \stackrel{\rightarrow}{\del} \!- \!\stackrel{\leftarrow}{\del} $.

So, in the free limit, our state~\eqref{eq:initial_state} is a linear superposition of double-trace primary states and their descendants, and those only; it is an exact two-particle state. Note that the normal ordering in~\eqref{eq:initial_state} removes the identity operator contribution from~\eqref{eq:GFFOPE}.

Another feature of the free limit is that the OPE coefficients in~\eqref{eq:GFFOPE} are known exactly~\cite{Heemskerk:2009pn}:%
\footnote{This is calculated by applying the Wick contraction to the correlator of four $\phi$ operators, and matching to the conformal partial wave expansion. 
}

\bne C_{n,l} = \sqrt{(1+ (-1)^l) c_n c_{n+l}}, \quad c_n := \frac{\Gamma^2 (\Delta_\phi + n) \Gamma (2\Delta_\phi + n -1)}{n!\, \Gamma^2 (\Delta_\phi) \Gamma(2\Delta_\phi + 2n - 1)} \label{eq:GFFOPEc}. \ene
So too are the dimensions of the double-trace operators $\phi_{n,l}$:
\bne \Delta_{n,l} = 2\Delta_\phi + 2n + l, \label{eq:GFFdim}\ene
where the $l$ is the spin, which must be even.

These formulas are specific to two dimensions, but the features of the GFF limit that we have mentioned, such as factorisation of correlators, the multipartite spectrum, and known conformal dimensions and OPE coefficients, hold in any dimension.

The takeaway is that, in the free limit, there is zero uncertainty in what the state~\eqref{eq:primd2} is, because all the conformal data is known exactly.
There is nothing to coarse-grain over
in the boundary state,
which matches $\rho_{\textrm{bulk}}$ staying pure because no black hole can form when $G_N = 0$. We need to go to finite $c$ to see black hole formation and non-trivial dynamics.

\paragraph{Corrections to the free approximation.}
Now we go to finite but large $c$. 
At finite $c$, the multitrace operator basis is overcomplete. 
Also, multitrace operators are no longer primary operators, because they no longer exactly commute with the dilatation operator, whereas primary operators do
(by definition).  

Light operators are those whose dimensions are finite as $c\to \infty$, $\Delta = O(1)$, and heavy operators are those whose dimensions are not.
All light operators are approximate multi-trace operators, and light multi-trace operators can now appear in the OPE~\eqref{eq:GFFOPE}, because bulk interactions allow for particle production. 
Some (but by no means all) heavy primary states are approximate multi-particle Fock states. An example is the stationary state that is dual to a pair of high-energy particles orbiting each other with a large separation due to large angular momentum. 

For a given holographic CFT, a primary or descendant operator, and an associated set of conformal data (e.g. the operator's dimension), either the GFF approximation is an accurate approximation of that data, or it is not.
This splits the set of operators into two groups, which we will call the integrable and the chaotic sectors. As we will see, the split depends on the choice of which particular set of conformal data to consider, and what we mean by accurate.

\textit{Dimensions.} Let us start our discussion of these sectors by considering the size of corrections to the free approximation of conformal dimensions, i.e. the anomalous dimensions. 

Let us focus on the family of double-trace operators $\phi^2_{n,l}$ built from a light single-trace primary $\phi$.
Going to general dimensions for a moment, and specialising to a holographic CFT, the contribution to the anomalous dimension of $\phi^2_{n,l}$ from stress tensor/graviton exchange is~\cite{kavirajUniversalAnomalousDimensions2015}, 
\bne \gamma_{n,l} \approx -G_N \frac{n^d}{l^{d-2}} \label{eq:gamnl} \ene
in the $l \gg n \gg 1$ regime. In the bulk, the anomalous dimensions equal the energy shift due to interactions, i.e.~\eqref{eq:gamnl} equals the shift due to graviton exchange, the gravitational binding energy.
For fixed $n$, increasing $l$ decreases the magnitude of the anomalous dimension~\eqref{eq:gamnl}, because the gravitational binding energy decreases when the particles are farther apart.
Vice versa, the same anomalous dimension's magnitude grows with increasing $n$, corresponding to a larger gravitational binding energy.
In the $n \gg l $ regime, we get similar behaviour: polynomial growth of $\gamma_{n,l}$ with increasing $n$ and polynomial decay with increasing $l$~\cite{fitzpatrickEffectiveConformalTheory2011}.

This implies that there are heavy, large spin primary states that are approximately the double-trace states $\phi^2_{n,l}\ket{0}$ with large $l$, for which the dimensions are known, since the free approximation improves the larger $l$ is. These are the primaries dual to a high-energy and angular momentum pair of particles orbiting each other. These primaries belong in the integrable sector, despite being heavy.
Specialising to $d=2$, the double-trace anomalous dimension in~\eqref{eq:gamnl} is $\gamma_{n,l} \approx n^2/c$. When $n \approx c$, the anomalous dimension is the same order as the GFF approximation for $\Delta_{n,l}$ (which is an allowed criterion for the GFF approximation of the dimension ceasing to be ``accurate"). Note that, at this value of $n$, the dimension of $\phi^2_{n,l}$ matches the $c$-scaling of the BTZ threshold $\Delta_{\tsc{BTZ}} > c/12$. 

We've so far only discussed double-trace operators. Let us consider a many-particle operator $:\!\!\phi^k\!\!:$ built from the light operator $\phi$. When $k \ll c$, the operator $:\!\!\phi^k\!\!:$ is a light primary and the GFF approximation for its dimension ($\Delta = k \Delta_\phi$) is accurate. As $k$ increases, so too does the magnitude of the anomalous dimension, because the gravitational self-interaction grows with particle number. When $k \gg c$, the free approximation has completely broken down, and a black hole has formed.

\textit{Energy eigenstate mixing.} The finite $c$ corrections to the GFF approximation also mix energy eigenstates. This starts to occur, as a rough estimate, for increasing operator dimension, when the anomalous dimensions are as large as the GFF approximation for the level spacing between primaries. At this point in the spectrum of conformal dimensions, we start to become ignorant of what the Hamiltonian's eigenvectors are. For the double-trace operators $\phi^2_{n,l}$, this occurs when~\eqref{eq:gamnl} is order one. This is a smaller value of $n$ than when $\gamma_{n,l} \approx \Delta_{n,l}$, showing that the point at which the GFF approximation ceases to be accurate depends on the data being approximated.

\textit{OPE coefficients.} There are also $1/c$ corrections to the GFF approximation of the OPE coefficients. 
There are some heavy primary operators for which we know effectively nothing about their OPE coefficients with light operators from the GFF approximation. 
Nonetheless, even for these heavy operators, we have statistical information about the OPE coefficients from the light primary data from symmetry and consistency conditions. For example, using crossing symmetry alone, the variance of the heavy-light-light (HLL) OPE coefficients is known~\cite{Pappadopulo:2012jk}, and equals 
\bne \overline{c_{\scriptscriptstyle LLH}^2}\sim \frac{\Delta_H^{2\Delta_L - 1}}{\rho_{\tsc{Cardy}}(\Delta_H)}, \quad \text{as }\Delta_H \to \infty . \label{eq:OPEav}\ene
The averaging in this formula is over all heavy operators (primaries and descendants) within a window of dimensions $[\Delta_a - \delta, \Delta_a +\delta]$, and how wide the window needs to be depends on whether the CFT is chaotic or integrable~\cite{collierUniversalDynamicsHeavy2020, belinRandomStatisticsOPE2021, anousOPEStatisticsHigherpoint2022}.
The statistics of higher moments of the OPE coefficients are approximately Gaussian, though there are also $e^{-S}$ suppressed non-Gaussianities (required by crossing symmetry~\cite{anousOPEStatisticsHigherpoint2022}). Note that~\eqref{eq:OPEav} is a $\Delta_H \to \infty$ asymptotic formula, and therefore not directly applicable to OPE coefficients of operators with dimensions in the small black hole range; at this point, we only wish to convey that there is some statistical knowledge of the UV data.

The takeaway is that, for a specified set of data (conformal dimension, couplings with light primaries, etc.), there are primaries whose data we know precisely, and primaries whose data we only know the statistical properties of.

\paragraph{Terra cognita and incognita.}
Let us now take a holographic CFT and partition the set of all the primary and descendant operators in the CFT into two subsets: those operators whose data are known (accessible to a low-energy observer) and the complement set. 

There are some issues with this bipartitioning. 
Firstly, ``known" is imprecise terminology, as all primaries have \textit{some} degree of uncertainty in their data, since even light primaries have $1/c$ corrections to their dimensions
and OPE coefficients 
(if they are not protected). A bipartition \textit{can} be precisely defined given a tolerance parameter for the size of the $1/c$ corrections to a specified set of data.
But such a tolerance parameter introduces dependence on an arbitrarily chosen quantity. 
Lastly, given such a bipartition, the separation between the two sets of operators is not sharp, as we do not expect large gaps in the size of corrections at the shared boundary of the sectors.

We will not address these issues further, because this bipartition of operators into those with known versus unknown data is only a heuristic. We are introducing it because it both motivates and will be useful in Sec.~\ref{sec:HowToCoarseGrain}, where we coarse-grain CFT states by integrating over unknown UV data. 

The primary states and descendants with known data span a Hilbert space $\cH_{\tsc L}$.
``$\mathrm{L}$'' denotes the fact that their data is determined by the conformal data of the \textit{light} single-trace primaries (see~\eqref{eq:GFFOPEc} and~\eqref{eq:GFFdim}). As discussed, not all states in $\cH_{\tsc L}$ are light, both because there are heavy descendants of light primaries, and because there are primaries dual to high-energy, stationary multi-particle states.
The primary states (and their descendants) whose associated data is unknowable from the light data span another Hilbert space $\cH_{\tsc H}$.
To overcome the $1/c$ suppression of corrections to the free approximation, all states in $\cH_{\tsc H}$ are necessarily heavy.
The bipartition of primaries and descendants splits the full Hilbert space into two:
\bne \cH = \cH_{\tsc L} \oplus \cH_{\tsc H}  .\label{eq:hdecomp} \ene
We call these the integrable and chaotic subspaces, though knowable/unknowable or accessible/inaccessible would also be appropriate names. Light/heavy would be appropriate were it not that there are some heavy states in $\cH_{\tsc L}$. Note that, as emphasised before, after projecting onto a microcanonical window in the small, unstable black hole mass range, most states in $\cH_{\tsc{H}}$ are \textit{not} black hole states.

Any descendant state is in the same chaotic or integrable subspace as the primary from which it descends, if we specify conformal dimension or OPE coefficients as the set of data whose uncertainty defines the bipartition. This is because conformal symmetry fixes one set of data in terms of the other. A primary and its descendants will not generally be in the same microcanonical window of dimensions, depending on the width of the window.

\paragraph{Particle-collision state.} 
Now we put this discussion in the context of our particle-collision state $\ket{\psi}$ given in~\eqref{eq:initial_state}. 
We showed in~\eqref{eq:psico} that the primary and descendant states with which $\ket{\psi}$ has the largest overlaps have dimensions centred on the CoM energy $\Delta \approx 2\omega$ and zero angular momentum. This $\omega$ is a tunable parameter, and we will consider three regimes. Note that since $\cH_{\tsc L}$ and $\cH_{\tsc H}$ are defined as the spans of two sets of energy eigenstates, time evolution cannot move the state $\ket{\psi}$ between the Hilbert subspaces.

When $\omega = O (1)$, there is insufficient energy to form a black hole. The primary and descendant states of dimension $\Delta = O(1)$ are not heavy enough to have large corrections to the free approximation, so they are all in the integrable sector $\cH_{\tsc L}$, thus so too is $\ket{\psi}$.  

When $\omega \gg c$, we are in the large AdS black hole regime. In the set of primary states with dimension $\Delta  \gg c$, there are always some in $\cH_{\tsc{L}}$, because, no matter how high the energy, there are always approximately stationary high-angular momentum multi-particle states in bulk. But we expect the majority of these super-heavy primary states, especially the low-spin ones which $\ket{\psi}$ has overlaps with, to be in $\cH_{\tsc H}$.

Lastly, we consider $\omega$ in the small, unstable black hole range, where $\omega$ scales with a non-zero power of $c$ less than one. Primary states cannot be dual to small AdS black hole microstates because one is stationary and the other is not. Primary states with $\Delta \approx \omega$ are heavy, and we expect low-spin ones to be in $\cH_{\tsc H}$, and therefore so too $\ket{\psi}$ (with non-zero but negligible overlaps with the energy eigenstates spanning $\cH_{\tsc{L}}$).
Our state $\ket{\psi}$ is initially an approximate two-particle state, by construction. This implies that, 
while none of the energy eigenstates in $\cH_{\tsc H}$ are approximate Fock states (by the definition of $\cH_{\tsc H}$), there must be non-stationary approximate Fock states in $\cH_{\tsc H}$. These are superpositions of non-Fock-like energy eigenstates with particular phases. Next, as time progresses, a small black hole horizon is formed in the bulk, and the particle interpretation of $\ket{\psi}$ breaks down. Later still, we again have an approximate Fock description once the black hole has evaporated into a gas of order $S_{\textrm{rad.}}$ weakly interacting Hawking quanta. The microcanonical dominance of radiation states implies that typical states in $\cH_{\tsc{H}}$ at this mean energy are approximate Fock states, though they are also highly complex linear superpositions of the energy eigenstates. The Hilbert subspace of unstable black hole microstates is inside of $\cH_{\tsc{H}}$, and it cannot be spanned by energy eigenstates. The above discussion suggests that $\cH_{\tsc H}$ contains both subspaces spanned by Fock and black hole states,
$ \cH_{\tsc H} \supset \cH_{\tsc {BH}} \oplus \cH_{\tsc {Fock}}$.
We expect this to hold at any sufficiently high energy to support black hole states, but in the small black hole mass range, we expect that any projector onto either $\cH_{\tsc {BH}}$ or $\cH_{\tsc {Fock}}$ will not commute with the Hamiltonian, because the primary and descendant states in $\cH_{\tsc H}$ cannot be either small black hole or non-stationary multi-particle states.

If we increase the impact parameter of the bulk particles in the pre-collision state~\eqref{eq:initial_state}, then this will increase the mean value of the spin in the boundary state. For sufficiently large impact parameter, $b \gg r_h(\omega)$, the particles will miss each other and not form a black hole, and we have a long-lived two-particle state.
The state will be high spin double-trace primary
rather than a high spin descendant of a low-spin double-trace primary: both the primary double-trace state $\phi^2_{n,l}\ket{0}$ and the descendant state $L_{-1}^l \phi^2_{n,0}\ket{0}$ have the same spin and dimension, at leading order in $c$, but $\phi^2_{n,l}\ket{0}$ describes two particles orbiting their mutual centre of mass, whereas $L_{-1}^l \phi^2_{n,0}\ket{0}$ describes two particles whose centre of mass is in motion but which are not orbiting each other.
From~\eqref{eq:gamnl}, we see that this will decrease the anomalous dimension for $d>2$, and so increase the overlap of our state~\eqref{eq:initial_state} with the energy eigenstates with free-particle descriptions. 
We expect the value of the angular momentum at the threshold of whether the two particles form a black hole or not to approximately match the value of the angular momenta of the energy eigenstates at the border between $\cH_{\tsc L}$ and $\cH_{\tsc H}$.

The takeaway of this subsection is that high-energy microcanonical windows of holographic CFTs have a natural but heuristic bipartition into two kinds of primary states: those in $\cH_{\tsc L}$ whose data (dimensions, OPE coefficients, wavefunction coefficients) are known, like the state dual to two trans-Planckian energy bulk particles orbiting each other, and those in $\cH_{\tsc H}$ whose data is known only statistically, because they overlap with black hole microstates. The uncertainty in the data in $\cH_{\tsc H}$ induces uncertainty in what the wavefunction coefficients are for the state~\eqref{eq:primd2} in the spin-dimension eigenbasis. This uncertainty naturally leads to a coarse-graining of exact states, which we will explore in Sec.~\ref{sec:HowToCoarseGrain}.

\subsection{Other ways to create an AdS black hole microstate} \label{sec:OtherWaysToSmash}
The construction we have focused on, using a pair of light operator insertions, is not the only way of constructing boundary states dual to small, unstable AdS black holes.
Here, for comparison, we non-exhaustively list other constructions.
\begin{enumerate}

\item 
\textit{Turning on a boundary source.} 
In~\cite{Bhattacharyya:2009uu}, the authors form black holes by turning on a non-normalisable mode at the boundary, corresponding to turning on a source in the boundary CFT. In their setup, there are two dimensionless parameters: the amplitude of the source and the length of time it is turned on (in AdS units), and the authors determine when small AdS black holes are formed in the parameter space. Note that collapse states engineered from the Euclidean path integral are time-reflection symmetric.

\item \textit{Inserting dust particles.} The collapse state of~\cite{Anous2016} is defined by inserting many primary operators around a circle
\bne \label{eq:dustV} \ket{\cV} = \lim_{n\to \infty} \frac{1}{\cN_n} \prod^n_{k=1} \psi (e_k, \overline{e}_k) \ket{0}, \qquad e_k := (1-\sigma)e^{2\pi i (k-1)/n}\, , \ene 
which is dual to inserting bulk particles which collapse to form a black hole. The operators are distributed evenly around the circle $|z| = 1-\sigma$, with $\sigma$ a UV regulator. 
The resulting bulk geometry is AdS-Vaidya. 

\item \textit{A pair of conical defects.} 
Another operator insertion approach is to insert two \textit{heavy} operators dual to conical defects that collide and form a black hole~\cite{banerjeeSignaturesBulkBlack2025}. For the OPE coefficients of these operators, see~\cite{Grabovsky:2024jwf, Abajian:2023bqv}.

\item \textit{Fermionic star.} The fermionic star state of~\cite{Arsiwalla:2010bt} is created by a composite boundary operator built from a light fermionic single-trace operator $\Psi$
\bne \Phi := \Psi \, \prod_i \del_i \Psi\, \prod_{i,j} \del_i \del_j \Psi \dots \ene
This is analogous to filling an atom's electron energy shells.
For sufficiently high operator dimension, the fermionic star will collapse to form a black hole.
The star is created with the insertion of a single local operator, in contrast to the other operator insertion constructions discussed so far.

\item \textit{Constrained random microstate.} 
For a microcanonical ensemble with an energy at which AdS black holes dominate the ensemble, a Haar-random state with components $\braket{E_i |\psi(0)} = U_{i1}$ will be a black hole microstate with high probability. 
This does not work for small, unstable black holes because they do not dominate the ensemble.
We can remedy this by picking a random state within the ensemble with the constraint that it has macroscopic properties consistent with it being a black hole. In Sec.~\ref{sec:detect_BH}, we will discuss one of these properties and how it distinguishes (also small, unstable) AdS black holes from radiation states.

\item \textit{AdS nonlinear instabilities.} 
Generic excitations of AdS of total energy above the black hole threshold are unstable to forming black holes because of the AdS confining potential~\cite{Bizon:2011gg, Dias:2011ss, diasNonlinearStabilityAsymptotically2012}, with an AdS-scale collapse time. For a review, see~\cite{Horowitz:2014hja}. Thus, we do not need fine-tuning in the boundary state to create a small AdS black hole; any high-energy state with an initially diffuse distribution of energy in the bulk suffices. However, this is not very practical, as we do not know the timescale for black hole formation. Moreover, generically, only a fraction of the matter will collapse into a black hole.
\end{enumerate}

\section{How to detect an AdS black hole} \label{sec:detect_BH}
To understand black hole evaporation from the CFT point of view, it is important to be able distinguish a black hole from other high entropy macrostates in the microcanonical ensemble, such as thermal gas, especially when the energy is low enough that the black hole is entropically subdominant.
The presence of quasi-normal modes is one way to detect the presence of a horizon \cite{Horowitz:1999jd}, or the fact that black hole out-of-time-ordered correlators (OTOCs) are maximally chaotic \cite{Shenker:2013pqa, Hernandez:2025seq}. 

We can also use boundary entanglement entropy as a probe.
As a small AdS black hole evaporates, it emits thermal Hawking radiation. It will evaporate away fully if it is in the mass range~\eqref{eq:mass_bound}, leaving behind a ball of thermal gas trapped by the AdS potential.
What happens to the state in the CFT? 
First of all, the bulk dual remains semiclassical throughout the process and does not become, for example, a superposition of geometric states.%
\footnote{
Typical states in the microcanonical ensemble cannot be superpositions of mass distributions like
\bne \ket{\psi} = \frac{1}{\sqrt{2}} (\ket{\text{BH}(x_1)} + \ket{\text{BH}(x_2)}). \ene
Such states have order $N^0$ variances in simple observables, such as the boundary stress tensor.  Such large variances are easily detectable. Also, in comparison, the typical variance of an operator in a microcanonical ensemble is
\bne \int d\mu_{\text{Haar}} (U) \text{Var}_{U\ket{\varphi}} (\cO)\approx \frac{\Tr (\cO^2)}{(e^S)^2} \lesssim e^{-S}. \label{eq:HaarVar}\ene
The Haar-averaged variance is exponentially suppressed, so only atypical states have order $N^0$ variances. Entropic suppression of variances in black hole microstates was also considered in~\cite{Balasubramanian:2007qv}, with similar conclusions.}
Geometric states can be distinguished for generic quantum states using boundary entanglement entropy, because they obey holographic entropy inequalities, such as monogamy of mutual information, which generic quantum states need not obey \cite{Hayden:2011ag}. 

From the bulk, forming a black hole means localising the available mass as much as possible. What could be a clear signature of this process from the CFT point of view? In this section, we will argue for why this is equivalent to minimising generic two-point functions in the CFT. Two-point functions are more sensitive probes of the state than one-point functions, because they are non-local and can probe the bulk IR region, whereas one-point functions are primarily sensitive to phenomena near the AdS UV boundary.

We will compare two-point functions of spacelike separated boundary probe operators between black holes and spherically symmetric balls of (backreacted) thermal gas in AdS, keeping the ADM mass fixed, first without and then with internal compact dimensions.
We assume that both the black hole and the thermal gas are centred on $r=0$ in global coordinates, and that the black hole is sufficiently heavy to be either long-lived or stable, such that both spacetimes are approximately stationary on AdS timescales. 

The AdS$_{d+1}$-Schwarzschild metric is
\begin{equation}
    ds^2 = \ell_{\text{AdS}}^2\left(- f(r)dt^2 + \frac{dr^2}{f(r)} + r^2 d\Omega_{d-1}^2\right), \quad f(r) = 1+ r^2 - \mu/r^{d-2}.
\end{equation}
The (dimensionless) $\mu$-parameter and the mass of the black hole are related through
\begin{equation}
    M \ell_{\text{AdS}} = \frac{(d-1)\Omega_{d-1}}{16 \pi G_N^{d+1}}\mu \ell_{\text{AdS}}^{d-1} .
    \label{eq: mass}
\end{equation}

For the thermal gas, we consider a perfect fluid ansatz:
\begin{equation}
    T^{\mu \nu} = (p+\rho)u^\mu u^\nu + p g^{\mu \nu}, \quad u^\mu = \frac{\delta^\mu_t }{\sqrt{-g_{tt}}}.
\end{equation}
The backreaction of the thermal gas depends on the equation of state $p = w \rho$. For thermal gases, the equation of state parameter $w$ lies between 0 and $1/d$, where $w=0$ describes a thermal gas made up of only cold dust particles and $w=1/d$ describes relativistic matter.
This stress tensor induces a backreaction on top of empty AdS. For the backreacted geometry, we take the spherically symmetric ansatz: 
\begin{equation}
    ds^2 = \ell_{\text{AdS}}^2\left(- e^{2\nu(r)}f(r)dt^2 + \frac{dr^2}{f(r)} + r^2 d\Omega_{d-1}^2\right), \quad f(r) = 1+r^2-m(r)/r^{d-2}.
    \label{eq: ansatz thermal gas}
\end{equation}
One can compute the stress tensor that corresponds to this background, using Einstein's equations
\begin{equation}
    T^{\mu \nu} = R^{\mu \nu} - \frac{R-2\Lambda}{2}g^{\mu \nu} , \quad \Lambda = -\frac{d(d-1)}{2\ell_{\text{AdS}}^2},
\end{equation}
and the energy density is related to the function $m(r)$ through
\begin{equation}
    \rho(r) = \frac{d-1}{2\ell_{\text{AdS}}^2}\frac{m'(r)}{r^{d-1}},
\end{equation}
such that positive energy density implies that $m(r)$ is a monotonic function: $m'(r) \geq 0$. Since at $r \to \infty$ we asymptote to the black hole metric, we have $m(r) \leq \mu$ and thus, for all $r$,
\begin{equation}
    g_{rr}^{\text{gas}} \leq g_{rr}^{\text{BH}}.
\end{equation}

\subsection{Two-point functions}
\label{sec: 2pt func}
Consider an asymptotically AdS$_{d+1}$ spacetime, a massive bulk scalar field $\phi$, and its boundary dual $\cO$:
\bne \cO(x) = \lim_{r \to \infty} r^{\Delta} \phi(x,r). \ene
Taking  the mass of $\phi$ to be in the range $1 \ll m_\phi  \ell_\text{AdS} \ll \ell_\text{AdS}/ \ell_\text{p}$, the two-point function can be computed in the geodesic approximation as
\bne \langle \cO(x_1) \cO (x_2) \rangle \approx e^{-m_{\phi} L(x_{1},x_2)} .\ene
$L$ is the length of the minimal geodesic that connects the two points on the boundary. 
We assume that the bulk state is approximately static, such that the bulk geodesic is a curve on a global time slice.
For this to be true, the black hole needs to be heavy enough to be thermodynamically stable, or long-lived in AdS units if unstable.
Using the rotational symmetry shared by both the ball of gas and AdS-Schwarzschild, we can reduce the calculation of geodesic lengths to a two-dimensional problem, with $(r,\theta)$-coordinates.
On this two-dimensional hypersurface through which the geodesic travels, the induced metric reads
\bne ds^2 = \frac{dr^2}{f(r)} + r^2 d\theta^2. \label{eq:emblf} \ene
Consider two emblackening factors $f_{\text{BH}}$ and $f_{\text{gas}}$, which describe the backreaction of a black hole and a ball of thermal gas with equal centre of mass.
Let us denote the curve of the black hole geodesic as $\gamma_{\tsc{BH}} =(R_{\tsc{BH}}(\lambda),\Theta_{\tsc{BH}}(\lambda))$ and the geodesic curve in the thermal gas geometry as $\gamma_{\text{gas}} = ( R_{\text{gas}}(\lambda),\Theta_{\text{gas}}(\lambda))$. Define $L_A (\gamma_B)$ to be the length of a curve $\gamma_{B}$ as measured by metric~\eqref{eq:emblf} using emblackening factor $f_A$, i.e.
\bne 
L_{A} (\gamma_{B}) 
:= \int d\lambda \sqrt{\frac{1}{f_{A}(R
_{B})} \dot{R}_{B}^2 + R^2_{B} \dot{\Theta}_{B} ^2} \,.\label{eq:Larg1}
\ene
In the previous section, it was found that $f_{\text{gas}} \geq f_{\tsc{BH}}$ for all $r$. Therefore, $L_{\tsc{BH}}(\gamma_{\tsc{BH}}) \geq L_{\text{gas}} (\gamma_{\tsc{BH}})$. 
We also have $L_{\text{gas}} (\gamma_{\tsc{BH}}) \geq L_{\text{gas}} (\gamma_{\text{gas}})$ because, from its definition, the minimal curve in the thermal gas geometry is the geodesic $\gamma_{\text{gas}}$. So we have that $L_{\tsc{BH}}(\gamma_{\tsc{BH}}) \geq L_{\text{gas}} (\gamma_{\text{gas}})$. 

Since $L(\theta_1,\theta_2)$, the geodesic length between two points $\theta_1$ and $\theta_2$ anchored on the boundary, is larger for the black hole background than for the thermal gas background, 
the boundary two-point function $\langle \cO(\theta_1) \cO(\theta_2)\rangle \approx e^{-m_\phi L(\theta_1,\theta_2)}$ is smallest for the black hole background. This agrees with our supposition that boundary two-point functions are minimised when energy in the bulk is concentrated as much as possible. 

\subsection{Ten-dimensional black holes}
\label{sec: 10dBH}

We now generalise to asymptotically AdS$_{d+1} \times \mathcal M^{D-d}$ spacetimes.%
\footnote{One reason that it is important to consider such spacetime is that all supersymmetric top-down realisations of AdS/CFT have internal bulk dimensions, as there are no interacting SCFTs in $d>6$, and the critical dimensions of superstring theory and M-theory are ten and eleven-dimensional, respectively. The results of the previous subsection are still important, both from the bottom-up perspective and because the absence of top-down realisations of AdS/CFT without internal compact dimensions may be a lamp-post effect of superstring theories, amongst other potential theories of quantum gravity.}
For fixed ADM mass, one solution is the uplifted (AdS$_{d+1}$-Schwarzschild)$\times \mathcal{M}^{D-d}$ geometry, which we refer to as the delocalised black hole. But, for sufficiently low mass, one expects the entropically favoured black hole to be localised on the compact space (and have a horizon radius that is much smaller than $\ell_{\textrm AdS}$). We refer to this as the localised black hole. The uplifted solution (AdS$_{d+1}$-Schwarzschild)$\times \mathcal{M}^{D-d}$ dynamically localises via Gregory-Laflamme~\cite{BanDou98, Gregory:1993vy}. For AdS$_5\times S^5$, it was shown numerically in \cite{DiaSan16} that the Gregory-Laflamme instability sets in for black holes with $M \ell_{\text{AdS}}/N^2 \lesssim 0.173$ and the localised black holes dominate entropically over the smeared ones from $M \ell_{\text{AdS}}/N^2 \lesssim 0.225$.

As mentioned before, for the thermal gas, it is entropically favourable to spread out over the compact internal dimensions, in contrast to the black hole. The cost in energy is only from the weak gravitational self-interaction (in contrast to spreading out in the AdS radial direction). So, we will assume that the thermal gas is spread out over the compact internal dimensions.

There is a generalisation of the geodesic approximation of boundary two-point funtions for bulk spacetimes with internal dimensions.
Consider an asymptotically AdS$_{d+1} \times S^{n}$ spacetime.%
\footnote{The following argument trivially generalises to AdS times any compact internal manifold. 
}
Let $(\vec x,r)$ be the coordinates on the AdS$_{d+1}$ factor, with $r$ the holographic radial direction, and $\Omega$ the coordinates on the $S^{n}$ factor. The boundary singlet operator $\cO$ is dual to the zero mode of bulk field $\phi$:
\bne \cO(\vec x) = \lim_{r \to \infty} r^{\Delta} \int d^{n} \Omega \;\phi(\vec x,r, \Omega) .\ene
We are interested in the two-point function
\bne \langle \cO(\vec x_1) \cO (\vec x_2) \rangle \approx \int d^{n}\Omega_1 \int d^{n}\Omega_2 \, e^{-m_{\phi} L(\vec x_{1}, \vec x_2, \Omega_{1}, \Omega_2)} .\label{eq:singtpfunc}\ene
By another saddlepoint approximation,
\bne \log \langle \cO(\vec x_1) \cO (\vec x_2) \rangle \approx  -m_{\phi} \tilde L(\vec x_1,\vec x_2),\ene
where
\bne \tilde L(\vec x_1,\vec x_2)
:= \min_{\Omega_1, \Omega_2} L(\vec x_{1},\vec x_2, \Omega_{1}, \Omega_2)  \ene
So, still only one geodesic is important for approximating the two-point function~\eqref{eq:singtpfunc}, even if the bulk has internal compact dimensions.

Now we compare boundary-anchored geodesics for a small black hole and a ball of gas of the same total mass. 
We take the mass to be such that the black hole is small, long-lived and localised on the internal space, i.e., for AdS$_5 \times S^5$, in the range $N^{4/9} \ll M \ell_{\text{AdS}} \ll N^{2}$ (see Fig.~\ref{fig:hierarchy}).
Let $g$ be the unperturbed AdS$\times \mathcal M$ metric and $\delta g$ the metric perturbation from either the black hole or the gas. 
Since the length of any spatial curve $X (\lambda)$ is
\bne L = \int d\lambda \sqrt{(g_{\mu\nu} + \delta g_{\mu\nu})\dot X^\mu \dot X^\nu}, \ene
it suffices to show that the spatial components of $\delta g$ are larger for one metric perturbation than another. 
The minimal length geodesic typically stays away from regions of large metric perturbation, such as near-horizon regions.
With this assumption, we can work in the linearised regime. 

While for angular geodesic motion on the AdS factor there is competition between the near-horizon divergence and the large-$r$ hyperbolic growth of the spatial metric in global coordinates, for the $S^n$ factor, there is only the near-horizon divergence and thus no reason for the minimal geodesic not to sit as far away from the black hole as possible. 
So, taking the black hole to be localised on the north pole of the sphere $S^n$, the minimal geodesic will stay at the south pole.

\begin{figure}
    \centering
\includegraphics[width=0.8\linewidth]{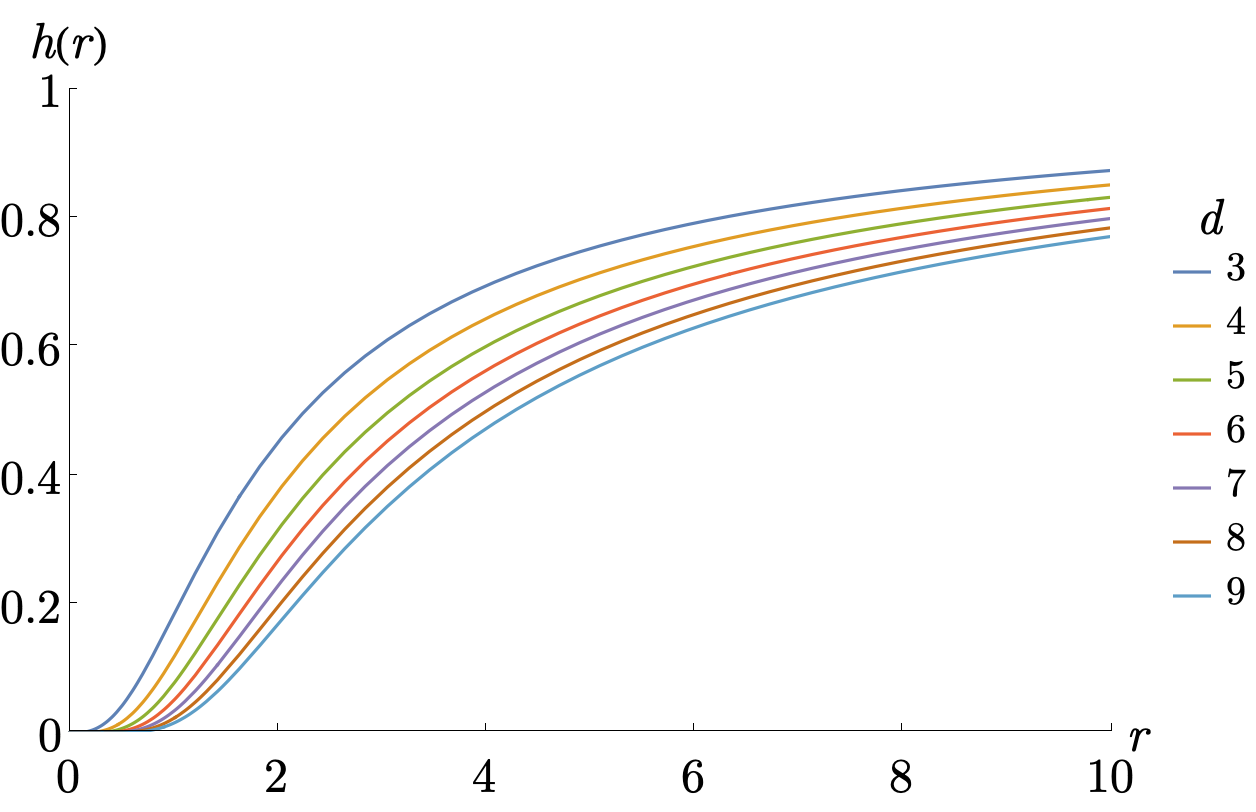}
    \caption{The mass function $h(r)$ for the thermal gas from \eqref{eq: therm sol}, plotted for $3 \leq d \leq 9$. In all dimensions, the mass function equals zero at $r=0$ and one as $r\to \infty$. The mass function at fixed radius $r$  decreases as the dimension $d$ increases.}
    \label{fig:thermalplot}
\end{figure}

For the ball of thermal gas, the pressure profile is related to the $m(r)$ and $\nu(r)$ functions in the metric ansatz~\eqref{eq: ansatz thermal gas}
as
\begin{equation}
\begin{split}
p(r)\ell_{\text{AdS}}^2 &= \frac{(d-1)\nu'(r) (1+r^2-m(r)/r^{d-2})}{r} - \frac{d-1}{2}\frac{m'(r)}{r^{d-1}}, \\
\rightarrow \quad \nu'(r)& = \frac{(1+w)m'(r)}{2r^{d-2}(1+r^2-m(r)/r^{d-2})}.
\end{split}
\end{equation}
For the equation of state $w=-1$, $\nu(r)$ has to be a constant which can be reabsorbed in the definition of $t$. For $w \geq -1$, $\nu(r)$ is also a monotonic function: $\nu'(r) \geq 0$. The coordinate $t$ can be defined such that as $r \to \infty$, $\nu(r) \to 0$. This means that at finite $r$,
\begin{equation}
    e^{2\nu(r)} \leq 1. 
\end{equation}
The backreacted thermal gas geometry thus has the properties
\begin{equation}
    \lim_{r \to \infty} m(r) = \mu, \quad m'(r) \geq 0, \quad  \lim_{r \to \infty} \nu(r) = 0, \quad \nu'(r) \geq 0.
\end{equation}
The last equation to solve is the conservation equation, which reads
\begin{equation}
\begin{split}
&-2 r w \left(r^{d+2}+r^d-r^2 m(r)\right) m''(r)-r^3 w (w+1) m'(r)^2\\
&+m'(r) \left(2 r^d \left(r^2 ((d-2) w-1)+(d-1) w\right)-r^2 (3 d w+d-4 w-2)
   m(r)\right)= 0.
   \end{split}
\end{equation}
This conservation equation is not analytically solvable, though it has been studied numerically~\cite{Vaganov}. 
Let us work in the small AdS black hole mass range, $\mu\ll 1$, as then the conservation equation can be solved perturbatively. 
We define the normalised mass function as $h(r) = m(r)/\mu$. 
Imposing the boundary conditions $h(0) = 0$ and $h(\infty) = 1$ gives
\begin{equation}
   h(r)= \frac{r^d \Gamma \left(\frac{w+1}{2 w}\right) \, _2F_1\left(\frac{d}{2},\frac{w+1}{2 w};\frac{d}{2}+1;-r^2\right)}{\Gamma \left(\frac{d}{2}+1\right) \Gamma
   \left(\frac{-d w+w+1}{2 w}\right)}\,, \quad \nu(r) =-\frac{\mu  w d \left(r^2+1\right)^{-\frac{w+1}{2 w}} \Gamma \left(\frac{w+1}{2 w}\right)}{2 \Gamma \left(\frac{d}{2}+1\right) \Gamma \left(\frac{-d w+w+1}{2 w}\right)} + \mathcal{O}(\mu^2).
   \label{eq: therm sol}
\end{equation}
The normalised mass function $h(r)$ for the radiation equation of state $w=1/d$ is shown in Fig. \ref{fig:thermalplot}. 

The backreaction of a small, localised, $D$-dim black hole on AdS$_{d+1} \times S^{D-d}$ is a complicated geometry, as can be seen from the asymptotic behaviour found numerically in \cite{DiaSan16}. 
However, since the minimal length boundary-anchored geodesic will stay at the south pole of the sphere if the black hole is at the north pole, this reduces the dimensionality of the problem. We just need the $(d+1)$-dim induced metric at the south pole.
For this, we take the same metric ansatz~\eqref{eq: ansatz thermal gas} as for the thermal gas, with an effective mass function $m_{\tsc{BH}}(r)$:
\begin{equation}
    f(r) = 1 + r^2 - \frac{m_{\tsc{BH}}(r)}{r^{d-2}}.
\end{equation} 

At small radii, $r \ll 1$, the thermal gas mass function grows as $r^d$, from \eqref{eq: therm sol}. For the localised black hole, at small radii and on the sphere's south pole, the emblackening factor is finite. It's expected to deviate from the empty AdS value, meaning that the effective mass function grows as $r^{d-2}$. The effective mass function is thus larger for the small black hole than for the smeared thermal gas for $r \ll 1$. 

The higher spherical harmonics of the localised black hole's metric perturbation fall off with radial distance, due to their KK masses. The zero mode of the metric perturbation is the same as for the delocalised black hole solution with the same ADM mass. So, for $r \gg 1$, 
\begin{equation}
     m_{\tsc{BH}}(r) \approx \mu.
     \label{eq: SBH O1}
\end{equation}
In contrast, the value of the thermal gas' normalised mass function $h(r)$ approaches one slowly, as can be seen in Fig. \ref{fig:thermalplot}, indicating that $h_{\tsc{BH}}(r) > h_{\textrm{gas}} (r)$ for $r \gg 1$.

We have shown that $h_{\tsc{BH}}(r) > h_{\textrm{gas}} (r)$ for both $r \gg 1$ and $r \ll 1$. Let us assume that it holds for all $r$. 
This is sufficient to guarantee that the minimal geodesic is longer for the small, localised black hole in $AdS_{d+1}\times S^{D-d}$ than for the thermal gas with the same ADM mass. The minimal geodesic in the delocalised black hole geometry is larger than both the thermal gas and the localised black hole. Therefore, we find a hierarchy in the magnitude of generic two-point functions in the different (approximately stationary) backgrounds in the microcanonical ensemble: the two-point function in the delocalised black hole geometry is the smallest, it is largest for the thermal gas, and we expect the two-point function in the localised black hole geometry to sit in between these values. 

\subsection{Two-point functions and (partial) deconfinement}
\label{sec: deconfinement}
Black hole formation is dual to deconfinement in holographic gauge theories \cite{Witten98,AhaGub00}. More recently, it has been argued that the CFT states dual to small, localised black holes in asymptotically AdS$\times \mathcal M$ spacetimes can be described by partial deconfinement \cite{Ber06,AspBer09,HanMal17,Ber18,HanIsh19,HanJev19}. In this section, we will study whether the results of the previous section can be understood in the language of deconfinement. We found that, in asymptotically AdS spacetimes, there is a hierarchy in the magnitude of generic simple boundary two-point functions in the microcanonical ensemble. 
Does this same hierarchy of generic two-point functions also exist in deconfined, confined and partially deconfined states? 

\subsubsection{Partial deconfinement review}

First, we review some of the notions of confinement and (partial) deconfinement in a Gaussian matrix model, following the discussion of \cite{HanJev19}. This is a simple model, but it is straightforward to generalise to other theories like $\cN=4$ SYM. 
Consider the Hamiltonian of a Gaussian matrix model with a gauge-singlet constraint:
\begin{equation}
    \hat{H} = \frac{1}{2}\sum_{I=1}^D \Tr(\hat{X}_I^2 + \hat{P}_I^2), \quad \hat{X}_{I,ij} = \sum_{\alpha=1}^{N^2-1} \hat{X}_{I\alpha} \tau^\alpha_{ij},\quad \hat{P}_{I,ij} = \sum_\alpha \hat{P}_{I\alpha} \tau^\alpha_{ij}.
    \label{eq: toy model Gaussian}
\end{equation}
$I$ is a spacetime index taking values in $\{0,\dots,D\}$, the matrices are $N\times N$, and $\tau^\alpha$ are the SU$(N)$ generators with $\Tr(\tau^\alpha \tau^\beta) = \delta^{\alpha \beta}$. We impose canonical commutation relations $[\hat{X}_{I\alpha},\hat{P}_{J\beta}] = i \delta_{IJ} \delta_{\alpha \beta}$. We can define creation and annihilation operators as
\begin{equation}
    \hat{A}_{I\alpha} = \frac{\hat{X}_{I\alpha}+ i \hat{P}_{I\alpha}}{\sqrt{2}}, \quad \hat{A}_{I\alpha}^\dagger = \frac{\hat{X}_{I\alpha} - i \hat{P}_{I\alpha}}{\sqrt{2}}, \quad [\hat{A}_{I\alpha},\hat{A}_{J\beta}^\dagger] = \delta_{IJ}\delta_{\alpha \beta}.
\end{equation}
SU$(N)$-invariant Fock-states are created by acting with creation operators on the vacuum and taking traces, e.g.\footnote{For heavy operators with $O(N)$ energy, different trace structures are not automatically orthogonal in the large $N$ limit due to trace relations. Therefore, the operators that create orthogonal microstates are not just the different trace structures but the Schur polynomials, which are generally complicated linear combinations of different trace structures \cite{KocKim24}. 
}
\begin{equation}
    \Tr(\hat{A}_I^\dagger \hat{A}_J^\dagger \hat{A}_K^\dagger \dots )\ket{0}.
\end{equation}
where $\ket{0}$ is the vacuum state $\hat{A}_I \ket{0} = 0$.
One can also create SU$(M)$ invariant Fock-states (with $M<N$) by taking a subset of the generators that together generate SU$(M)$ and exciting only with these generators:
\begin{equation}
    \Tr_{\{\alpha_1, \dots \alpha_{M^2-1}\}}(\hat{A}_I^\dagger \hat{A}_J^\dagger \hat{A}_K^\dagger \dots )\ket{0}.
    \label{eq: SU(M) state}
\end{equation}
Here the notation $\Tr_{\{\alpha_1, \dots \alpha_{M^2-1}\}}$ means that we only include the $M^2-1$ generators of the subgroup SU$(M)\subset$ SU$(N)$.

One can gauge fix, as in \cite{AzeHan09}, to a gauge that diagonalises as many $X_I$ as simultaneously possible. One of the interpretations of the matrix entries is the D-brane effective theory picture \cite{Witten96}. Here, the diagonal entries $X_I^{ii}$ describe the position of the $i$-th D0-brane in $\mathds{R}^9$, and the off-diagonal entries $X_I^{ij}$ describe how many open strings are excited between the $i$-th and $j$-th D0-brane. 
The black hole is a bound state of many D-branes and open strings, and even in the gauge where the matrices $X_I$ are as simultaneously diagonal as possible, states that are dual to these black holes form large non-commutative blocks. Large black holes in AdS are conjectured to be dual to these deconfined states, where the $X_I$ matrices have almost all $N^2$ entries excited. These states are created by acting on the vacuum with a few long traces of $O(N^2)$ creation operators. On the other hand, states that are dual to thermal gas are created by acting with a few short traces, which describe only a bunch of D-branes connected by open strings. These states are called confined.
Lastly, there is a notion of partial deconfinement, which is conjectured to describe small black holes in the dual picture. In a partially deconfined state, the $X_I$ are not fully non-commutative but have an $M \times M$ block in them, with $M<N$. These states are created by acting with a few long traces, of length $O(M^2)$, but only including the creation operators of the $M^2-1$ generators that together generate SU$(M)$, as in \eqref{eq: SU(M) state}. Individual particles are described by small blocks of $O(1)$ size.

\subsubsection{Two-point functions}

\paragraph{\boldmath$N=3$.} Now that we have reviewed the basics of the partial deconfinement literature, we can study two-point functions in the simplest toy example where we can have some features of confined and (partially) deconfined states. This is the free Gaussian model \eqref{eq: toy model Gaussian} with $N=3$. We have $D>1$, such that the $X_I$ are not simultaneously diagonalizable, but we restrict here to states created by only one of the creation operators $\hat{A}^\dagger_I$, such that we can drop the index $I$.
We have
\begin{equation}
    \hat{A}  = \sum_{\alpha=1}^8 \hat{A}_\alpha \tau^\alpha = \begin{pmatrix}
        \hat{A}_3 + \frac{1}{\sqrt{3}}\hat{A}_8 & \hat{A}_1 - i \hat{A}_2 & \hat{A}_4 - i \hat{A}_5 \\
        \hat{A}_1 + i \hat{A}_2 & -\hat{A}_3 + \frac{1}{\sqrt{3}}\hat{A}_8 & \hat{A}_6 - i \hat{A}_7 \\ \hat{A}_4 + i \hat{A}_5 & \hat{A}_6 + i \hat{A}_7 & - \frac{2}{\sqrt{3}}\hat{A}_8
    \end{pmatrix}, \quad [\hat{A}_\alpha, \hat{A}_\beta^\dagger] = \delta_{\alpha \beta}.
\end{equation}
Let us first compare a deconfined state to a confined state with four creation operators:
\begin{equation}
    \begin{split}
    \ket{\psi_{\text{deconfined}}} &= \cN^{-1} \Tr((\hat{A}^\dagger)^4)\ket{0},\quad \ket{\psi_{\text{confined}}}  = \frac{(\hat{A}_3^\dagger)^2(\hat{A}_8^\dagger)^2}{2}\ket{0}.
    \end{split}
\end{equation}
When expanded out, the deconfined state has 36 different terms. The confined state is created by acting only with the diagonal generators.%
\footnote{Note that the two states are not orthogonal: they have an overlap $\braket{\psi_{\text{deconfined}}|\psi_{\text{confined}}} = \frac{1}{2\sqrt{10}}$. However, for heavier states, one expects this overlap to decrease, as the number of terms in the deconfined state will grow quickly.}  
Probe operators are created by a finite number of short traces, which describe in the dual picture a small collection of D-branes connected by open strings. In this SU(3) example, probe operators are built up out of only the diagonal generators $\hat{A}_{i}$ and $\hat{A}_{i}^\dagger$ for $i=3,8$.\footnote{In the case of large $N$, also a few nearly diagonal generators can be included in the probe operators.} An example of such a (Hermitian) operator is 
\begin{equation}
    \mathcal{O} =\, :\!\left(\hat{A}_3^\dagger + \hat{A}_3\right)\left(\hat{A}_8^\dagger + \hat{A}_8\right)\!:
\end{equation}
As this example illustrates, generic probe operators depend only on the number operators $\hat{N}_i := \hat{A}_i^\dagger \hat{A}_i$ of the diagonal generators.\footnote{In this toy example, we are ignoring the spatial dependence of the operator $\mathcal{O}$. When incorporating this correctly, the small-distance behaviour of the two-point function should be the same for different states. Here we are studying the $s$-wave two-point function, which should encode the long-distance correlators.}
We compute the expectation value of the number operators $\hat{N}_i$ for $i = 3,8$ in both of these states and find
\begin{equation}
\langle \hat{N}_i \rangle_{\text{deconfined}} = \frac{1}{2}, \quad \langle \hat{N}_i \rangle_{\text{confined}} = 2, \quad i = 3,8.
\end{equation}
We also have partially deconfined states for SU$(3)$. These states are obtained by only acting with the creation operators $\{\hat{A}_1^\dagger,\hat{A}_2^\dagger,\hat{A}_3^\dagger\}$, as these are the three generators of SU(2). For example,
\begin{equation}
    \ket{\psi_{\text{partially deconfined}}} = \cN^{-1}\Tr_{\{1,2,3\}}((\hat{A}^\dagger)^4)\ket{0} = \frac{\left((\hat{A}_1^\dagger)^2+(\hat{A}_2^\dagger)^2+(\hat{A}_3^\dagger)^2\right)^2}{2\sqrt{30}}\ket{0}.
\end{equation}
In contrast to the confined and deconfined states, the expectation values of the number operators for $\hat{A}_3$ and $\hat{A}_8$ in the partially confined state are not equal:
\begin{equation}
    \langle \hat{N}_3 \rangle_{\text{partially deconfined}} = \frac{4}{3}, \quad \langle \hat{N}_8 \rangle_{\text{partially deconfined}} = 0.
\end{equation}
Now we compare the correlators of typical, simple probe operators in these states. Such operators are sums of products of the number operators $\hat{N}_3$ and $\hat{N}_8$. 
Heuristically, correlators of typical operators depend less on the expectation values of the individual number operators than on their average:
\begin{equation}
\begin{split}
    \frac{1}{2}\langle \hat{N}_3+\hat{N}_8 \rangle_{\text{partially deconfined}}=\frac{2}{3}.
\end{split}
\end{equation}
In this toy model and with this heuristic, the confined state will typically have the largest correlators, followed by the partially deconfined state, then the deconfined state. This matches our expectations from the bulk: thermal gas gives rise to the largest two-point functions, followed by the small localised black holes, then delocalised black holes smeared over the internal space.

\paragraph{Large \boldmath{$N$}.} Moving on, let us consider large $N$ and states that are created by acting with $E$ creation operators on the vacuum.
Small probe operators are created by a few short traces, meaning that they are built up of almost only the diagonal generators. We will thus compute the expectation value of only nearly-diagonal number operators $\hat{N}_i := \hat{A}_i^\dagger \hat{A}_i$, of which there are approximately $N$. 
Deconfined states are created by acting with a few long traces on the vacuum, and the $E$ creation operators are approximately equally distributed among the $N^2-1$ generators. The expectation values of the number operators in the deconfined state are
\begin{equation}
\langle \hat{N}_i \rangle_{\text{deconfined}} \approx \frac{E}{N^2}.
\end{equation}
In confined states, the $E$ creation operators are also equally distributed, but now only over the  $O(N)$ (nearly) diagonal generators. In a typical confined state, the expectation value of the number operators of the nearly-diagonal generators are
\begin{equation}
\langle \hat{N}_i \rangle_{\text{confined}} \approx \frac{E}{N}.
\end{equation}
The number operators for the other, non-diagonal generators have zero expectation value. 

A partially deconfined state is created by a few long traces acting on the vacuum but restricted to $M^2-1$ generators of SU($M$) with $M<N$. 
Of these generators, approximately $M$ are nearly-diagonal and thus are allowed constituents of probe operators. 
The expectation values for the corresponding approximately $M$ number operators are
\begin{equation}
\langle \hat{N}_i\rangle_{\text{partially deconfined}}\approx \frac{E}{M^2}.
\end{equation}
The other $N^2-M^2$ generators of SU$(N)$ have zero number operator expectation value in the partially deconfined state. 
The probe operators are built up out of all of the approximately $N$ nearly-diagonal generators, not only the approximately $M$ that are excited in the partially deconfined state. 
As for the $N=3$ case, for estimating and comparing correlators of generic probes between states, what is more appropriate to compare is not the individual number operator expectation values but rather their average:
\begin{equation}
\frac{1}{N}\sum_i\langle \hat{N}_i\rangle_{\text{partially deconfined}} \approx \frac{E}{NM}.
\end{equation}
Since $N>M>1$, the two-point function of generic probe operators will be largest for the confined state, then for the partially deconfined state and smallest for the completely deconfined state. This again matches the bulk expectation.

To conclude, in this section, we have argued that generic two-point functions are a useful diagnostic for whether a state at fixed energy is a thermal gas state or a black hole state. 
In the bulk, the energy density becomes more spatially concentrated during the collapse procress, and then less concentrated during evaporation.
We have argued that the different points in this process can be distinguished by the magnitude of the boundary two-point functions.
For large AdS-Schwarzschild black holes, the minimal lengths of boundary-anchored geodesics are larger than for the backreacted thermal gas geometry with the same asymptotic mass, meaning that the corresponding two-point functions will be smaller. This is unaffected by extra compact internal dimensions. However, these compact dimensions cause small black holes to localise on the internal space, and we have argued that the minimal geodesic length in this geometry is still larger than for the thermal gas, but smaller than in the smeared black hole background. 
From the CFT side, we also found evidence for smaller two-point functions in states dual to black holes. In holographic gauge theories, deconfined states are dual to large AdS black holes and confined states to thermal gas. Small and localised black holes are conjectured to be dual to partially deconfined states. In a free matrix model toy example, we showed how the hierarchy of the two-point functions in these states with varying levels of deconfinement matches bulk expectations.

The toy model we have used is a free theory, unlike the strongly coupled theories dual to semiclassical Einstein gravity. At zero coupling, confined and (partially) deconfined states already exist in the model and we can compute things analytically. One of the features that is expected to disappear at finite coupling is the $1/N$ suppression of the two-point function in the deconfined state compared to the confined state. 
It is intriguing to see that we get the same hierarchy of two-point function as expected from the bulk side, despite our toy model being free.  It would be interesting to see if this hierarchy persists at strong coupling.

There are other probes for black hole formation one can consider, besides correlation functions of local operators.
Polyakov loops are an order parameter for (de-)confinement \cite{Polyakov78}. 
Since we are studying the black hole and thermal gas in the microcanonical ensemble, an immediate connection with the Polyakov loop is difficult to make. The Polyakov loop is computed in Euclidean time and thus by construction in the canonical ensemble. While it could be interesting to see whether, in this ensemble, a nonzero Polyakov loop would also correspond to smaller simple two-point functions, the corresponding computation to do on the bulk side would be to compare the black hole and thermal gas at equal temperatures and not at equal mass.

\section{How to coarse-grain an AdS black hole} \label{sec:HowToCoarseGrain}

We wish to connect the duality between semiclassical gravity and ensemble averaging with the black hole information problem in the AdS context.

\begin{figure}
    \centering
    \includegraphics[width=0.95\linewidth]{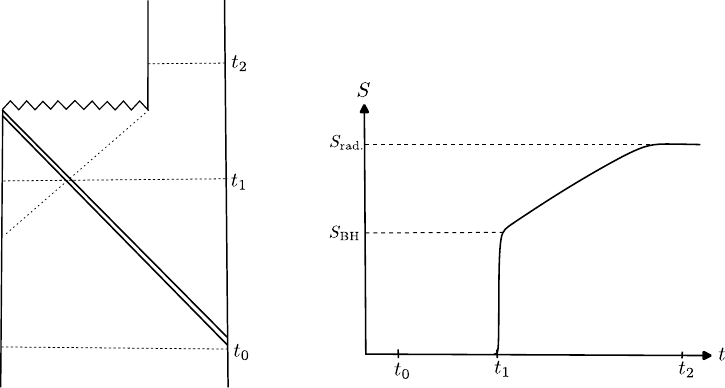}
    \caption{Left: The creation of a small, unstable AdS black hole from a pure initial state, such as a collapsing star or two colliding particles. Semiclassically, the black hole emits Hawking radiation and eventually evaporates away to leave behind a mixed state of thermal radiation. Right: The expected entropy curve of the semiclassical bulk state reduced to the region outside the horizon. It rises rapidly when the horizon forms, and continues to increase while the Hawking radiation is emitted.}
    \label{fig:IntroFig}
\end{figure}

Let $\rho(t)$ be a pure, exact state in a holographic CFT dual to the formation of a large AdS black hole, or the formation and evaporation of a small, unstable AdS black hole, like in Fig.~\ref{fig:IntroFig}.
For a fixed bulk timeslice that intersects the boundary at time $t$, let $\rho_{\rm bulk}(t)$ be the bulk semiclassical approximation of the exact state $\rho(t)$, reduced to the spatial region exterior to any black hole horizon. 
The semiclassical state $\rho_{\rm bulk}$ will become mixed with time, when the horizon forms and Hawking radiation is produced. Therefore $\rho_{\rm bulk} \neq \rho$, because the latter stays pure forever due to boundary unitarity.
The state $\rho_{\textrm{bulk}}$ is naturally interpreted as a coarse-graining of the exact state, because the semiclassical approximation in the path integral formulation neglects off-shell, non-peturbative, and UV physics. 
The key question of this section is: what is the boundary dual of $\rho_{\textrm{bulk}}$? What coarse-graining of the exact state ($\rho \mapsto \overline{\rho}$) will make it match $\rho_{\textrm{bulk}}$? 

In this section, we will explore different coarse-graining maps of exact time-evolved density matrices:
\bne \rho(t) \mapsto \overline{\rho(t)}. \ene
All of these will be quantum channels: completely positive, trace-preserving (CPTP) linear maps.%
\footnote{Except for the maximal ignorance principle prescription discussed in Sec.~\ref{sec:ensembles_of_density_matrices}, which is CPTP but not linear.
}
Some but not all will involve averaging over ensembles, such as a set of CFT data, Hamiltonians, or density matrices. 

When we time-evolve a generic atypical initial state in a chaotic system, it thermalises, meaning that it becomes indistinguishable from other typical states in the ensemble, if we only use simple macroscopic observables. One definition of a coarse-grained state is the maximum entropy density matrix consistent with those observables.  

Our central task is to find coarse-graining maps for which $\overline{\rho}$ matches the semiclassical bulk state $\rho_{\textrm{bulk}}$. We will focus on matching the mixedness of the states, as this is at the core of the tension with unitarity.
In the semiclassical bulk description for the time evolution of our state~\eqref{eq:initial_state}, if the two particles collide to form a small AdS black hole, it will eventually evaporate to form a gas of mixed state Hawking radiation,%
\footnote{The final stage of the black hole's evaporation, when the horizon is Planckian, is outside the regime of validity of the gravitational EFT. Our only assumption is that the final stage does not purify the earlier radiation, i.e. there is no high-entropy remnant.}
as depicted in Fig.~\ref{fig:IntroFig}. In this figure, we show the expected bulk entropy based on the existence of an apparent horizon. It combines the entropy associated with the apparent horizon with that of a thermal gas in the bulk. However, one can also imagine other bulk coarse-graining maps based on different bulk quantities and which capture thermalisation in a more continuous way, as expected from e.g. \cite{BalBer112,BalBer11}, 
For a given coarse-graining map, if we do have a faithful match $\overline{\rho} \approx \rho_{\textrm{bulk}}$, then $\overline{\rho(t)}$ should evolve to a mixed state on a timescale set by the formation and evaporation of the black hole. 

We will quantify the time scale at which $\overline{\rho(t)}$ becomes mixed by calculating its purity.
Defining $N := \textrm{dim}(\cH)$, then, for a general density matrix, the purity $\Tr (\rho^2)$ equals $1$ iff $\rho$ is pure, $1/N$ iff it is maximally mixed, and it is bounded between those values.
We will use two definitions of mixing times. 
First, the time at which the state is no longer approximately pure:
\bne t_{\textrm{mix}} := \inf_t \{ \, t \; : \; 1-\Tr\left(\overline{\rho(t)}^2\right) > \epsilon \} . 
\label{eq:tmix}\ene
Second, when $\overline{\rho(t)}$ is approximately maximally mixed, as measured by its Hilbert-Schmidt distance from the maximally mixed state:%
\footnote{The Hilbert-Schmidt (2-norm) distance between $\overline{\rho}$ and the maximally mixed state is related to the purity of $\overline{\rho}$: 
\bne ||\overline{\rho} - \mathbbm{1}/N||_2 = \sqrt{\Tr (\overline{\rho}^2) - \frac{1}{N}} .\ene
Other measures of distance may give different mixing timescales, but we expect the same $N$-scaling. 
}
\bne t_{\textrm{max-mix}} = \inf_t \{ \, t \; : \; \Tr\left(\overline{\rho(t)}^2\right) - \frac{1}{N} < \epsilon \} . \ene 
This is the later time ($t_{\rm max-mix} > t_{\rm mix}$) at which there is maximal uncertainty about what the time-evolved state is. 
For the black-hole producing initial state~\eqref{eq:initial_state}, the purity of $\Tr (\overline{\rho(t)}^2)$ should rapidly start to decrease when the horizon begins to form, around the particle collision time, which is approximately half the AdS light-crossing time, so we are looking for a mixing time $t_{\rm mix}$ that is order one in AdS units.

There is not a single precise mixing time for us to match to, because the map from boundary to bulk time slices is one-to-many. To elaborate, on the left of Fig.~\ref{fig:IntroFig}, many bulk time slices end at boundary time $t_1$, 
and it is not a priori clear which $\rho_{\rm bulk}$ to match the coarse-grained boundary state $\overline{\rho(t_1)}$ to. Some of these bulk slices do not intersect the black hole horizon, and some do, so a single boundary $\overline{\rho(t_1)}$ cannot simultaneously capture these different bulk states' coarse-grained entropies. Rather, for a fixed single boundary time slice, there is in principle a different coarse-grained $\overline{\rho}$ for each choice of bulk slice and its associated $\rho_{\rm bulk}$.
However, restrictions to bulk time foliations whose time slices are geometrically ``nice" (bounded mean curvature) give state mixing times within a finite range, because all nice slices passing through a cut of the black hole horizon intersect the AdS boundary within an order one boundary time range. As a result, for a fixed boundary time, the bulk mixing times span a finite range. So, we will be satisfied if $\overline{\rho(t)}$ has a mixing time which matches bulk expectations for a fixed nice bulk time foliation (e.g. maximal volume or constant mean curvature slices) within an order one range. 

The value of $\overline{\rho(t)}$ should also be sensitive to the phases (in a fixed basis) of $\rho(t)$. This is important because time evolution only changes those phases, so a coarse-graining map that is phase-insensitive gives a time-independent $\overline{\rho}$. 
Note that, assuming ergodicity, the long-time average of $\overline{\rho(t)}$ is an entropically dominant gas of Hawking radiation, but $\overline{\rho(t)}$ should still be sensitive to the initial state at intermediate times. 

In Sec.~\ref{sec:Uncertainty}, we motivated the heuristic that there is a bipartition of the set of energy eigenstates into integrable and chaotic sectors, 
splitting the Hilbert space into $\cH = \cH_{\tsc L} \oplus \cH_{\tsc H}$. Where appropriate, we will apply this heuristic and coarse-grain the state by replacing unknown exact data from the chaotic sector with its ensemble average.

We will consider states $\rho$ dual to the formation of both small, unstable AdS black holes and large, stable ones. The structure of $\cH_{\tsc{H}}$ projected onto the microcanonical window is more complicated for the former: as discussed in Sec.~\ref{sec:Uncertainty}, that $\cH_{\tsc{H}}$ contains both small black hole and definite particle number states, but neither are primary or descendant states, because they are not stationary. For each type of coarse-graining map, we will start and focus on the minimal version of the prescription, and these give the coarse-grained states $\overline{\rho}$ that best match $\rho_{\rm bulk}$ for the large AdS black hole, because its $\cH_{\tsc{H}}$ has less structure. However, we will also discuss modifications and refinements of the prescriptions for matching expectations for small, evaporating black holes.

It will be useful to have some projection-related notation defined. The complementary pair%
\footnote{Complementary in the sense that the pair of projection operators sum to the identity operator.}
of projection operators onto the integrable and chaotic subspaces are $P_{\tsc H} : \cH \to \cH_{\tsc H}$ and $P_{\tsc L} : \cH \to \cH_{\tsc L}$.
The projectors satisfy the properties
\bne P_{\tsc H} + P_{\tsc L} = \mathbbm{1}, \quad P_{\tsc H} P_{\tsc L} = 0, \quad P_{\tsc \alpha}^\dagger = P_{\tsc \alpha}, \quad P_{\tsc \alpha}^2 = P_{\tsc \alpha} \ene
where $\alpha \in \{\mathrm{H},\mathrm{L} \}$.
Given a pure state $\ket{\psi}$, like our black hole-forming state~\eqref{eq:primd2}, the subspace probabilities are $p_\alpha = \bra{\psi}P_\alpha \ket{\psi}$. The state projections are 
\bne \ket{\psi_\alpha} = \frac{1}{\sqrt{p_{\alpha}}} P_{\tsc \alpha} \ket{\psi}.\ene 
Any density matrix, pure or mixed, can be projected into blocks,
\bne \rho = \begin{pmatrix} 
\rho_{\tsc L} & \rho_{\tsc L\tsc H} \\
\rho^\dagger_{\tsc L \tsc H} & \rho_{\tsc H} 
\end{pmatrix}, \ene
where, for $\alpha,\beta\in \{H,L\}$, $\rho_{\alpha \beta} = P_{\alpha} \rho P_{\beta}$ (and we have abbreviated $\rho_{\alpha\alpha} \to \rho_\alpha$).
Note that these block operators are not generally normalised density matrices, $\Tr (\rho_\alpha) \leq 1$. When $\rho = \ket{\psi}\bra{\psi}$, then $\Tr (\rho_\alpha) = p_\alpha$, and (no sum) $\rho_\alpha^2 = p_\alpha \rho_\alpha$ because $\rho_\alpha$ is rank one when $p_\alpha \neq 0$.

At the end of this section, in Table~\ref{tab:summtab}, we will summarise our findings: the coarse-graining maps, and whether the corresponding coarse-grained states $\overline{\rho(t)}$ have properties that match the semiclassical bulk approximation of the exact density matrix.

\subsection{Averaging over OPE coefficients} \label{sec:OPEAverage}
Recall that our pre-collision state in the spin-dimension eigenbasis, Eq.~\eqref{eq:primd}, is
\bne \ket{\psi} = \cN^{-1} \sum_{h,\bar{h} \in \text{primaries}} c_{\phi\phi\cO_{h,\bar{h}}} \sum_{m, \bar{m}} \psi_{h,\bar{h},m,\bar m} \ket{h+m,\bar{h} + \bar{m}} \label{eq:primd3}.\ene 
This is a sum over primary states and their descendants. We showed that the dimensions in~\eqref{eq:primd3} are centred around the bulk CoM energy $\Delta \approx 2\omega$, and we calculated the kinematic-dependent factors $\psi$.

From~\eqref{eq:primd3}, our density matrix is quadratic in the OPE coefficients:
\bne \begin{split} \rho &= (\ket{\psi_{\tsc L}} + \ket{\psi_{\tsc H}})(\bra{\psi_{\tsc L}} + \bra{\psi_{\tsc H}}) \\
&= \sum_{i,j} c_{\phi\phi i} c_{\phi\phi j}(\dots) + \sum_{i,a} c_{\phi\phi i} c_{\phi\phi a} (\dots)+ \sum_{a,b} c_{\phi\phi a} c_{\phi\phi b} (\dots). 
\label{eq:quadraticOPE}
\end{split} \ene
We use $i, j, \dots$ to denote primaries whose OPE coefficient $c_{\phi\phi i}$ with $\phi$ is known (the integrable sector), and $a, b, \dots$ to denote primaries for which the corresponding coefficient is not known (the chaotic sector).

There is a natural coarse-graining of this $\rho$, coming from the uncertainty of the OPE coefficients for an observer with access only to low-energy data. 
The coarse-graining map of this subsection replaces unknown products of exact OPE coefficients, such as $c_{\phi\phi a} c_{\phi\phi b}$ in~\eqref{eq:quadraticOPE}, with their known average (within the window of primaries with nearby conformal dimensions). Note that it is not necessary to introduce an explicit ensemble of conformal data in this prescription, i.e. an ensemble of CFTs (which may only satisfy the conformal bootstrap conditions on average~\cite{belinApproximateCFTsRandom2024}). 

Now we will average~\eqref{eq:quadraticOPE} sector by sector.
First, by assumption, the $c_{\phi\phi i}$ OPE coefficients are known, so
$\overline{c_{\phi\phi i}c_{\phi\phi j}} = c_{\phi\phi i}c_{\phi\phi j}$ and therefore $\overline{\rho_{\tsc L}} = \rho_{\tsc L}$.
Next, using $\overline{c_{\phi\phi a}} = 0$,%
\footnote{
To justify $\overline{c_{\phi\phi a}} = 0$, note that $c_{\phi\phi a}$ does not have a definite sign, because if $O_a$ is a primary operator, then so is its negative, and they have opposite sign OPE coefficients. 
Without knowing $\sgn(c_{\phi\phi a})$ (which is UV data), we cannot choose a basis of primary operators to flip $\sgn(c_{\phi\phi a})$ and so give the signs a non-zero mean.}
we have
\bne \overline{c_{\phi\phi i}c_{\phi\phi a}} = c_{\phi\phi i} \overline {c_{\phi\phi a}} = 0, \ene
and therefore
\bne \overline{\ket{\psi_{\tsc L}} \bra{\psi_{\tsc H}}} = \overline{\ket{\psi_{\tsc H}} \bra{\psi_{\tsc L}}} = 0 .\ene
In other words, the averaging kills the off-diagonal blocks in $\overline{\rho}$, and we are left with
\bne \overline{\rho} = \begin{pmatrix} \rho_{\tsc L} & 0 \\ 0 & \overline{\rho_{\tsc H} } \end{pmatrix} . \label{eq:rhobardiag} \ene
Then, because $\rho_{\tsc L} \, \overline{\rho_{\tsc H}} = 0$, as states in $\cH_{\tsc L}$ have zero overlap with states in $\cH_{\tsc H}$, the purity of the averaged state is the sum
\bne \label{eq:rhe} \Tr (\overline{\rho}^2) = \Tr (\rho_{\tsc L}^2) + \Tr (\overline{\rho_{\tsc H}}^2). \ene
This is not a sum of purities, because $\rho_{\tsc L}$ and $\rho_{\tsc H}$ are not normalised. Using that $\rho_{\tsc L}^2 = p_{\tsc L} \rho_{\tsc L}$ (because $\rho_{\tsc L}$ is rank one if $\rho$ is pure) we can write the first term as $(1-p_{\tsc H})^2$. 
If we define the normalised density matrix $\tilde{\rho}_{\tsc H} := \overline{\rho_{\tsc H}}/p_{\tsc H}$, then we can write~\eqref{eq:rhe} as a function of $p_{\tsc H}$ and the purity of $\tilde{\rho}_{\tsc H}$:
\bne \Tr (\overline{\rho}^2) = (1-p_{\tsc H})^2 + p_{\tsc H}^2 \Tr (\tilde{\rho}_{\tsc H}^2) .\ene
This is generically less than one, because, even if $\Tr(\tilde{\rho}_{\tsc H}^2)=1$, unless $p_{\tsc H}= 0$ or $1$, $\Tr(\overline{\rho}^2)$ is smaller than one and thus $\overline{\rho}$ is mixed.
Let us now consider two extremes of this formula. Firstly, if $p_{\tsc H} \approx 0$, then 
\bne \Tr (\overline{\rho}^2) = 1 - O(p_{\tsc H}). \label{eq:approx_pur}  \ene This shows that when $\ket{\psi(0)}$ is predominantly in the $\cH_{\tsc L}$ subspace, then $\overline{\rho}$ is approximately pure. This is a reflection of the fact that the $c_{\phi\phi i}$ coefficients for the Fock states are known exactly, so there is no uncertainty in what $\rho_{\tsc L}$ is. 

In the other extreme, $p_{\tsc H} \approx 1$, in which case $\Tr (\overline{\rho}^2) \approx \Tr (\overline{\rho_{\tsc H}}^2)$. Recall that 
\bne \rho_{\tsc{H}} = \sum_{a,b } c_{\phi\phi a} c_{\phi\phi b} (\dots).\ene 
Let us assume that the average value of the pair of OPE coefficients is of the form
\bne \overline{c_{\phi\phi a}c_{\phi\phi b}} = \frac{f(\Delta_a,\Delta_\phi)\delta_{ab}}{e^{S(\Delta_a)}} \label{eq:LLHav} \ene
where $f$ is an $O(c^0)$ smooth function and $e^{S(\Delta_a)}$  is the number of primary states both in $\cH_{\tsc H}$ and with dimensions in a window centred on $\Delta_a$. 
Eq.~\eqref{eq:LLHav} is 
proven~\cite{Pappadopulo:2012jk} (using crossing symmetry) in the infinite energy limit, $\Delta_a \to \infty$, with the asymptotic form of $f$ given in~\eqref{eq:OPEav}, but it is an unproven extrapolation when $\Delta_a$ is in the small, unstable AdS black hole range, e.g. $c^{1/4} \ll \Delta_a \ll c^{2/3}$ for AdS$_3 \times$S$^3$ (see Fig.~\ref{fig:hierarchy0}).%
\footnote{In~\cite{Pappadopulo:2012jk}, the average is over all operators (primaries and descendants). See~\cite{collierUniversalDynamicsHeavy2020} for the result when averaging over only Virasoro primaries in 2d.}
\footnote{Such a result, if true, would not be dissimilar from the extension of the exponentiation of Virasoro blocks from external operator dimensions of order $c$ to order $c^\alpha$ with $0<\alpha <1$~\cite{Alkalaev:2024knk}.}

The exponential suppression in~\eqref{eq:LLHav} is necessary for $\Tr (\overline{\rho}) = 1$, given the exponentially large number of states being summed over. The Kronecker delta follows from the assumption that the signs of $c_{\phi\phi a}$ and $c_{\phi\phi b}$ are uncorrelated for $a\neq b$.

Eq.~\eqref{eq:LLHav} implies that averaging $\rho_{\tsc H}$ over the OPE coefficients diagonalises it, and that those diagonal entries are both a smooth function of $\Delta_a$ and also exponentially suppressed.
Were $\rho_{\tsc H}$ linear in $c_{\phi\phi a}$, its average would vanish; it is because $\rho_{\tsc H}$ is quadratic in the OPE coefficients that the average does not vanish, and that the coefficients' variance $\overline{c^2}$ plays a role.
A remarkable thing is that, even though the matrix elements of $\rho_{\tsc{H}}$ depend on unknown OPE coefficients and are $e^{-S}$ suppressed, we still know the average value of the on-diagonal elements.

Now we average $\rho_{\tsc H}$,  giving the diagonal density matrix
\bne \overline{\rho_{\tsc H}} = \cN^{-2} \sum_{a, m,\bar m}  \overline{c_{\phi\phi a}^2}\, |\psi_{h_a, \bar h_a, m, \bar m}|^2 \ket{h_a+m,\bar{h}_a + \bar{m}} \bra{h_a+m,\bar{h}_a + \bar{m}} .\label{eq:rhohav}\ene
$a$ indexes the primary states in $\cH_{\tsc H}$.  
Let us calculate the purity of~\eqref{eq:rhohav}. 
From eq.~\eqref{eq:psico}, recall that the $\psi$ coefficients have a Gaussian factor of width $\sqrt{\omega}$ centred around $h+m \approx \omega$. 
This Gaussian causes the state to have overlaps with primary states within a window of conformal dimensions centred on $\Delta = 2\omega$. 
We also use the fact that the $\overline{c_{\phi\phi a}^2}$ factor is approximately constant within that window if it varies more slowly than the Gaussian.%
\footnote{From Eq.~\eqref{eq:OPEav}, we know it does in the $\Delta_a \to \infty$ limit, and we assume it does too for lighter $\Delta_a$, in the small AdS black hole range.}
Then, $\overline{\rho_{\tsc H}}$ 
is a diagonal density matrix with
$e^S$ approximately equal non-zero elements, 
\bne
\overline{\rho_{\tsc H}}
\;\approx\;
\cN^{-2}\,\overline{c^2}\;
\diag\!\big(
0,\ldots,0,
\underbrace{1,\ldots,1}_{e^{S}},
0,\ldots,0
\big).
\ene
The purity of such a density matrix is
\bne \Tr (\overline{\rho}^2) \approx \Tr (\overline{\rho_{\tsc H}}^2) = e^{-S}. \label{eq:mixpu} \ene
So, $\overline{\rho}$ is maximally mixed within the window.

Let us give a physical interpretation of the results we have so far. First, recall that $2\omega$ equals both the centre of mass energy of the bulk particle collision and, through the Gaussian in~\eqref{eq:psico}, twice the mean value of the conformal weights in the boundary state. If $\omega$ is only sufficiently high to excite light bulk states, $\omega = \Theta (c^0)$,
the dominant contribution to $\ket{\psi}$ is from approximate-Fock states, whose CFT data are known, and this gives a pure $\overline{\rho}$, as seen in~\eqref{eq:approx_pur}. Correspondingly, in the bulk, we do not have sufficient energy to form a black hole; instead, post-collision, we evolve towards typical radiation states. A low-energy observer can, in principle, distinguish between such states composed of low-energy excitations with $O(c^0)$ particles.
Next, if we increase $\omega$ to above the black hole threshold, $\omega \gg c$, we predominantly have primary states in $\ket{\psi}$ which are not approximate-Fock states.
Because we do not know the OPE coefficients of these states exactly, only their statistical properties, the coarse-grained density matrix is mixed, and we get the purity~\eqref{eq:mixpu}, which matches what we expect from the large black hole we form in the bulk. As we vary $\omega$ between the two extremes, we get an interpolation between the purities~\eqref{eq:approx_pur} and~\eqref{eq:mixpu}. The precise form of the interpolation depends on the statistics of OPE coefficients for heavy primaries of intermediate dimension around the large AdS black hole threshold and lower, as well as the choice of which OPE coefficients to average.

We have shown that replacing products of OPE coefficients with their average gives a coarse-grained state $\overline{\rho(0)}$ that is pure when the energy $\omega$ is too low to form a black hole, and maximally mixed when there is sufficient energy to form a large black hole. This roughly matches what we expect from the semiclassical state $\rho_{\rm bulk}$. But the $\overline{\rho_{\tsc{H}}}$ in~\eqref{eq:rhohav} does not depend on the phases of the $\psi$ wavefunction coefficients, so the purity $\Tr (\overline{\rho(0)}^2)$ is the same for any pair of pure states that only differ by their phases in the spin-dimension eigenbasis.     

\subsubsection{Time-dependence and mapping between ensembles}

We have not yet discussed time-dependence; we have only calculated $\Tr (\overline{\rho(0)}^2)$. If we naively define $\overline{\rho(t)}$ by time-evolving the OPE coefficient-averaged $\overline{\rho(0)}$, without any further averaging, then the purity will be constant (which does not match semiclassical bulk expectations). This is because
\textit{any} coarse-graining prescription which does not average over the Hamiltonian 
\bne \rho(t) \mapsto \overline{\rho (t) } = e^{-iHt} \overline{\rho(0)} e^{iHt}, \label{eq:proba} \ene
will give a time-independent purity $\Tr (\overline{\rho(t)}^2)$. As a further mismatch with semiclassical bulk expectations, we have also found in Eq.~\eqref{eq:mixpu} that $\overline{\rho(t)}$ is mixed at $t=0$ when $\omega \gtrsim c$, so we do not get that $\overline{\rho (t)}$ only becomes mixed \textit{after} the black hole has formed. 

To get a time-dependent purity for the averaged density matrix, we need an ensemble of Hamiltonians. 
With a joint probability distribution $p(H,c)$ on Hamiltonians and OPE coefficients, the time-dependent coarse-grained density matrix is
\bne \overline{\rho(t)} =  \int dH dc\, p(H,c) \, \rho(t)\\
\label{eq:rhohba} \ene 
If the CFT Hamiltonian and the OPE coefficients are statistically independent%
\footnote{
The CFT's Hamiltonian and OPE coefficients are not completely independent, as there are known correlations between OPE coefficients and densities of states~\cite{Post:2024itb}. However, we expect the effects of such correlations to be subleading.}
then $p(H,c) = p(H) p(c)$, and
\bne \overline{\rho(t)} = \int dH p (H)\left( e^{-iH t} \left( \int dc\,p(c) \,\rho (0) \right) e^{iH t} \right).\\
\label{eq:rhohb} \ene 
Two components of the coarse-graining in~\eqref{eq:rhohb} are relevant for the purity $\Tr (\overline{\rho(t)}^2)$, both ultimately coming from the uncertainty in what the theory's dynamics are. There is the coarse-graining of the initial state in the $\overline{\rho(0)} = \int p(c) \, \rho (0)$ component, which we explored in the first half of this subsection, and also coarse-graining from averaging the time evolution. 

Let us see what this implies for $\overline{\rho_{\tsc H} (t)}$.  
In the large AdS black hole regime, $\omega \gtrsim c$, and for a $\rho_{\tsc H}$ whose energy width is narrower than the bandwidth of the Hamiltonian ensemble, $p(H)$ reduces to a Haar measure on the Hamiltonian-diagonalising unitary, and the RMT's energy level distribution. In the next two subsections, we will explore what coarse-grained states $\overline{\rho(t)}$ one finds from ensembles of Hamiltonians similar to this, determine quantitative results for its purity and mixing timescale, and compare to expectations from the semiclassical bulk dynamics.

\subsection{Averaging over Hamiltonians: uncertainty in eigenstates}
\label{sec:Averaging over Hamiltonians: uncertainty in eigenstates}
Let us consider the coarse-graining of a density matrix by averaging its time evolution over an ensemble of Hamiltonians.
Ensemble averaging over Hamiltonians is ubiquitous in the quantum chaos and random matrix theory (RMT) literature~\cite{dalessioQuantumChaosEigenstate2016}. Starting with \cite{CotJen21}, substantial evidence has accumulated that the spectrum of strongly coupled CFT's exhibit a form of random matrix universality. 
In this subsection, we will focus on averaging over $U$, and leave averaging over the Hamiltonian's eigenvalues with an ensemble $\mu (\Lambda)$ to the next subsection, Sec.~\ref{sec:EnergyLevels}.
A given ensemble $\mu(U)$ of unitary matrices that rotate the Hamiltonian  $H = U \Lambda U^\dagger$ reflects the uncertainty in the fine-grained details of the Hamiltonian, in particular, what the eigenstates are. 

Consider an arbitrary time-evolved $N \times N$ density matrix 
\bne \rho(t) = U e^{i\Lambda t} U^\dagger \rho(0) U e^{-i\Lambda t} U^\dagger. \label{eq:rhoex41} \ene
The coarse-graining map of this subsection is
\bne \overline{\rho(t)} = \int d\mu (U) \rho(t) \label{eq:twirlH},\ene
where $\mu$ is a to-be-specified measure on $SU(N)$.%
\footnote{ 
In quantum information parlance, time evolution is a quantum channel, and averaging over the unitary group like in~\eqref{eq:twirlH} is twirling that channel.}

\subsubsection{Unitary-invariant ensemble: the Haar measure} \label{sec:UnitaryInvariant}
First, we consider the simplest non-trivial measure on $SU(N)$, the unitary-invariant Haar measure. This is the maximal ignorance measure on what the Hamiltonian eigenstates are, as one expects within a high-energy microcanonical window of a chaotic system if the only conserved charge is energy. 

\begin{figure}
    \centering
    \begin{overpic}[width=0.85\linewidth]{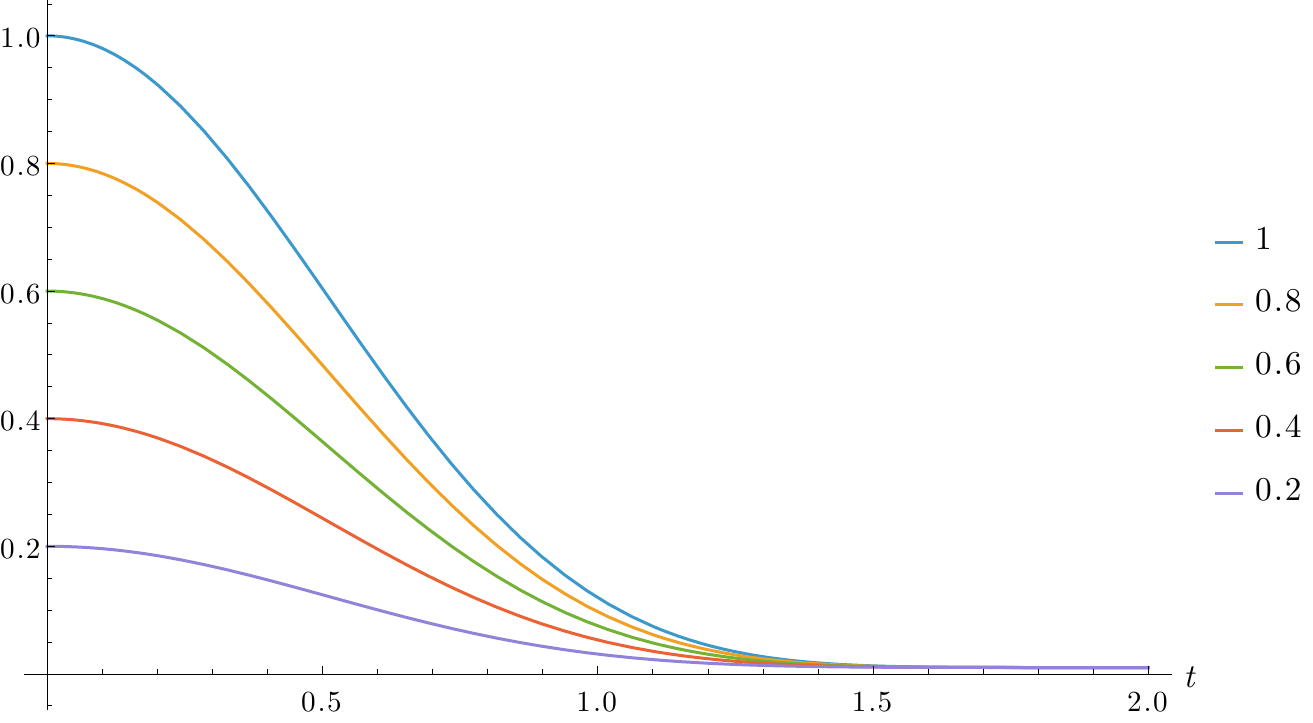}
    \rule{0pt}{1.2cm}
    \put(-1,57){$\Tr(\overline{\rho(t)}^2)$}
    \put(92,41){$\Tr (\rho(0)^2)$}
    \end{overpic}
    \caption{A plot of~\eqref{eq:pur} with $N=100$ for different values of initial purity $\Tr (\rho(0)^2)$, approximating the SFF $g$ with~\eqref{eq:discSFF} and $E_{\rm max} = 2$.
    }
    \label{fig:puris}
\end{figure}

Introducing the normalised infinite temperature ($\beta =0$) spectral form factor (SFF)
\bne g(t) := \frac{|\!\Tr (e^{iHt})|^2}{N^2}, \ene
the Haar-averaged density matrix is, using the Weingarten functions,
\bne \label{eq:avhpu} \overline{\rho (t)} 
= \frac{(N^2 g(t) -1)\rho(0) + N(1-g(t))\mathbbm{1}}{N^2 -1} .\ene
$\mathbbm{1}$ is the identity operator. Eq.~\eqref{eq:avhpu} shows that our coarse-graining map~\eqref{eq:twirlH} is a depolarising channel, because $\overline{\rho}$ is of the form 
\bne \overline{\rho} = p\, \rho + \frac{(1-p)}{N} \mathbbm{1}
\qquad \text{with} \qquad
 p(t) = \frac{N^2 g(t) -1}{N^2 -1}. \label{eq:depleq}\ene
The purity of the coarse-grained state in~\eqref{eq:avhpu} is
\bne \label{eq:pur} \Tr \left ( \overline{\rho (t)}^2 \right) = \frac{1}{N} + \left( \frac{N^2 g -1}{N^2 -1} \right)^2 \left (\Tr (\rho(0)^2)- \frac{1}{N} \right). \ene
This purity is sensitive to the initial state, though only through the purity of $\rho(0)$, 
because averaging over the Haar measure removes all other information about the initial state. 
If the initial state is already maximally mixed, then $\Tr (\rho(0)^2) =\Tr (\overline{\rho (t)}^2 ) = 1/N$, showing that it stays maximally mixed. 
Otherwise,~\eqref{eq:pur} is time-dependent through the time-dependence of the $\beta =0$ spectral form factor $g(t)$. The state is no longer pure on an order $N^0$ timescale $t_{\rm mix}$.
From~\eqref{eq:pur}, the $t_{\rm max-mix}$ at which $\overline{\rho}$ is approximately maximally mixed is when $(N^2 g -1)^2 \ll N^{3}$ (assuming a pure initial state), which for generic SFFs gives an $N$-dependent $t_{\rm max-mix}$. 
Fig.~\ref{fig:puris} shows~\eqref{eq:pur}, for different values of $\Tr (\rho(0)^2)$, approximating $g$ with the average disconnected SFF for
Wigner's large-$N$ semicircular distribution $\rho(E) \propto \sqrt{E_{\textrm{max}}^2 - E^2}$, which is 
\bne \overline{g_d} = \left(\frac{2 J_1(E_{\textrm{max}} t)}{E_{\textrm{max}} t}\right)^2 .\label{eq:discSFF}\ene

\subsubsection{Distinguishable and indistinguishable sectors} \label{sec:DisAndIndis}

Now we will add a little more structure to the coarse-graining map of Sec.~\ref{sec:UnitaryInvariant}. We add a subsector to the Hamiltonian for which we know the eigenstates exactly.  

Take an arbitrary state $\rho$ and decompose it into the two $\cH_{\tsc L}$ and $\cH_{\tsc H}$ subsectors
\bne \rho = \begin{pmatrix} 
\rho_{\tsc L} & \rho_{\tsc L\tsc H} \\
\rho^\dagger_{\tsc L \tsc H} & \rho_{\tsc H} 
\end{pmatrix}. \ene
Let us time-evolve it with an ensemble of Hamiltonians
\bne H = \Lambda_{\tsc L} \oplus U_{\tsc H} \Lambda_{\tsc H} U^\dagger_{\tsc H} \label{eq:Hansa} \ene
and let the measure on $U_{\tsc H}$ be Haar random. This ensemble is a refinement of the prescription in Sec.~\ref{sec:UnitaryInvariant} and reduces to it when $|\cH_{\tsc L}| = 0$. The ensemble models a microcanonical window which contains both primary states with known and unknown wavefunction coefficients (in a fiducial basis).
The Haar measure on $U_{\tsc H}$ assumes that there is no structure in the chaotic sector. 
This ensemble will dynamically mix pure states in the chaotic sector ($p_{\tsc{H}} \approx 1$) but not those in the integrable sector ($p_{\tsc{L}} \approx 1$).

If we start with a state $\ket{\psi(t)}$ in $\cH_{\tsc H}$, it will never evolve to a state in $\cH_{\tsc L}$ with the ensemble~\eqref{eq:Hansa}. This is consistent with $\ket{\psi(t)}$ being dual (at different times) to both a pair of particles and the black hole they form post-collision.
These two states differ only by phases in the energy eigenbasis (and the two-particle state has a transient, approximate Fock state description), but both states are in $\cH_{\tsc H}$. 

Let $N_{\tsc H}$ be the dimension of $\cH_{\tsc H}$, and define the chaotic sector SFF%
\footnote{This can also be written in terms of the projector onto the chaotic sector as follows,
\bne g_{\tsc H}(t) = \left| \frac{\Tr (P_{\tsc H} e^{iHt})}{\Tr (P_{\tsc H})} \right |^2. \ene}
\bne g_{\tsc H}(t) := \frac{|\!\Tr_{\cH_{\tsc H}} (e^{iHt})|^2}{N_{\tsc H}^2}. \ene
For~\eqref{eq:Hansa}, the results for the averages of the density matrix blocks are%
\footnote{
The calculation of $\overline{\rho_{\tsc L\tsc H}(t)}$ uses the identity
\bne \int dU U X U^\dagger \equiv \frac{\Tr (X)}{N} \mathbbm{1} .\ene
}
\bne \overline{\rho_{\tsc L}(t)} = \rho_{\tsc L}(t), \ene
\bne \overline{\rho_{\tsc H}(t)} = \eqref{eq:avhpu}|_{(\rho\, \mapsto \rho_{\tsc H},\, N \mapsto N_{\tsc H},\, g \mapsto g_{\tsc H}  )}\, , \ene
and 
\bne \overline{\rho_{\tsc L \tsc H}(t)} = e^{i\Lambda_{\tsc L} t} \rho_{\tsc L\tsc H}(0) \overline{U_{\tsc H} e^{-i\Lambda_{\tsc H} t} U_{\tsc H}^\dagger} = e^{i\Lambda_{\tsc L} t} \rho_{\tsc L\tsc H}(0)  \frac{\Tr (e^{-i\Lambda_{\tsc H} t})}{N_{\tsc H}}\, . \ene 
The resulting purity of $\overline{\rho}$ is 
\bne \begin{split} \Tr (\overline{\rho(t)}^2) &= \Tr (\rho_{\tsc L}(t)^2) + 2 \Tr (\overline{\rho_{\tsc L\tsc H}(t)} \;\overline{\rho_{\tsc L\tsc H}(t)}^\dagger) + \Tr(\overline{\rho_{\tsc H}(t)}^2) \\ 
&= \Tr (\rho_{\tsc L}(0)^2) + 2 g_{\tsc H}(t) \Tr (\rho_{\tsc L\tsc H}(0)\rho_{\tsc L\tsc H}(0)^\dagger)  + \eqref{eq:pur}|_{(\rho\, \mapsto \rho_{\tsc H},\, N \mapsto N_{\tsc H},\, g \mapsto g_{\tsc H} )}. 
\label{eq:notpu}
\end{split} \ene
This is time-dependent and has a greater sensitivity on the initial state than~\eqref{eq:pur}, which it reduces to if only $\rho_{\tsc H} (0) \neq 0$. If only $\rho_{\tsc L} (0) \neq 0$, then the purity~\eqref{eq:notpu} is constant. The larger the fraction $\rho$ that is in the integrable sector, the longer the mixing time.

If the initial state is fully in the integrable sector (only $\rho_{\tsc L} (0) \neq 0$) then the purity~\eqref{eq:notpu} is constant. Otherwise, the purity is time-dependent and has a greater sensitivity on the initial state than~\eqref{eq:pur}, which it reduces to if the initial state is fully in the chaotic sector (only $\rho_{\tsc H} (0) \neq 0$). The more that $\rho(0)$ is in the integrable sector, the longer the mixing time.

As an illustrative example, consider an arbitrary pure initial state and its projections onto the $\cH_{\tsc{L}}$ and $\cH_{\tsc{H}}$ sectors:
\bne \ket{\psi(0)} = \sqrt{p_{\tsc{L}}} \ket{\psi_{\tsc L}} + \sqrt{1-p_{\tsc{L}}}\ket{\psi_{\tsc H}}. \label{eq:exampur}\ene
The parameter $p_{\tsc{L}}$ controls how much of the initial state is in $\cH_{\tsc{L}}$ versus  $\cH_{\tsc{H}}$. 

Plugging~\eqref{eq:exampur} into~\eqref{eq:notpu} gives
\bne \Tr (\overline{\rho(t)}^2) = p_{\tsc{L}}^2 + 2 p_{\tsc{L}}(1-p_{\tsc{L}}) g_{\tsc H} (t) + (1-p_{\tsc{L}})^2 \left(\frac{1}{N_{\tsc H}} + \left( \frac{N_{\tsc H}^2 g_{\tsc H} (t) -1}{N_{\tsc H}^2 -1} \right)^2 \left (1- \frac{1}{N_{\tsc H}} \right)\right). \label{eq:specp} \ene

\begin{figure}
    \centering
    \begin{overpic}[width=0.85\linewidth]{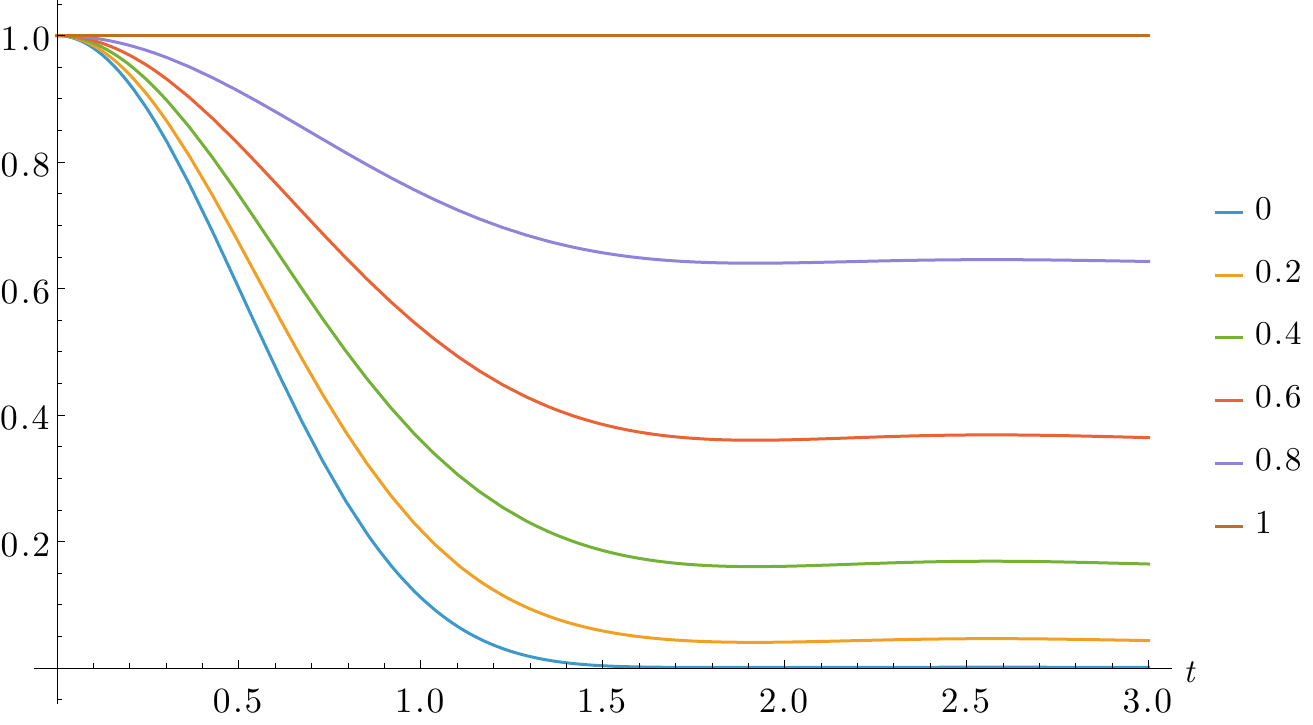}
    \put(-1,57){$\Tr(\overline{\rho(t)}^2)$}
    \put(95,43){$p_{\tsc{L}}$}
    \end{overpic}
    \caption{A plot of~\eqref{eq:specp}, which is the purity of the time-evolved initial state~\eqref{eq:exampur} averaged over a mixed ensemble~\eqref{eq:Hansa} of Hamiltonian rotation unitaries on the integrable and chaotic sectors. The plot uses $N_{\tsc H} = 100$ and 
    approximates the $g_{\tsc{H}}$ with~\eqref{eq:discSFF} and $E_{\rm max} = 2$. 
    }
\label{fig:ChaoticIntegrablePurities}
\end{figure}

We plot this in Fig.~\ref{fig:ChaoticIntegrablePurities} for different values of $p_{\tsc{L}}$. When $p_{\tsc{L}}=0$, corresponding to a black hole, the initial state is in the chaotic sector, $\ket{\psi(0)} \in \cH_{\tsc H}$, and the state rapidly mixes, and asymptotes to the maximally mixed state with purity $1/N_{\tsc H}$. If we change the initial state so that $p_{\tsc{L}}$ is larger, the time-evolved state mixes more slowly and is less mixed as $t\to \infty$. When $p_{\tsc{L}}=1$, the initial state is in the integrable subspace, $\ket{\psi(0)} \in \cH_{\tsc L}$, and $\overline{\rho(t)}$ stays pure. This reflects the fact that we know the time evolution of states in the integrable subspace exactly.

Let us put this in the context of colliding particles in AdS. If the kinematics are such that a black hole forms, then the boundary state is far from a Fock state, its wavefunction coefficients in a fiducial basis are essentially unknown, and $p_{\tsc{L}} \approx 0$. Next, if we increase the impact parameter, then eventually the particles will orbit their mutual centre of mass rather than form a black hole, so the boundary state is an approximate Fock state whose dynamics are known, and $p_{\tsc{L}} \approx 1$. 

\paragraph{Phase dependency.} The purities~\eqref{eq:pur} and~\eqref{eq:specp} are insensitive to the phases of the overlaps $\braket{i|\psi (0)}$ of the initial state, and this highlights a general feature: it is difficult, for a given coarse-graining map $\rho \mapsto \overline{\rho}$, for $\Tr (\overline{\rho}^2)$ to be sensitive to the phases in the initial state. 
Note that the purity of the \textit{exact} density matrix depends on only the absolute values of the matrix elements of $\rho$ and not their phases:
\bne \Tr \rho^2 = \sum_{i,j} |\rho_{ij} |^2 . \ene
So, \textit{any} coarse-graining map for which $|\overline{\rho}_{ij}|$ is insensitive to the phase of $\rho_{ij}$ will give a purity $\Tr(\overline{\rho}^2)$ that is similiarly insensitive. This is a problem because, in thermalisation, the matrix elements of atypical initial states $\rho_{ij} (0)$ and those of the typical states they evolve towards, $\rho_{ij}(0) e^{iE_{ij}t}$, differ only in their phases. Then we get the same $\Tr (\overline{\rho}^2)$ for an atypical macrostate, like an unstable AdS black hole, and a typical state, like the gas of Hawking radiation the same black hole would evaporate into. So, phase-independent coarse-graining maps give $\overline{\rho}$'s which are not always consistent with $\Tr (\overline{\rho}^2) \approx \Tr (\rho_{\textrm{bulk}}^2)$. 

\subsubsection{Further refinements}

\paragraph{Locality.} Both the boundary CFT and the bulk EFT are local QFTs, and, since locality is essential in the bulk description of black hole formation and evaporation, the ensemble of Hamiltonians we average over should account for this. The Haar measure we have used on the Hamiltonian-diagonalising unitaries treats all states equally and so disregards all physical structure in the theory, including locality. In the RMT literature, there have been refinements of Hamiltonian ensembles to account for k-locality in many-body systems, such as the so-called embedded random matrix ensembles~\cite{kota2014embedded}. Recently, and more closely related to our motivation, the authors of~\cite{shiLocalDynamicsStructure2023} studied the imprint of locality on energy eigenstate statistics in a toy model. They found inconsistencies between using the Haar measure and the time-dependence of R\'enyi entropies expected from locality, and proposed a different measure. A refinement of the coarse-graining prescription of this subsection would be to use a measure $\mu (U)$ like those in~\cite{kota2014embedded, shiLocalDynamicsStructure2023} that accounts for locality.

\paragraph{Charge sectors.} We can further refine this coarse-graining prescription by adding more sectors. To motivate this, note that the Hamiltonian of a CFT on a sphere commutes with the rotation generators, and so is block-diagonal with each block corresponding to an $SO(d)$ irrep. If the CFT has additional symmetries and charges, then there is a further decomposition of each $SO(d)$ block into smaller blocks labelled by those charges. Then, we average each block separately, with the justification that microstates with different global charges are distinguishable. Each block can have its own independent Hamiltonian statistics, and we expect rich behaviour in such a refinement of the coarse-graining prescription.

\paragraph{Thouless energy-width windows.} Is it justified to apply RMT statistics to our particle-collision state, which has energy width $\sigma_E \sim \sqrt{\omega}$? One is justified in using RMT for energy windows narrower than the Thouless energy. In some systems, the Thouless energy is $E_{\textrm{Th.}} \sim D/L^2 \sim \beta/L^2$ where $L$ is the characteristic size of the system. For a BTZ black hole, $M \sim c \beta^{-2}$, so $E_{\textrm{Th.}} \sim \sqrt{c/M} \sim 1$. 
This further supports adding more sectors to our model, in this case, to group energy eigenstates into windows of unit order width, and apply RMT statistics separately to each of these narrower windows.
\subsection{Averaging over Hamiltonians: uncertainty in energy levels} \label{sec:EnergyLevels}
In this subsection, we consider another coarse-graining map interpretable as ignorance of the microscopic Hamiltonian, an averaging over a distribution of energy levels:
\bne \overline{\rho(t)} = \int d\mu (E_1, \dots, E_N) \left( \sum_{ij} \rho_{ij}(0) e^{i(E_i -E_j)t} \ket{i}\bra{j} \right). \label{eq:energy}\ene
The ensemble encodes and depends on the degree of uncertainty in what the Hamiltonian's eigenvalues are. This uncertainty leads to an uncertainty in what the $e^{i(E_i- E_j)t}$ phases are in the off-diagonal elements of $\rho(t)$, and therefore a time-growing mixedness of $\overline{\rho(t)}$. 

\subsubsection{Pure initial state} \label{sec:EnergyLevels1}
Take an arbitrary pure state $\ket{\varphi}$, time evolve it, average the resulting density matrix over energy levels as prescribed by Eq.~\eqref{eq:energy}, and split the result into diagonal and off-diagonal parts in the energy eigenbasis to get
\bne \begin{split}
\overline{\rho(t)} 
&= \sum_{i}|\!\braket{\varphi |i}\!|^2 \ket{i}\bra{i} + \sum_{i\neq j} \braket{i|\varphi} \braket{\varphi|j} \overline{e^{i(E_i-E_j)t}} \ket{i}\bra{j} .
\end{split} \label{eq:averh} \ene
The purity of this averaged state is sensitive to the overlaps $|\!\braket{i|\varphi}\!|^2$,
but we can determine its approximate value for a typical $\ket{\varphi}$.
The typical overlap between a Haar-random $\ket{\varphi}$ and an energy eigenstate is
\bne \braket{i|\varphi} = \frac{1}{\sqrt{N}} e^{i\theta_i} \ene
where $N$ is the Hilbert space dimension and $e^{i\theta_i}$ is a random phase. Using this, the typical purity of~\eqref{eq:averh} is
\bne \begin{split}
    \Tr (\overline{\rho(t)}^2) &= \sum_{i, j} \frac{1}{N^2} \left|\overline{e^{i(E_i-E_j)t}}\right|^2 \\
    &= \frac{1}{N} + \frac{N(N-1)}{N^2} \left|\overline{e^{i(E_1-E_2)t}}\right|^2.\label{eq:enpur}
\end{split} \ene
To reach the second line, we split the sum into diagonal and off-diagonal terms and used that, given the symmetries, the off-diagonal terms are identical. 
We can connect the purity~\eqref{eq:enpur} to the averaged SFF using the relation
\bne \overline{g (t)} = \frac{1}{N} + \frac{N-1}{N} \overline{e^{i(E_1-E_2)t}} \label{eq:useqn}\ene  
to get
\bne \Tr (\overline{\rho(t)}^2) = |\overline{g}|^2 + \frac{(1-|\overline{g}|)^2}{N-1} .\label{eq:enpur2} \ene
\begin{figure}
\centering
\begin{overpic}[width=0.85\linewidth]{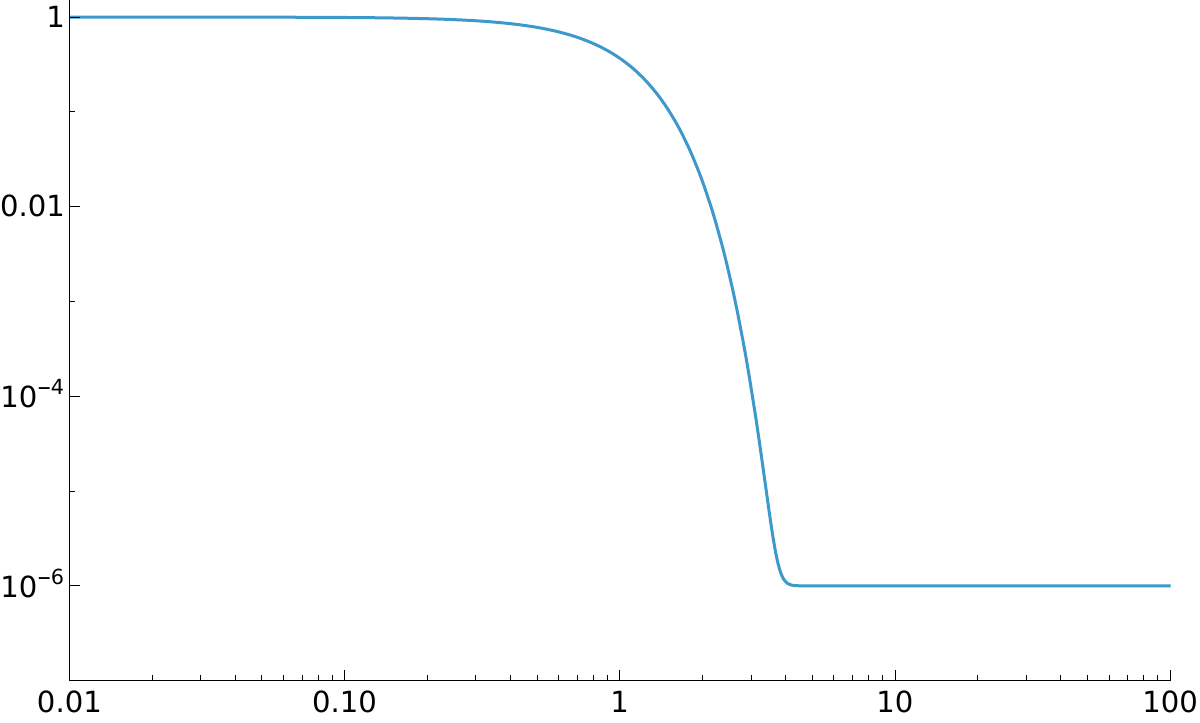}
  \put(0,62){$\Tr(\overline{\rho(t)}^2)$}
  \put(99,4){$t$}
\end{overpic}
\caption{A plot of the purity~\eqref{eq:enpur2} for $N = 10^6$, using the SFF~\eqref{eq:gdapprox} with $\Delta E = 1$. }
\label{fig:TimeAvgPurity}
\end{figure}
As time progresses, uncertainty in the energy levels induces mixedness in $\overline{\rho(t)}$. 
At early times, $t = \Theta( N^0)$, $\bar g$ is dominated by its disconnected contribution
\bne \overline{g(t)} \approx \bar{g}_{\textrm{d}} := \frac{\left|\overline{\Tr(e^{iHt})}\right|^2}{N^2} .\label{eq:gdisc} \ene
Take this to be a trace over a microcanonical window centred on mean energy $\overline{E}$ and with $N$ states. Let us assume that the marginal probability distribution $\rho(E)$ of any single energy eigenvalue is Gaussian distributed, so that%
\footnote{Using Wigner's semicircular distribution gives the Bessel function formula~\eqref{eq:discSFF} familiar from RMT. An argument not to use this distribution is that our states are in a microcanonical window away from the edges of the spectrum, so should be insensitive to those global features. We use a Gaussian rather than a top hat function for $\rho(E)$ because realistic states, like our particle-collision state, have tails outside of the window. } 
\bne |\overline{g}_{\textrm{d}}| = \left|\frac{1}{\sqrt{2\pi(\Delta E)^2}}\int_{-\infty}^{\infty} dE e^{-\frac{1}{2}\left(\frac{E-\overline{E}}{\Delta E}\right)^2}\, e^{iEt}\right| = e^{-\frac{1}{2}(\Delta E)^2\, t^2} .\label{eq:gdapprox}  \ene
The width $\Delta E$ corresponds to the energy resolution of the low-energy observer, so we take it to be $N$-independent. 

In Fig.~\ref{fig:TimeAvgPurity}, we plot the purity~\eqref{eq:enpur2} using~\eqref{eq:gdisc} and~\eqref{eq:gdapprox}.
As can be seen, the averaged state starts pure at $t=0$. Using~\eqref{eq:gdapprox}, the timescale at which the purity~\eqref{eq:enpur2} shows that $\overline{\rho(t)}$ is starting to become mixed is 
\bne t_{\textrm{mix}} \approx \frac{1}{\Delta E}, \ene 
which is order $N^0$.
After that, the purity continues to decrease. The purity is approximately at its minimal value of $1/N$ 
when $\overline{g} = \Theta( N^{-1/2})$, giving 
$t_{\rm{max-mix}} = \Theta (\sqrt{\log N})$.
This $t_{\rm{max-mix}}$ is still much earlier than the SFF dip time $t_{\tsc {dip}} = \Theta (N^{1/2})$, meaning that $\overline{\rho}$ is maximally mixed long before the connected piece of $\overline{g}$ starts to dominate. Also, since $\overline{g}_c \lesssim N^{-1}$, $\overline{\rho}$ stays maximally mixed at late times:  
\bne \Tr (\overline{\rho (t)}^2) \approx \frac{1}{N}, \qquad t > t_{\rm max-mix} \approx \sqrt{\log N} .\ene

The time $t_{\rm max-mix}$ is parametrically larger than what we expect from the black hole horizon formation and ringdown timescales. The coarse-grained state is maximally mixed much later than both the thermalisation and scrambling times, whose $N$-scalings are $N^0$ and $\log S = \log(\log N)$ respectively. It is closer to the Schwarzschild evaporation time $t_{\rm{evap.}}/r_h = \Theta (S)$, though the physics are seemingly unrelated.
It takes a long time for the state to become fully mixed because nearest-neighbour energy eigenstates need to dephase, and their energy separation is of order $1/N$. From the bulk perspective, it is not clear whether or how an observer in a single copy of the spacetime would be able to capture this timescale.

One might wonder whether our large $t_{\rm max-mix}$ is an artifact of our choice of measure for the distance between $\overline{\rho}$ and the maximally mixed state. We will not answer that fully, but we do know that we would not reduce $t_{\rm max-mix}$ if we used von Neumann entropy to quantify how close we are to the maximally mixed state, rather than the purity.
This is because the entropy of $\overline{\rho}$ cannot be near-maximal if its purity is not near-minimal. This follows from the inequality%
\footnote{To derive this, we use
\bne \log N - S(\rho) = S_{\rm rel}(\rho ||\frac{\mathbbm 1}{N}) ,\ene
then the quantum Pinsker inequality
\bne S_{\rm rel}(\rho ||\sigma) \geq \frac{1}{2\log 2} ||\rho -\sigma||_1^2, \ene
monotonicity of the Schatten norm
\bne||X||_1 \geq ||X||_2, \ene
and
\bne \left|\left|\rho-\frac{\mathbbm 1}{N}\right|\right|_2^2 = \Tr (\rho^2) - \frac{1}{N}. \ene
}
\bne \log N - S(\rho) \geq \frac{1}{2\log 2} \left(\Tr (\rho^2) - \frac{1}{N}\right). \ene

The purity formula~\eqref{eq:enpur2} is the same as a result derived in a different model in previous work (eq.~(2.28) in~\cite{deboerPageCurvesReplica2023}). The decreasing purity shown in Fig.~\ref{fig:TimeAvgPurity} quantifies the uncertainty in what the initially pure state has evolved to. 

\subsubsection{General initial state} \label{sec:EnergyLevels2}
Now we consider a general, possibly mixed initial density matrix $\rho(0)$, as opposed to a pure one. This is to explore what happens if the prescription coarse-grains both the initial state and the time-evolution.

The averaged, time-evolved density matrix is
\bne \overline{\rho (t)} = \sum_{ij} \rho_{ij} (0) \overline{e^{iE_{ij}t}}\ket{i}\bra{j}. \ene
We can rewrite this, using~\eqref{eq:useqn},
to get
\bne \overline{\rho (t)} = \frac{N(1-\overline{g(t)})}{N-1}\rho_{\mathrm{diag.}}(0) + \frac{N \overline{g (t)}-1}{N-1} \rho (0), \label{eq:rhoav}\ene 
where
$\rho_{\mathrm{diag.}} = \sum_i \rho_{ii}(0) \ket{i}\bra{i}$. The state~\eqref{eq:rhoav} has purity
\bne \Tr \overline{\rho(t)}^2 = \Tr \rho_{\mathrm{diag.}}(0)^2 + \left(\frac{N\overline{g(t)}-1}{N-1} \right)^2 \left(\Tr \rho(0)^2 -  \Tr \rho_{\mathrm{diag.}}(0)^2 \right), \ene
using that $\Tr (\rho_{\mathrm{diag.}}\, \rho) = \Tr(\rho_{\mathrm{diag.}}^2)$. 
The mixing time scale is approximately the same as that in~\eqref{eq:enpur2}.
In contrast to~\eqref{eq:enpur2}, the purity is sensitive to the initial state, though only through the purity of the initial state.  

We have shown that averaging over energy levels gives a coarse-grained state $\overline{\rho(t)}$ whose purity begins to decrease on a timescale that matches the black hole horizon formation timescale of two colliding particles. In the first version of the prescription we explored, $\overline{\rho(t)}$ is always initially pure, which is contrary to what we expect if the initial state is dual to a black hole. We also found that the start and endpoints of the mixing are parametrically separated in time, $t_{\rm max-mix} \gg t_{\rm mix}$, which is contrary to bulk expectations and suggests a need for further refinements of the prescription. 

\subsection{Averaging over time windows} \label{sec:time_windows}
Let us consider a new coarse-graining map where we average the state over time windows. The physical interpretation is that realistic observers perform measurements with finite resolution clocks. Time-averaging is a coarse-graining prescription because it suppresses the contribution of high-frequency modes. 
To define the map, take the average of $\rho(t)$ over a time window of width $2\Delta t$:
\bne \overline{\rho(t)} := \frac{1}{2\Delta t} \int_{t-\Delta t}^{t+\Delta t} \rho (t') dt' . \label{eq:timeav1}\ene
Consider a general time-evolved state
\bne \ket{\psi(t)} = \sum_i c_i e^{i E_i t} \ket{E_i}. \ene
In the energy eigenbasis, the coarse-grained density matrix is
\bne \overline{\rho(t)} = \sum_{ij} c_i c_j^* e^{i E_{ij} t} \operatorname{sinc}
 (E_{ij} \Delta t) \ket{E_i}\bra{E_j} \ene
 where $E_{ij} := E_i - E_j$ and $\operatorname{sinc}(x) = \sin(x)/x$.
The time-averaging has the effect of suppressing off-diagonal elements of $\overline{\rho(t)}$ with energy differences $E_{ij} \Delta t \gg 1$. The purity of this $\overline{\rho(t)}$ is
\bne \Tr (\overline{\rho(t)}^2) = \sum_{ij} |c_i|^2 |c_j|^2 \operatorname{sinc}^2 (E_{ij} \Delta t). \label{eq:avgtp}\ene
This purity is sensitive to the magnitudes of $c_i$ in the initial state, but not their phases.
The larger $\Delta t$ is, the smaller the purity, with $\lim_{\Delta t \to \infty} \Tr (\overline{\rho(t)}^2) = \sum_i |c_i|^4$, which is the inverse participation ratio of $\ket{\psi(0)}$. For fixed $\Delta t$, the wider $\ket{\psi(0)}$ is in the energy eigenbasis, the smaller the purity. So, for states within a particular microcanonical window, energy eigenstates have $\Tr (\overline{\rho(t)}^2) = 1$, whereas typical states have mixed $\overline{\rho(t)}$ if $\Delta t$ is larger than the smallest inverse level spacing.

The purity~\eqref{eq:avgtp} is time-independent.
It can be made time-dependent with a window width $\Delta t$ that changes as time progresses, such as $\Delta t(t) \approx \Delta t (0) + \sqrt{D t}$, where $D$ is a temporal diffusion constant. Physical justifications for this include accumulating uncertainty in the observer's clock or that a weak signal requires measurements over longer time periods.

A second natural way to coarse-grain $\rho(t)$ by integrating over a time window is to define
\bne \overline{\rho(t)} := \frac{1}{t} \int_0^t \rho(t') dt' . \label{eq:2ndtw}\ene
This state obeys the second law of thermodynamics: the entropy of $\overline{\rho(t)}$ increases monotonically with $t$, and also asymptotes to the microcanonical entropy if the system is ergodic~\cite{dalessioQuantumChaosEigenstate2016}. But $\overline{\rho (0)}$ is pure if $\rho(0)$ is pure, so~\eqref{eq:2ndtw} never coarse-grains the initial state, even if it is dual to a black hole microstate. This particular deficiency can be ameliorated by combining the two prescriptions:
\bne \overline{\rho (t)} := \frac{1}{t + 2\Delta t} \int_{-\Delta t}^{t+\Delta t} \rho (t') dt' .\ene
A lesson here is that there are many coarse-graining maps, even within a given family of prescriptions, and, if a given prescription for coarse-graining $\rho$ does not satisfy a particular desired property, then this can often be remedied with a modification of the prescription.

\subsection{Other ways to coarse-grain}

\subsubsection{Ensembles of density matrices.} \label{sec:ensembles_of_density_matrices}
One perspective is that semiclassical gravity emerges from a principle of maximum ignorance, meaning that the maximum entropy ensemble of density matrices conditioned on a set of low-energy data, such as few-body correlators, reproduces semiclassical physics, such as on-shell actions of multiboundary wormhole geometries~\cite{deBoer:2023vsm}.

If we have a map from a given $\rho(t)$ to a probability measure on a set of density matrices,  $\rho(t) \mapsto \mu_{\rho(t)} (\rho')$, then we define the coarse-grained $\rho(t)$ as
\bne \overline{\rho(t)} = \int d\mu_{\rho(t)}(\rho') \,\rho' .\ene
Note that this coarse-graining map is non-linear in the state, unlike those considered so far, because the measure can depend on the state that we are averaging.
A natural measure to consider is the maximal entropy one
\bne S\left (\mu_{\rho(t)}\right) = \max_{\mu'_{\rho(t)}} S\left(\mu'_{\rho(t)}\right), \qquad S(\mu) := - \int d\mu (\rho') \log \mu (\rho') ,\label{eq:maxig} \ene
subject to the constraint that the expectation values of a set of observables $\{\cO\}$ agree for the exact and coarse-grained density matrix, up to some tolerance $\delta$:
\bne \forall \cO_i \in \{\cO\}:\quad  \left(\tr (\overline{\rho(t)} \cO_i) - \tr (\rho(t) \cO_i)\right) \in [-\delta, \delta]
\label{eq:obsco} \ene
The physical meaning of the tolerance is the measurement resolution of the observer.  
In the context of holography and semiclassical gravity, it is reasonable to say that a low-energy observer cannot distinguish observables that are non-perturbatively close in $G_N$, such as the energy level spacings of black hole microstates within a given microcanonical window, i.e. $\delta = O(e^{-1/G_N})$. 
We can constrain $\tr (\overline{\rho(t)} \cO_i)$ to match the semiclassical computation of $\langle \cO_i (t) \rangle$, up to non-perturbative corrections. Note that, given a $\mu_{\rho(t)}$, we can compute other coarse-grained quantities, including those that are non-linear in $\rho(t)$, such as $\Tr (\overline{\rho(t)^n})$. 

Let us consider the state $\ket{\psi(t)}$ analysed in Sec.~\ref{sec:initial_state_hh_basis}, which, in the semiclassical bulk approximation, forms an unstable AdS black hole which consequently evaporates. After the black hole has formed but before it has evaporated ($t \in [\pi/2,\pi/2 + t_{\mathrm{evap.}}]$), few-body operators cannot distinguish $\rho(t) = \ket{\psi(t)}\bra{\psi(t)}$ from other black hole microstates with the same macroscopic properties, so $\mu_{\rho(t)}$ is uniform and $\overline{\rho(t)}$ is maximally mixed: 
\bne \overline {\rho(t)} = \frac{1}{e^{S_{\mathrm{BH}}}} \mathbbm{1}.\ene
This $\mathbbm{1}$ is the projection operator onto the subspace of black hole microstates.

There is a similar coarse-graining map which defines $\overline{\rho}$ as the largest entropy density matrix which satisfies the constraint~\eqref{eq:obsco}. This map was used to define a coarse-grained notion of vN entropy called simple entropy in~\cite{engelhardtDecodingApparentHorizon2018}, but it does not allow us to average arbitrary functions of the density matrix like~\eqref{eq:maxig} does, say, if we wished to define coarse-grained R\'enyi entropies, because the prescription does not give us a measure on the set of density matrices like~\eqref{eq:maxig} does.

A trivial way for the coarse-graining map~\eqref{eq:obsco} to give a $\overline{\rho}$ which matches any given density matrix on a set of quantities (like the observables in~\eqref{eq:obsco}, or an entropy curve like in Fig.~\ref{fig:IntroFig}) is to condition the maximisation of $S(\mu_{\rho})$ on matching those quantities. This is guaranteed to work and give the desired matching between density matrices, but it is feeding in the desired output as input. What would be most interesting is if the entropy curve in Fig.~\ref{fig:IntroFig} can be matched after only conditioning $\mu(\rho)$ on simple observables like in~\eqref{eq:obsco}.

\subsubsection{Diagonal approximation} \label{sec:DiagonalApprox}
Another coarse-graining map is the diagonal approximation~\cite{polkovnikovMicroscopicDiagonalEntropy2011}, which has been considered in the context of holography and semiclassical gravity in~\cite{ubaldoAdS$_3$RMT$_2$Duality2023, Dong:2023bfy, Penington:2024jmt}, and, like us, as an AdS/CFT coarse-graining map in~\cite{chandraCoarseGrainingPure2022}. 
Given an orthonormal basis $\{ \ket{n} \}$ and density matrix $\rho$, the diagonal approximation is a projection onto the diagonal components of $\rho$:
\bne \rho \mapsto \overline{\rho} = \rho_{\mathrm{diag.}} = \sum_n \ket{n} \!\braket{n|\rho |n} \!\bra{n} . \label{eq:diaga} \ene
A density matrix that is diagonal in a fixed basis behaves like a classical probabilistic mixture for observables that are also diagonal in that basis.
The coarse-graining map~\eqref{eq:diaga}, and the entropy of $\rho_{\rm diag.}$ for fixed $\rho$, is highly sensitive to the choice of basis in which to perform the diagonalisation.

The physical interpretation of the diagonal approximation is that, if we consider a time-evolved $\rho(t)$, then typically the off-diagonal elements of $\rho (t) $ rapidly dephase. Then the off-diagonal elements cancel each other in the expectation value of observables $\Tr (\rho(t) \cO)$, so $\rho(t)$ and $\rho_{\mathrm{diag.}}(t)$ are indistinguishable: 
\bne \Tr (\rho(t) \cO) \approx \Tr (\rho_{\mathrm{diag.}}(t) \cO). \label{eq:ETHapp}\ene 
For observables $\cO$ that obey the ETH and when $\{\ket{n}\}$ is the energy eigenbasis,~\eqref{eq:ETHapp} is an exact equality in the thermodynamic limit.
The matching of expectation values is something that this coarse-graining map has in common with the coarse-graining in Sec.~\ref{sec:ensembles_of_density_matrices}. Notice that the diagonal approximation in the energy eigenbasis has the same effect as the OPE averaging in \ref{sec:OPEAverage}.

Which eigenbasis it is natural to perform the diagonalisation in depends on the state and set of observables. For any basis, if $\rho(t)$ is pure and equal to $\ket{\psi (t)} \bra{\psi(t)}$, then $\rho_{\mathrm{diag.}} (t)$ is insensitive to the phases of $\ket{\psi(t)}$ in that basis. 

If $\{ \ket{n} \}$ is the energy eigenbasis, then $\rho_{\mathrm{diag.}} (t)$ is static, and equal to $\lim_{\Delta t \to \infty} \overline{\rho(t)}$ for the averaging-over-time-window coarse-graining map considered in Sec.~\ref{sec:time_windows}. A typical black hole microstate will give a highly mixed $\rho_{\mathrm{diag.}}$ if it has overlaps with many energy eigenstates, which is a reasonable assumption, but so too will any typical microcanonical state, which is contrary to the behaviour expected when non-black hole states dominate the ensemble. Furthermore, the purity is constant, which is also contrary to the sought-after behaviour.

If $\{ \ket{n} \}$ is \textit{not} the energy eigenbasis, but for example a basis of pointer states, then $\rho_{\mathrm{diag.}} (t)$ is time-dependent. At $t=0$, $\rho_{\mathrm{diag.}}(0)$ is mixed, unless $\rho (0)$ is one of the basis states $\ket{n} $, in which case the timescale of mixing is inversely proportional to the spread of $\ket{n}$ in the energy eigenbasis, $t_{\text{mix}} \sim \sigma_{E,n}^{-1}$, where 
\bne \sigma_{E,n}^2 = \braket{n|H^2|n} - \braket{n|H|n}^2.\ene

\subsection{Entropies and replica wormholes} \label{sec:rep_wormholes}
We can extend the calculation of the coarse-grained purity to define coarse-grained von Neumann entropies, and connect to replica wormholes and unitarity restoration. The von Neumann entropies can be calculated via the replica trick from both Tsallis and R\'enyi entropies.%
\footnote{A side comment: the von Neumann, R\'enyi, and Tsallis entropies are all limiting cases of a two-parameter family of entropy measures called unified entropies~\cite{hu2006generalized}.}
\footnote{Tsallis and R\'enyi entropies are computed from multiple replica copies of the state, and, unlike ordinary observables, can distinguish between mixed states and probability distributions of pure states, because they are non-linear in the density matrix.}
Tsallis entropies are defined as
\bne S_n (\rho) := \frac{1}{n-1}(1-\Tr (\rho^n)),\ene
and the von Neumann entropy equals 
\bne S_{\textrm{vN}} (\rho) = -\Tr \rho \log \rho = \lim_{n\to 1} S_{n}(\rho). \ene
The purity $\Tr (\rho^2)$ is related to the $S_2$ Tsallis entropy. We work with Tsallis entropies rather than the R\'enyi entropies $(1-n) S_n^{\textrm{(R\'enyi)}} = \log \Tr \rho^n$ because the logarithm makes the R\'enyi entropies difficult to average directly.

We define the coarse-grained von Neumann entropy as the $n\to 1$ limit of a suitably defined coarse-graining of the Tsallis entropies:
\bne S^{(\rm{coarse})}_{\rm vN} = \lim_{n\to 1} S^{(\rm{coarse})}_{n} .\ene  
There are two natural definitions of coarse-grained Tsallis entropies given a density matrix coarse-graining map. 
First, the moment-averaged entropy
\bne S_{n}^{(\textrm {coarse} )} (\rho) := \frac{1}{1-n} \left(\Tr (\overline{\rho^n})-1 \right) \label{eq:cogte}, \ene
and second, the entropy of the averaged state $S_n (\overline{\rho})$.%
\footnote{Note that the vN entropy of the averaged state is always greater than the averaged vN entropy, $S_{\rm vN} (\overline{\rho}) \geq \overline{S_{\rm vN}(\rho)}$, for any kind of averaging, because $S_{\rm vN}$ is concave in $\rho$. In contrast, the Tsallis and R\'enyi entropies are only concave functions of $\rho$ for $0<n<1$~\cite{hu2006generalized}. For a review of the different types of quantum entropies and their properties, see~\cite{Chehade:2019}.} 
In general, these do not have the same $n\to 1$ limit. The latter equals the term in the former coming from the disconnected component of the averaged density matrix moment $\overline{\rho^n} = \overline{\rho}^n + \dots$, and $\Tr(\overline{\rho^n}) \neq \Tr (\overline{\rho}^n)$ generally. 

$S_{n} (\overline{\rho})$ trivially equals $S_{\rm vN} (\overline{\rho})$ in the $n\to 1$ limit, and it is generally easier to calculate than~\eqref{eq:cogte}. To give an example, for the coarse-grained $\overline{\rho}$ in~\eqref{eq:avhpu},
\bne S_{n} (\overline{\rho}) = 
\frac{1}{n-1} \left( 1- \sum_{i=1}^N \left(p(t) \lambda_i + \frac{1-p(t)}{N}\right)^n \right), 
\ene
where $p(t)$ is given in~\eqref{eq:depleq} and $\lambda_i$ are the eigenvalues of $\rho(0)$. 

Now we connect this discussion to unitarity and replica wormholes. Consider the OPE coefficient averaging prescription in Sec.~\ref{sec:OPEAverage}. Each factor of $\rho$ in $\rho^n$ is a sum of \textit{pairs} of OPE coefficients (see Eq.~\eqref{eq:quadraticOPE}). When we average $\rho^n$, because of the approximately Gaussian statistics of the OPE coefficients, it equals a sum of Wick contractions between $2n$ OPE coefficients. Diagrammatically, for $n=3$:
\bne \includegraphics
[width=\linewidth]
{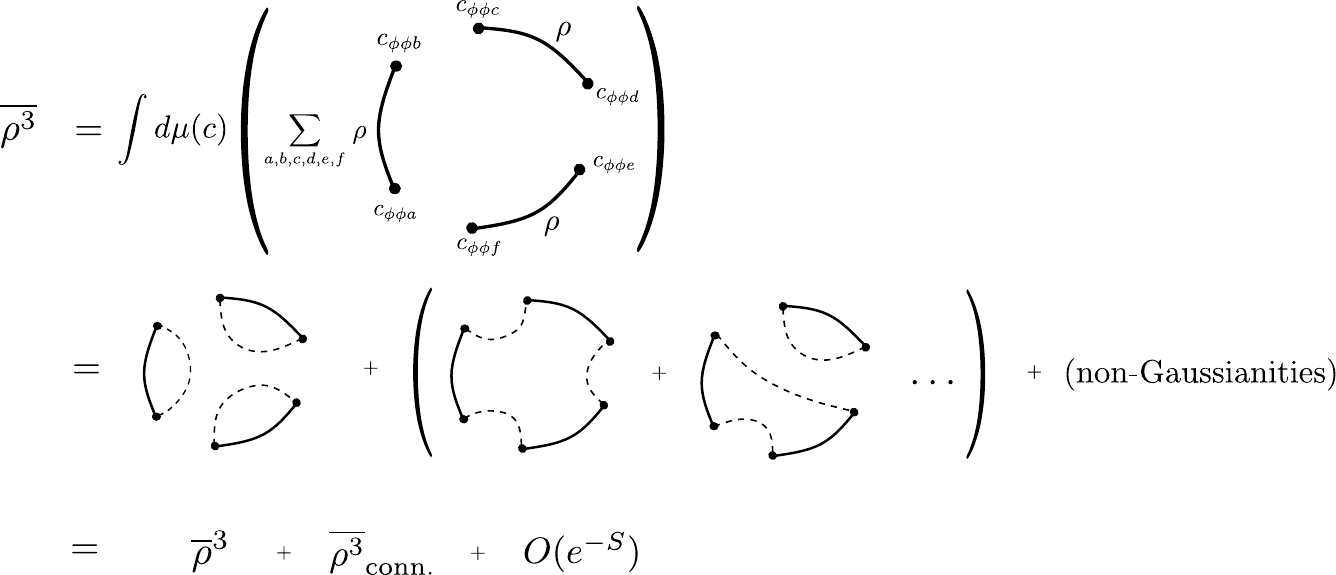} \label{eq:repwh}\ene
Each solid line represents a copy of the density matrix $\rho$, the pair of dots represents each copy's two OPE coefficients, and the dashed lines represent the Wick contractions between OPE coefficients.

There are several aspects of~\eqref{eq:repwh} to discuss.  The averaged $\overline{\rho^n}$ splits into a disconnected contribution equal to $\overline{\rho}^n$, a sum of connected contributions from Wick contractions between different copies of $\rho$, and contributions from the $e^{-S}$ suppressed non-Gaussian corrections to the statistics of the OPE coefficients. There is only one completely connected diagram, and many partially connected diagrams, just as in~\cite{Penington:2019kki}, though one difference is that the non-planar diagrams in~\eqref{eq:repwh} are not suppressed.  The connected diagrams are reminiscent of replica wormhole-like geometries that connect the replica copies of $\rho$.
The non-Gaussian corrections are $e^{-S}$ suppressed, so they are negligible for individual matrix elements of $\overline{\rho^n}$, but note that small corrections to matrix elements can give leading order changes to $\Tr (\overline{\rho^n})$, because of the sum over $e^S$ elements. So, if we neglect the non-Gaussian corrections, then we do not have $S^{(\rm{coarse})}_{\rm vN}(\rho) \approx S^{(\rm{exact})}_{\rm vN}(\rho)$ prima facie. Nonetheless, the remarkable feature of replica wormholes and the gravitational path integral is that off-shell and UV corrections can be neglected, and we still get an $S^{(\rm{coarse})}_{\rm vN}(\rho)$ consistent with unitarity~\cite{Almheiri:2020cfm}. 
Analogously, even if our OPE coefficient averaging makes $\rho$ highly mixed, $S_{\rm vN}(\overline{\rho}) \gg 0$, we expect the replica wormhole-like connected diagrams in~\eqref{eq:repwh} to ``rescue" unitarity, meaning $S^{(\rm{coarse})}_{\rm vN}(\rho) \approx 0$ if $\rho$ is pure, \textit{even if we neglect non-Gaussian, non-perturbative corrections to $\overline{\rho^n}$}. See also~\cite{baoRyuTakayanagiFormulaMultiBoundary2025} on the connection between replica wormholes and OPE coefficient averaging.

Beyond OPE coefficient averaging, we might expect the same mechanism to occur for any of our coarse-graining maps that average over information with approximate Gaussian statistics, such as the averaging over unitaries with a Haar measure of Sec.~\ref{sec:UnitaryInvariant}, because they will have the same Wick contraction diagrammatic structure as in~\eqref{eq:repwh}. 
Indeed, as evidence to back up this claim, in~\cite{deboerPageCurvesReplica2023} we showed explicitly that this mechanism holds in a setup similar to Sec.~\ref{sec:UnitaryInvariant}, where we averaged over Hamiltonian rotation unitaries in a dynamical random matrix model of black hole evaporation. In fact, one of the main points of~\cite{deboerPageCurvesReplica2023} was to interpret the resolution of the information paradox as the possibility to distinguish mixed states from statistical mixtures of pure states using replica computations: when the coarse-grained density matrix is a classical mixture of pure states, single-copy low-energy observables cannot distinguish this from an intrinsically mixed state. However, by taking replicas of the state, semiclassical observers \textit{can} distinguish these two and unitarity is rescued.

\subsection{Summary table}\label{sec:summ_tab}

\begin{table}[htbp]
\centering
\footnotesize
\begin{threeparttable}
\begin{tabular}{|M{3cm}|M{1.8cm}|M{1.2cm}|M{1.4cm}|M{1.4cm}|M{1.4cm}|M{1.4cm}|}
\hline
& \multicolumn{2}{c|}{\parbox{3cm}{\centering 
\vspace{1mm} Is $\overline{\rho(t)}$ sensitive to this in $\rho(t)$?}}
&  
\multicolumn{2}{c|}{\parbox{2.8cm}{\centering 
\vspace{1mm} 
Purity of $\overline{\rho(t)}$}}
& \multicolumn{2}{c|}{\parbox{2.8cm}{\centering 
Purity of $\overline{\rho(0)}$
}} \\ 
\cline{2-7}
& Eigenvalues & Phases & Sensitive  to $\rho(t)$'s phases? & Mixing time $t_{\textrm{mix}}$ & BH $\Rightarrow$ mixed? & Not BH $\Rightarrow$ pure? \\
\Xhline{5\arrayrulewidth}
\textbf{Ensembles of OPE coefficients } (Sec.~\ref{sec:OPEAverage}) & Yes & Yes & No & $t_{\rm d.o.}$  & Yes & Yes \\
\hline
\multicolumn{3}{|l|}{\textbf{Ensembles of Hamiltonian rotations}} & & & & \\ \hline
Unitary-invariant (Sec.~\ref{sec:UnitaryInvariant}) & Yes\tnote{a} & No & No & $t_{\rm d.o.}$ & No\tnote{b} & Yes\tnote{b} \\
\hline
Unitary-invariant on a Hilbert subspace (Sec.~\ref{sec:DisAndIndis}) & Yes\tnote{c}  & No & No &  $t_{\rm d.o.}$ & No\tnote{b} & Yes\tnote{b} \\
\hline
\multicolumn{2}{|l|}{\textbf{Ensembles of energy levels}} & & & & & \\ \hline
Pure initial state (Sec.~\ref{sec:EnergyLevels1}) & No & No & No & $t_{\rm d.o.}$ & No\tnote{b} & Yes\tnote{b}\\
\hline
General mixed initial state (Sec.~\ref{sec:EnergyLevels2}) & Yes & Yes & No & $t_{\rm d.o.}$ & N.A.\tnote{d} & N.A.\tnote{d} \\
\hline
 \textbf{Averaging over time windows 
} (Sec.~\ref{sec:time_windows})
& Yes & Yes & No & 
$\infty$\tnote{e} & Yes\tnote{f} 
& 
Yes\tnote{f} \\
\hline
\textbf{Maximum entropy $\bm{\mu (\rho)}$ 
} (Sec.~\ref{sec:ensembles_of_density_matrices})
& Yes & Yes & Yes 
& $t_{\text{indist.}}$ & Yes & Yes \\
\hline
\multicolumn{2}{|l|}{\textbf{Diagonal approximation
} 
 } & & & & & \\ \hline
Energy eigenbasis (Sec.~\ref{sec:DiagonalApprox}) & Yes & No & No & $\infty$ & 
Yes
& No\tnote{g} \\
\hline
\end{tabular}
\caption{A summary of the coarse-graining maps 
analysed in Sec.~\ref{sec:HowToCoarseGrain} for a general time-evolved $\rho(t)$.
The importance of these properties is explained at the start of Sec.~\ref{sec:HowToCoarseGrain}: they are necessary conditions for the coarse-grained boundary state to match the bulk semiclassical state in the context of the formation and evaporation of a small AdS black hole. The mixing time column entries are explained in the main body of Sec.~\ref{sec:summ_tab}.
}
\label{tab:summtab}
\begin{tablenotes}[flushleft]
\item[a] Yes, but sensitive only to the purity of $\rho (0)$.
\item[b] Because the averaged initial density matrix is pure for all initial pure states. 
\item[c] Yes, but sensitive only to $\Tr (P_{\tsc L}\rho)$, the fraction of the state that is in 
$\cH_{\tsc L}$.
\item[d] This envisions an initial coarse-graining step of $\rho(0)$ before averaging the time evolution, such as combining with Sec.~\ref{sec:OPEAverage}. When and how $\rho(0)$ is mixed depends on that step of the prescription.
\item[e] Unless one modifies the coarse-graining map to make the time window width time-dependent.
\item[f] Yes, if $\sigma_{\tsc E} \, \Delta t \gg 1$ for the black hole state and $\sigma_{\tsc E}\, \Delta t \ll 1$ for the non-black hole state, where $\sigma_{E}$ is spread of the state in the energy eigenbasis. Since a black hole state and an evaporated black hole state differ only by phases, we cannot have both at the same time.
\item[g] No, for a typical non-black hole state, because the diagonal approximation in a given eigenbasis makes a pure state mixed unless it is a basis state. 
\end{tablenotes}
\end{threeparttable}
\end{table}

In Table~\ref{tab:summtab}, we summarise the coarse-graining maps we have considered in this section.
In the columns, we list the properties that we wish the maps to satisfy. These properties were explained and motivated at the start of this section: they are necessary conditions for the coarse-grained boundary state to match the bulk semiclassical state in the context of the formation of AdS black holes. We want our coarse-grained boundary state to be pure when there is no black hole, and mixed when there is one (see Fig.~\ref{fig:IntroFig}). The importance of the sensitivity to phases is that this is all that differs between the pre-collision initial state~\eqref{eq:initial_state} and the black hole that is formed after time evolution (in the energy eigenbasis). We are looking for an $O(1)$ mixing time when a black hole is imminently being formed, such as for the initial state~\eqref{eq:initial_state}.

To explain the notation in the mixing time column, the entries $t_{\rm d.o.}$ denote the decay onset time of the infinite temperature spectral form factor, which is typically $O(1)$ with respect to the Hilbert space dimension (see Sec.~\ref{sec:EnergyLevels1}), and $t_{\textrm{indist.}}$ is when $\rho(t)$ becomes indistinguishable from typical density matrices in the window. The $t_{\rm mix} = \infty$ entries indicate that the purity of $\overline{\rho(t)}$ is constant. See Eq.~\eqref{eq:tmix} for the definition of $t_{\rm mix}$.

Not all of the properties in the table are independent of each other. For example, insensitivity of $\overline{\rho(t)}$ to the phases of $\rho(t)$ implies insensitivity of its purity $\Tr(\overline{\rho(t)}^2)$ (though not vice versa). 

In the table, $\rho(0)$ is a density matrix with support on a microcanonical window with sufficiently high mean energy to include AdS black hole states. 
Beyond that, we do not assume what bulk initial state $\rho(0)$ corresponds to, because we are interested in evaluating the coarse-graining maps on their ability to give different qualitatively different $\overline{\rho}$ depending on the whether the initial state is a black hole or not, and, if not, on whether the state is a pre-black hole formation state like~\eqref{eq:initial_state}. 

The entries in the table are only for the minimal versions of each type of coarse-graining map, for lack of space. 
See the relevant subsections for discussion of modifications and refinements.
The list is not exhaustive, and there is scope for exploration of other types of coarse-graining maps. 
To our knowledge, this paper is the first time that these maps have been applied to the Lorentzian, dynamic setup of black hole formation and evaporation in AdS/CFT. For further discussion, we refer the reader to Sec. \ref{sec: discussion}.

\subsection{Small black holes}
\label{sec: small black holes}
Most of the results in the previous subsections apply to large black holes only. 
Large black holes dominate the microcanonical ensemble, so one can coarse-grain in a structureless way, like when we used the Haar measure in Sec.~\ref{sec:Averaging over Hamiltonians: uncertainty in eigenstates}.
For microcanonical ensembles with energy in the small, unstable black hole range, there is still a notion of coarse-graining needed to capture the black hole entropy and semiclassical production of mixed state Hawking radiation, but small black hole microstates are atypical within their ensembles. Supposing that the initial state is a small black hole, time evolution takes us through partially evaporated microstates before reaching the typical radiation states, because of bulk locality. Any notion of coarse-graining black hole evaporation must account for this richer structure in the microcanonical ensemble.  

Now we discuss a model for the Hilbert space structure of a microcanonical window with energy $E$ in the small black hole range. In the model, the microcanonical Hilbert space $\cH$ is further decomposed into Hilbert spaces $\cH^{(n)}$ as follows:
\begin{equation}
    \cH = \bigoplus_{n=0}^E \cH^{(n)}, \quad \cH^{(n)} =  \cH_{\tsc{BH}}^{(E-n)} \otimes \cH_{\tsc{rad}}^{(n)}.
    \label{eq:Hblockdec}
\end{equation}
Here $\cH_{\tsc{BH}}^{(E-n)}$ and $\cH_{\tsc{rad}}^{(n)}$ are the microcanonical Hilbert spaces of the black hole with energy $E-n$ and the Hawking radiation with energy $n$ respectively.\footnote{In a physical setup, the Hilbert space cannot be decomposed exactly into black hole and radiation Hilbert spaces, but for our toy model we will take this decomposition to be exact.}
\footnote{The width of the black hole and radiation microcanonical Hilbert spaces is implicitly taken to equal one, for simplicity, such that $n$ takes integer values between zero and $E$.} 
We write the Hilbert space dimensions as
\begin{equation}
    \dim \cH_{\tsc{BH}}^{(n)} \equiv N_{\tsc{BH}}^{(n)}, \quad \dim \cH_{\tsc{rad}}^{(n)} \equiv N_{\tsc{rad}}^{(n)}, 
\end{equation}
and $N^{(n)} \equiv N_{\tsc{BH}}^{(E-n)} N_{\tsc{rad}}^{(n)}$. Since $\cH^{(0)}_{\tsc{rad}}$ and $\cH^{(0)}_{\tsc{BH}}$ only contain the vacuum state, their dimension is one.
Lastly, we define $N \equiv \sum_{n=0}^E N^{(n)}$. The dimension of each Hilbert space increases with increasing energy.
For a black hole initial state to evaporate, the evaporated states must be entropically dominant:
\begin{equation}
    N^{(n)} \ll N^{(n+1)}
\end{equation}
for all $n \in \{0, \dots , E\}$.
A pure state in the Hilbert space $\cH$ can be decomposed into a sum of its normalised projections
\begin{equation}
    \ket{\psi} = \bigoplus_{n=0}^E \sqrt{p_n}\ket{\psi^{(n)}}
    \label{eq: pure state}
\end{equation}
A natural way to implement instantaneous coarse-graining over the black hole microstates is to unitarily rotate the black hole microstates, within each microcanonical subspace, and then average over this unitary matrix with the Haar measure. Since each microcanonical subspace is distinguishable, the unitary matrix is different for each $\cH^{(n)}_{\tsc{BH}}$ and we label the Haar unitary acting on $\cH^{(n)}_{\tsc{BH}}$ as $U^{(n)}_{\tsc{BH}}$.
\bne \begin{split} \overline{\ket{\psi}\bra{\psi}} &= \int \prod_{i=0}^E dU_{\tsc{BH}}^{(i)} \bigoplus_{n,m=0}^E \sqrt{p_n p_m} \left(U_{\tsc{BH}}^{(E-n)}\otimes \mathds{1}_{\mathrm{rad}}^{(n)}\right)\ket{\psi^{(n)}} \bra{\psi^{(m)}} \left(U_{\tsc{BH}}^{(E-m)\dagger}\otimes \mathds{1}_{\mathrm{rad}}^{(m)}\right)\\
&= \bigoplus_{n=0}^E  \frac{p_n}{N^{(E-n)}_{\tsc{BH}}} \, \mathds{1}_{\tsc{BH}}^{(E-n)} \otimes \rho_{\textrm{rad}}^{(n)},\qquad \qquad  \rho_{\textrm{rad}}^{(n)}:=\Tr_{\cH_{\tsc{BH}}^{(E-n)}}\left(\ket{\psi^{(n)}} \bra{\psi^{(n)}}\right).
\label{eq:avgpur}
\end{split} \ene
The averaging eliminates the $n\neq m$ off-diagonal blocks because the set of unitaries $U_{\tsc{BH}}^{(i)}$ are mutually independent, and we used the identity that, for any $\ket{\varphi} \in \cH_A \otimes \cH_B$,
\bne \int d U_A (U_A \otimes \mathds{1}_B)\, \ket{\varphi} \bra{\varphi}\, (U_A^\dagger \otimes \mathds{1}_B)  = \frac{\mathds{1}_A}{N_A} \otimes \Tr_A (\ket{\varphi}\bra{\varphi}). \ene
The averaged state~\eqref{eq:avgpur} has purity
\begin{equation}
    \Tr(\overline{\rho}^2) = \sum_{n=0}^E \frac{p_n^2}{N_{\tsc{BH}}^{(E-n)}} \Tr\left(\left(\rho_{\textrm{rad}}^{(n)}\right)^2\right).
\end{equation}
If $\ket{\psi} $ is a product state in $\cH^{(n)}$, then the purity of the coarse-grained state is $\frac{1}{N_{\tsc{BH}}^{(E-n)}}$, corresponding to the maximally mixed state on the black hole degrees of freedom within that subspace. If $\ket{\psi}$ is a typical state in $\cH^{(n)}$, then $\rho_{\textrm{rad}}^{(n)}$ will be approximately maximally mixed if the energy evaporated away $n$ is sufficiently low that $N_{\textrm{rad}}^{(n)} \ll N_{\textrm{BH}}^{(E-n)}$, by Page's theorem, and so $\overline{\rho}$ will be maximally mixed (within the $\cH^{(n)}$ subspace). After the Page time, when $n$ is sufficiently large that $N_{\textrm{rad}}^{(n)} \gg N_{\textrm{BH}}^{(E-n)}$, then $\rho_{\textrm{rad}}^{(n)}$ will not be maximally mixed. In the extreme limit, when the black hole is fully evaporated and $\ket{\psi}$ is a superposition of radiation states, i.e. $p_E = 1$, we find $\Tr(\overline{\rho}^2) = 1$, telling us that the coarse-grained state is still pure.
So, we have found a prescription that coarse-grains and mixes $\rho$ if it is a small black hole (with the expected purity), and leaves radiation states unaffected.

If we also wish to view the radiation states as chaotic and needing to be coarse-grained over, such that they can not be distinguished by simple observers within $\cH^{(n)}_{\tsc{rad}}$, we can introduce independent unitaries $U_{\tsc{rad}}^{(n)}$. The computations are straightforward to extend to this situation, and the coarse-grained density matrix reads
\begin{equation}
    \overline{\rho} = \bigoplus_{n=0}^E \frac{p_n}{N^{(n)}} \mathds{1}_{(n)}, \quad \mathds{1}_{(n)} \in \mathds{R}^{(N^{(n)} \times N^{(n)})}.
\end{equation}
At late times and assuming ergodic time evolution, we expect that $p_n \approx N^{(n)}/N$, so the coarse-grained density matrix is maximally mixed in the entire microcanonical window.\footnote{Note that we did not assume unitary invariance within each window $\cH^{(n)}$, only in $\cH^{(E-n)}_{\tsc{BH}}$ and $\cH^{(n)}_{\tsc{rad}}$ separately, but the resulting coarse-grained density matrix is still maximally mixed in the entire window.}

We can also write down a prescription for averaging dynamically, by coarse-graining over the eigenstates and eigenvalues of the Hamiltonian. 
In App. \ref{ap: toy model} we construct such an explicit toy model and describe how to coarse-grain the Hamiltonian in this energy window, where we assume that the radiation states are not coarse-grained over.
The Hamiltonian describes the dynamics both within and between the sectors $\cH^{(n)}$, and we coarse-grain only over the internal black hole dynamics, both on the eigenstates and the eigenvalues of the Hamiltonian. We have performed simulations to check that the purity behaves as expected and that ergodicity is satisfied, both when the initial state is a black hole microstate and a radiation microstate. Again, changing this setup to also include coarse-graining over the radiation dynamics is straightforward.

\section{Discussion}
\label{sec: discussion}
We close with a discussion of our results and comments on possible future directions.

In this paper, we have explored how semiclassical AdS black hole dynamics arise from a coarse-graining of the boundary CFT.
A key motivation of this work is to resolve the black hole information paradox in a clean, bathless AdS/CFT setup. When the AdS boundary has reflecting boundary conditions and the boundary CFT is \textit{not} coupled to a bath for the radiation to escape into, the information problem is a tension between the production of mixed state Hawking radiation and the unitarity of the exact dynamics.
Part of the resolution is that, while the exact state does remain pure, the mixed semiclassical state is a coarse-grained state.

In Sec.~\ref{sec:colling_particles}, we constructed exact boundary CFT states dual to two particles colliding to form an AdS black hole that, depending on the CoM energy, does or does not subsequently evaporate. These states are a sandbox for understanding how the semiclassical black hole dynamics emerge from the unitary evolution of the forever-pure exact boundary state.
In the spin-dimension eigenbasis, we showed that the particle-collision state's wavefunction coefficients~\eqref{eq:psico} are sharply peaked on dimensions centred around the CoM energy of the collision and zero angular momentum, and that the phases are initially highly correlated. As time progresses, the state dephases and becomes indistinguishable from other states in the microcanonical window, first from other small black hole states, then from radiation states. 

Coarse-graining arises naturally, because our CFT states describing black hole formation and evaporation are only known insofar as the relevant CFT data is known (Sec.~\ref{sec:Uncertainty}). From the perspective of a low-energy observer with only partial, statistical information about the UV CFT data, there is uncertainty in the initial state and the subsequent dynamics, and this induces uncertainty in the time-evolved state. 
A useful heuristic is to split the Hilbert space into those primary states whose conformal data is known (the integrable sector), and those whose data is known only statistically (the chaotic sector). 
The chaotic sector contains both approximate Fock states (like the two-particle initial state) and small unstable black hole states. These are related by time evolution and therefore differ only by relative phases in the energy eigenbasis, though neither are approximate energy eigenstates.

In Sec. \ref{sec:detect_BH}, we explored how black hole states can be distinguished from thermal gas states using boundary two-point functions. In Sec. \ref{sec: 2pt func} we showed that proper lengths of curves are larger in AdS-Schwarzschild black hole backgrounds compared to backreacted thermal gas with the same total energy, causing boundary two-point functions to be smaller. When there are internal compact dimensions, black holes below a certain mass are expected to localise on the internal dimensions due to Gregory-Laflamme instabilities. In this case, in Sec. \ref{sec: 10dBH}, we argued that the relevant bulk geodesic for boundary singlet two-point functions sits at the antipodal point of the internal space from the black hole, and that the metric backreaction there is still larger than for the thermal gas, but smaller than for a delocalised black hole (uplifted (AdS-Schwarzschild)$\times \mathcal M$). This causes the geodesic lengths to have a hierarchy: largest for the delocalised black hole, then the localised black hole, and smallest for the thermal gas (all with the same total mass). The magnitude of the boundary two-point functions is in the reverse order. In Sec. \ref{sec: deconfinement} we connected this to partial deconfinement. In CFT gauge theories, deconfined states are conjectured to be dual to large AdS black holes, partially deconfined states to small localised black holes and confined states to thermal gas. We argued that, in the free limit, generic two-point functions are ordered in the same way as expected from the dual picture: largest for confined states and smallest for deconfined states, with partially deconfined states in between.

In Sec.~\ref{sec:HowToCoarseGrain}, we sharpened the notion that the semiclassical bulk state is a coarse-grained state by exploring numerous inequivalent prescriptions for averaging the exact state and comparing the result with the semiclassical bulk state. The prescriptions include averaging over the partially known data (e.g. OPE coefficients, operator dimensions, wavefunction coefficients) and temporal coarse-graining. The prescriptions and whether their coarse-grained states share key properties with the semiclassical black hole state are summarised in Table~\ref{tab:summtab}. 

We found that many of the coarse-graining maps share important features. Some of the coarse-graining maps are instantaneous diagonal projections, though exactly what the eigenbasis of the projection is differs between the prescriptions.
Some other coarse-graining maps leverage the time dependence coming from the spectral form factor (SFF) of the Hamiltonian. This introduces a time dependence into the purity of the coarse-grained state. While $t_{\text{mix}}$ matches semiclassical expectations, the time it takes for the density matrix to become approximately maximally mixed does not match any expected or clear semiclassical timescale.
It is also not clear to us what could be a dual bulk computation, within a single copy of the spacetime, that could capture this timescale.  

As far as matching the listed properties, every coarse-graining map in the table is partially successful. Only the prescription from Sec.~\ref{sec:ensembles_of_density_matrices} satisfies all the properties considered. All the rest fail to satisfy the property that the purity $\Tr( \overline{\rho(t)}^2)$ is sensitive to the phases in $\rho(t)$, for the reason explained at the end of Sec.~\ref{sec:EnergyLevels1}, and some fail to satisfy other properties too. 
But it is premature to claim that Sec.~\ref{sec:ensembles_of_density_matrices}'s coarse-graining map is the ``correct" one, 
because the table does not include some of the prescription modifications discussed earlier in the section, which can change the properties of a given coarse-graining map, and also because the map in Sec.~\ref{sec:ensembles_of_density_matrices} is only guaranteed to work if the desired output is given as input in the prescription. 
For a given prescription, we interpret an inability to satisfy a property as a shortcoming that indicates a need for refinement, not a need for the prescription to be disregarded, though our analysis indicates that some coarse-graining maps are more naturally effective than others.

Our setup differs from some holographic models of black hole evaporation in that the mixedness of the state does not arise from tracing out a subsystem, but from coarse-graining over data and degrees of freedom that are inaccessible to a low-energy observer. 
This is how the information in the exponentially complex correlations of the exact state is effectively lost.
Remarkably, insofar as calculating unitarity-consistent entropy curves, unitarity can be ``rescued" by including replica wormhole-like connected diagrams when averaging moments of density matrices (see Sec.~\ref{sec:rep_wormholes} and~\cite{deboerPageCurvesReplica2023}), despite working in a large $N$ approximation that drops non-Gaussian corrections.

In Sec. \ref{sec: small black holes}, we discussed how the small black hole regime requires extra substructure, and we have given two possible coarse-graining prescriptions. The first coarse-graining procedure acts instantaneously and mixes the black hole microstates, such that the resulting coarse-grained density of a black hole state is mixed, even if the exact state is pure. The second coarse-graining map is dynamical and acts by coarse-graining over the Hamiltonian, and is worked out further in App. \ref{ap: toy model}. 

\paragraph{Outlook and future research.} 
We now discuss the outlook and potential avenues for future research. 

There are a couple of direct and accessible extensions of our work worth investigating. Firstly, when we determined the exact particle-collision state in the spin-dimension eigenbasis in Sec.~\ref{sec:initial_state_hh_basis}, we restricted ourselves to $1+1$ boundary dimensions. It would be worthwhile to generalise to higher dimensions where there are thermodynamically unstable AdS black holes without the need to take the compact dimensions into account, though we do not expect qualitatively different results. Secondly, we did not do a careful quantitative analysis of the effect of non-zero impact parameter on the exact state~\eqref{eq:primd} or on the coarse-grained density matrix, though we discussed it qualitatively in Sec.~\ref{sec:creating_small_BH}. The threshold impact parameter at which the particles are just close enough to form a black hole is of particular interest. The primary operators with dimension and spin corresponding to that impact parameter are expected to be at the threshold at which corrections to the GFF approximation become large (e.g. the double-trace operator $\phi^2_{n,l}$ ceases to be an approximate primary operator): the boundary between the integrable and chaotic primary operator sectors. 

It would be instructive to compare the entropies of the exact and semiclassical bulk states with the black hole interior traced out in our single boundary, bathless AdS/CFT setup. This would be more aligned with standard discussions of the Page curve, which consider the entropy of the radiation exterior to the black hole horizon~\cite{Page:2013dx}. 
However, tracing out spatial regions is a semiclassical construction, and therefore difficult to define precisely for exact states. Even if one puts that issue aside, it is not clear how to map a spatial partitioning of the bulk Hilbert space into a corresponding coarse-grained partition of the CFT Hilbert space, necessary to calculate radiation entropy curves. It may be a partitioning into low and high spatial momentum modes, given UV/IR duality. Such a partitioning was explored in~\cite{agonCoarseGrainedQuantum2018} as a way of coarse-graining density matrices. Other frameworks for tackling the problem could be operator subalgebras and entanglement wedges.

It would be interesting to see whether there is a more general principle at play causing two-point functions in a black hole background to have the smallest magnitude possible at fixed energies. Another natural boundary probe to consider would be one-point functions. As is shown in \cite{MalGri21}, one-point functions can capture the proper time from the event horizon to the singularity in a phase by coupling the dual field to the simplest higher-derivative coupling to the gravitational field. 
This can also be computed in the backreacted thermal gas background, and it would be valuable to explore whether a hierarchy also exists for the one-point functions. Since one-point functions are typically dominated by the near-boundary behaviour, which is, to leading order, the same for black holes and the thermal gas, we expect this to be a next-to-leading order effect. Another issue is the fact that these computations are best suited for Euclidean time, where the singularity only appears as a complex saddle and not in the real integral. How to compute the backreacted one-point function in Lorentzian time is less clear. 
As mentioned before, another natural extension of the analysis performed in Sec. \ref{sec: deconfinement} is to study whether two-point functions in the microcanonical ensemble are also smallest in deconfined states at strong coupling, which is the regime dual to semiclassical gravity.

It is worth further study of whether one could add further refinements to the coarse-graining maps in App.~\ref{ap: toy model}, such that we can capture all the expected time scales in one prescription. These could be either the time scales based on the presence of an apparent horizon as in Fig. \ref{fig:IntroFig}, or maybe more continuous increases of entropy as expected from e.g. \cite{BalBer112,BalBer11}.

There may be fruitful interplay between small AdS black holes and $T\overline{T}$ deformations. In the holographic context, these deformations bring the Dirichlet boundary to a finite cutoff~\cite{McGough:2016lol}, and potentially stabilise small AdS black holes. Then, small AdS black holes would be \textit{typical} states within a microcanonical window in a $T\overline{T}$-deformed holographic CFT.

From a high-level perspective, there is substantial scope for further study of small AdS black holes, both in their own right and in their role within the clean, environment-free, holographic formulation of the information problem that is central to this paper. Small AdS black holes have numerous mysterious features, some of which we have discussed and explored in Sec.~\ref{sec:detect_BH}. We also believe that the construction of exact CFT states dual to the formation and evaporation of small, unstable AdS black holes, such as our two-particles-colliding construction in Sec.~\ref{sec:creating_small_BH}, is a useful starting point for the study of many aspects of black holes, including dynamics, singularities, and the final Planck-scale stage of evaporation. 

\acknowledgments

We would like to thank Vijay Balasubramanian, David Berenstein, Marius Gebershagen, Felix Haehl and Shreya Vardhan for useful discussions. JdB is supported by the European Research Council under the European Union’s Seventh Framework Programme (FP7/2007-2013), ERC Grant agreement ADG 834878. The work of JH is supported by the Dutch Black Hole Consortium with project number NWA.1292.19.202 of the research programme NWA which is (partly) financed by the Dutch Research Council (NWO), and the work of AR is supported by FWO-Vlaanderen projects G012222N and G0A2226N, by the VUB Research Council through the Strategic Research Program High-Energy Physics, and by FWO-Vlaanderen through a Senior Postdoctoral Fellowship 1223125N.

\appendix

\section{Toy model for small, evaporating black holes in AdS}
\label{ap: toy model}
Consider a microcanonical ensemble with a mean energy in the small, unstable black hole range, such that the ensemble is dominated by radiation states.
Since there are black holes in such a microcanonical window, there should still be a notion of coarse-graining which accounts for the black hole entropy and the tension of the semiclassical evaporation with unitarity. Not the entire microcanonical Hilbert space should be coarse-grained over, because the entropically dominant radiation microstates are approximate Fock states which can distinguished from each other.
In this appendix, we construct a toy model for the Hamiltonian describing the dynamics in the Hilbert space for small black holes and obtain the apparent mixing of the density matrix that one expects in semiclassical gravity.

\subsection{Set-up}
We take the Hilbert space decomposition to be the same as in Sec.~\ref{sec: small black holes} and propose a prescription to coarse-grain dynamically over the black hole degrees of freedom.
The microcanonical window around energy $E$ is decomposed into smaller Hilbert spaces that have energy $E-n$ in the black hole and energy $n$ in the radiation states. The Hamiltonian describes dynamics both within and between these sectors, and we coarse-grain only over the internal black hole dynamics. 
We will construct an ensemble of Hamiltonians and average over the time-evolved density matrix. After projection onto the microcanonical window, all eigenvalues of Hamiltonians in the ensemble are in the interval $[E-\Delta E/2, E+ \Delta E/2]$. We will work in the eigenbasis of a single draw $H_0$ from the ensemble. 
All of the Hamiltonians in the ensemble will be of the form
\begin{equation}
    H = E \mathds{1} + \Delta H,
\end{equation}
and for $H_0$ this $\Delta H$ will be a diagonal matrix where the diagonal values are in the microcanonical band $\pm \Delta E/2$. Since the $E \mathds{1}$ component commutes with all matrices, it does not affect the time evolution of the density matrix. Therefore, we only need to consider the ensemble for $\Delta H$. This $\Delta H$ will have two contributions, which we will call $H_{\tsc{mix}}$ and $H_{\tsc{hop}}$. The mixing Hamiltonian $H_{\tsc{mix}}$ will describe the dynamics within each $\cH^{(n)}$, and the hopping Hamiltonian $H_{\tsc{hop}}$ describes the interactions between the different subspaces. Firstly we describe $H_{\tsc{mix}}$. The $n$-th block $H_{\tsc{mix}}^{(n)}$ of this Hamiltonian acts on the $n$-th Hilbert subspace $\cH^{(n)}$ of~\eqref{eq:Hblockdec} as
\begin{equation}
    H_{\tsc{mix}} = \bigoplus_{n=0}^E H_{\tsc{mix}}^{(n)}, \quad H_{\tsc{mix}}^{(n)} := \bigoplus_{n=0}^E \left(H_{\tsc{mix,BH}}^{(E-n)} \otimes \mathds{1}^{(n)}_{\tsc{rad}} + \mathds{1}_{\tsc{BH}}^{(E-n)} \otimes H_{\tsc{mix,rad}}^{(n)}\right).
    \label{eq:Hdecomp}
\end{equation}
We assume that there is no uncertainty in $H_{\tsc{mix,rad}}^{(n)}$. Since the eigenstates and eigenvalues should be the same for all draws of the ensemble, we pick a single diagonal matrix with eigenvalues drawn from a normal distribution centred around zero and with standard deviation $\sigma$.

For $H_{\tsc{mix,BH}}^{(E-n)}$, unitary-invariant ensembles of Hamiltonians are natural, as one is not able to distinguish different black hole microstates. 
We will draw our mixing Hamiltonian from a unitary ensemble with Gaussian statistics: the Gaussian unitary ensemble (GUE). 
The probability measure of the GUE is
\begin{equation}
    \mu_{\tsc{GUE}}(X) \propto \exp\left(- \frac{1}{2\alpha^2}\Tr(X^2)\right).
\end{equation}
We choose the scale parameter $\alpha$ of the ensemble to be such that the eigenvalues of the matrices $H_{\tsc{mix,BH}}$ have the same standard deviation $\sigma$ as the $H_{\tsc{mix,rad}}^{(n)}$.%
\footnote{If the GUE-matrix is $M\times M$-dimensional, this requires a scale parameter of $\alpha = \sigma/\sqrt{M}$.} Using this construction, the matrix $H_{\tsc{mix}}^{(n)}$
has eigenvalues with variance $2\sigma^2$.\footnote{For the first and the last microcanonical subspaces $\cH^{(0)}$ and $\cH^{(E)}$, we have, respectively, either fully unevaporated black hole states or fully evaporated radiation states, so we take $H_{\tsc{mix,rad}}^{(0)} = 0$ 
and $H_{\tsc{mix,BH}}^{(0)}=0$, as they act on one-dimensional Hilbert spaces.
To ensure that the eigenvalues of the mixing Hamiltonians in these subspaces still have variance $2\sigma^2$, the GUE matrix $H_{\tsc{mix,BH}}^{(E)}$ and the diagonal entries of $H_{\tsc{mix,rad}}^{(E)}$ are drawn from the GUE and normal distribution respectively, such that they have eigenvalues with standard deviation $\sqrt{2}\sigma$ instead of $\sigma$.
}
The density matrix that is obtained by averaging over the ensemble, i.e.
\begin{equation}
\overline{\rho_{\tsc{mix}}(t)} := \int d\mu(H_{\tsc{mix}}) \, e^{-iH_{\tsc{mix}}t} \rho(0) e^{iH_{\tsc{mix}}t}, 
\end{equation}
with $H_{\tsc{mix, rad}}^{(n)}$ fixed in the ensemble and $H_{\tsc{mix,BH}}^{(E-n)}$ GUE-distributed, will create states that are maximally mixed in their black hole component.\footnote{If one assumes that also the radiation microstates are indistinguishable within $\cH^{(n)}_{\tsc{rad}}$, the Hamiltonian $H_{\tsc{mix,rad}}^{(n)}$ is not a single diagonal matrix but can also be drawn from e.g. the GUE, with scale parameter such that the eigenvalues have standard deviation $\sigma$. The results in the rest of this subsection are straightforward to generalise.} 
For example, if we start with the pure state 
\begin{equation}
    \ket{\psi^{(n)}} = \ket{\psi^{(E-n)}}_{\tsc{BH}} \otimes \ket{\psi^{(n)}}_{\tsc{rad}} \in \cH^{(n)},
\end{equation}
after enough time has passed, the averaged density matrix approaches
\begin{equation}
    \lim_{t \to \infty}\overline{\rho_{\tsc{mix}}(t)} = \frac{\mathds{1}  \otimes \ket{\widetilde{\psi}^{(n)}}_{\tsc{rad}}\bra{\widetilde{\psi}^{(n)}}_{\tsc{rad}}}
{N_{\tsc{BH}}^{(E-n)}} 
  ,
\end{equation}
where the identity matrix is $\left(N_{\tsc{BH}}^{(E-n)}\times N_{\tsc{BH}}^{(E-n)}\right)$-dimensional. This is the only nonzero block in the density matrix, all other entries are zero.
The time scale for the averaged density matrix to become approximately maximally mixed is set by the variance $2\sigma^2$. 
\begin{figure}
    \centering
\includegraphics[width=0.95\linewidth]{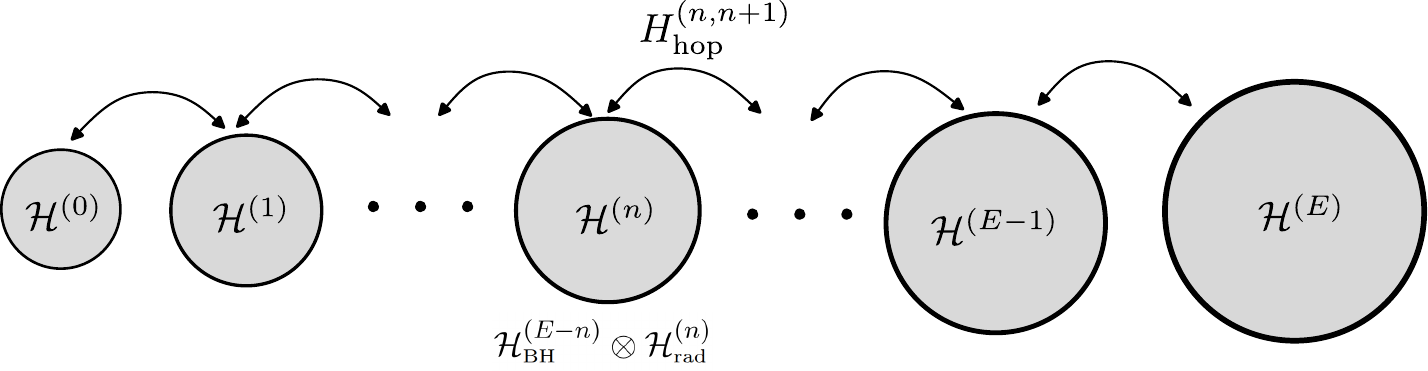}
    \caption{Our microcanonical Hilbert space structure represented as a chain, with $E$ in the small black hole range. Suppose our state starts on the left, in $\cH^{(0)} = \mathcal{H}_{\tsc{BH}}^{(E)}\otimes \ket{0}_{\tsc{rad}}$, where all the energy is in the black hole. It passes through the partially-evaporated subspaces $\cH^{(n)}$, where the black hole has energy $\approx E-n$. It ends on the right, in $\cH^{(E)} =\ket{0}_{\tsc{BH}} \otimes \mathcal{H}_{\tsc{rad}}^{(E)}$, the fully-evaporated-black-hole subspace. This has the highest entropy, because $|\cH^{(n)}|$ grows exponentially with $n$, due to the entropic dominance of the radiation states. }
    \label{Fig:HoppityHop}
\end{figure}

The mixing Hamiltonian~\eqref{eq:Hdecomp} alone does not allow for any evaporation: a state that starts in a subspace $\cH^{(n)}$ will remain 
there. 
We add a hopping Hamiltonian $H_{\tsc{hop}}$, which allows energy transfer between the black hole and its environment, and therefore states to hop from $\cH^{(n)}$ to $\cH^{(m)}$. Ensuring that only finite energy quanta can move from the black hole to the radiation means that $|m-n|$ is a finite (model-dependent) number, which we will take to be one. 
This means that $H_{\tsc{hop}}$ is a sum of overlapping Hamiltonians $H_{\tsc{hop}}^{(n,n+1)}$ that describe the hopping between $\cH^{(n)}$ and $\cH^{(n+1)}$, corresponding to the emission and absorption of Hawking quanta:
	\begin{equation}
	H_{\textrm{hop}} =
	\left(
	\scalebox{0.8}{% <-- scale factor here
		\tikz[baseline=(current bounding box.center), scale=1]{%
			% -------------------------------------------------
			% Sizes
			% -------------------------------------------------
			\def\sA{2.0}
			\def\ds{1.3}
			\pgfmathsetmacro{\sB}{\sA*\ds}
			\pgfmathsetmacro{\sC}{\sB*\ds}
			\pgfmathsetmacro{\sD}{\sC*\ds}
			\pgfmathsetmacro{\sE}{\sD*\ds}
			
			% -------------------------------------------------
			% Positions
			% -------------------------------------------------
			\def\xA{0} \def\yA{0}
			\pgfmathsetmacro{\xB}{0.85}  \pgfmathsetmacro{\yB}{-0.85}
			\pgfmathsetmacro{\xC}{\xA+\sA} \pgfmathsetmacro{\yC}{\yA-\sA}
			\pgfmathsetmacro{\xD}{\xB+\sB} \pgfmathsetmacro{\yD}{\yB-\sB}
			\pgfmathsetmacro{\xE}{\xC+\sC} \pgfmathsetmacro{\yE}{\yC-\sC}
			
			% -------------------------------------------------
			% Draw squares
			% -------------------------------------------------
			\fill[green!60,opacity=.6]  (\xA,\yA) rectangle ++(\sA,-\sA);
			\fill[red!60,opacity=.6]    (\xB,\yB) rectangle ++(\sB,-\sB);
			\fill[pink!60,opacity=.6]   (\xC,\yC) rectangle ++(\sC,-\sC);
			\fill[purple!60,opacity=.6] (\xD,\yD) rectangle ++(\sD,-\sD);
			\fill[teal!60,opacity=.6]   (\xE,\yE) rectangle ++(\sE,-\sE);
			
			% -------------------------------------------------
			% Overlaps (blue rectangles)
			% -------------------------------------------------
			\foreach \Xa/\Ya/\Sa/\Xb/\Yb/\Sb in {
				\xA/\yA/\sA/\xB/\yB/\sB,
				\xB/\yB/\sB/\xC/\yC/\sC,
				\xC/\yC/\sC/\xD/\yD/\sD,
				\xD/\yD/\sD/\xE/\yE/\sE
			}{
				\pgfmathsetmacro{\xmin}{max(\Xa,\Xb)}
				\pgfmathsetmacro{\xmax}{min(\Xa+\Sa,\Xb+\Sb)}
				\pgfmathsetmacro{\ymax}{min(\Ya,\Yb)}
				\pgfmathsetmacro{\ymin}{max(\Ya-\Sa,\Yb-\Sb)}
				\draw[blue,line width=0.5pt] (\xmin,\ymax) rectangle (\xmax,\ymin);
			}
			
			% -------------------------------------------------
			% Diagonal blue squares
			% -------------------------------------------------
			\pgfmathsetmacro{\sDiag}{min(\sA,\sB)-abs(\xB-\xA)}
			\pgfmathsetmacro{\sDiagDE}{min(\sD,\sE)-abs(\xD-\xE)}
			\draw[blue,line width=0.5pt] (\xA,\yA) rectangle ++(\xB,\yB);
			\draw[blue,line width=0.5pt] (\xE+\sDiagDE,\yE-\sDiagDE) rectangle ++(\sE-\sDiagDE,-\sE+\sDiagDE);
			
			% -------------------------------------------------
			% Labels
			% -------------------------------------------------
			\node[anchor=north east] at (\xA+\sA,\yA) {$H_{\mathrm{hop}}^{(0,1)}$};
			\node[anchor=north east] at (\xB+\sB,\yB) {$H_{\mathrm{hop}}^{(1,2)}$};
			\node[anchor=north east] at (\xC+\sC,\yC) {$H_{\mathrm{hop}}^{(2,3)}$};
			\node[anchor=north east] at (\xD+\sD,\yD) {$H_{\mathrm{hop}}^{(3,4)}$};
			\node[anchor=north east] at (\xE+\sE,\yE) {$H_{\mathrm{hop}}^{(4,5)}$};
			
			% -------------------------------------------------
			% Diagonal dots
			% -------------------------------------------------
			\foreach \i in {0.2,0.4,0.6}{
				\node at (\xE+1.0*\sE+\i,\yE-1.0*\sE-\i) {$\cdot$};
			}
		}
	}
	\,\right)
    \label{eq:hopping}
	\end{equation}
The solid coloured squares represent the hopping terms $H_{\tsc{hop}}^{(n,n+1)}$ and they have size $(N^{(n)} + N^{(n+1)})^2$. 
The blue squares represent the $\cH^{(n)}$ Hilbert subspaces, which the hopping Hamiltonian takes us between and are size $N^{(n)} \times N^{(n)}$. 
All other entries in $H_{\textrm{hop}}$ are zero. 

The $H_{\tsc{hop}}^{(n,n+1)}$ are each a single draw of some ensemble with unitary invariance, and the eigenvalues have zero mean and variance $2\sigma^2$.    
This ensemble does not necessarily have to be GUE; as long as it is unitarily invariant and the eigenvalues have zero mean and variance $2\sigma^2$, we expect ergodicity to hold, by which we mean that the late time average density matrix is equidistributed among all microstates: $ \lim_{t \to \infty}\overline{\rho(t)} \approx \frac{\mathds{1}}{N}.$\footnote{We verified that the eigenvalues of the hopping Hamiltonians do not have to be drawn from a chaotic ensemble for ergodicity to hold. We have repeated the numerics in Sec. \ref{sec: numerics}, but instead of drawing the hopping Hamiltonians from a GUE, we decompose them as $H_{\tsc{hop}}^{(n,n+1)} = U \Lambda U^\dagger$. Here we draw $U$ from the Haar ensemble and $\Lambda$ is a diagonal matrix with diagonal entries drawn from a Poisson distribution (multiplied by a random $\pm 1$). This ensures that the eigenvalues have zero mean, and we fix the mean of the Poisson distribution such that the standard deviation of the eigenvalues of $\Lambda$ is 10. This shows the same behaviour as drawing the hopping Hamiltonians from the GUE, such that its eigenvalues also have a standard deviation of 10.}

The averaged density matrix is defined as
\begin{equation}
    \overline{\rho(t)} := \int d\mu(H)\, e^{-iHt}\rho(0) e^{iHt},
\end{equation}
again with all components of the Hamiltonian fixed within the ensemble $\mu$, except for the GUE-distributed black hole mixing terms. 
We do not average over the hopping Hamiltonians, because the only uncertainty in the system should come from the chaotic internal black hole dynamics. However, one should still ensure that the eigenvalues of the Hamiltonians have some variance and are not highly degenerate, as this introduces extra symmetry which we don't expect to be there in a physical system and can hinder the ergodicity of the Hamiltonian. 
As mentioned before, after time evolution with an ergodic Hamiltonian, we expect the subspace probabilities $p_n = \Tr(P_n \rho P_n)$ (with $P_n$ the projection operator onto $\cH^{(n)}$) to approach $N^{(n)}/N$. When the radiation microstates are highly entropically dominant over the black hole microstates, $p_E$ approaches one and all other subspace probabilities approach zero.
We choose the variance in eigenvalues of all the mixing Hamiltonians and the hopping Hamiltonians to be the same, to ensure that the late-time density matrix is spread equally over all $N$ microstates. 
From numerical experimentation, the late-time subspace probabilities $p_n$ deviate from the expected values of $N^{(n)}/N$ when the variances are non-identical.
To corroborate this, in
\cite{YauYin25} it was shown that ergodicity is satisfied for a banded GUE matrix where each entry is drawn from a normal distribution with fixed variance and the banded GUE has width $W > N^{1/2+\epsilon}$. The structure of the full Hamiltonian in \cite{YauYin25} is as in Eq.~\eqref{eq:hopping} but with all subspaces the same size: $N^{(n)} = W$ for all $n$. Since the width of the diagonal blocks is equal, equal variance for each entry means that the diagonal blocks have the same variance in the eigenvalues as well. In our evaporating black hole model, the Hilbert spaces $\cH^{(n)}$ are increasing in size down the diagonal, as radiation states are entropically dominant. This means that the width of the banded Hamiltonian model is not constant, but increases with $n$. However, we take \cite{YauYin25} and our numerics as evidence that, as long as the variances of the eigenvalues of all diagonal blocks are approximately equal, the $p_n$ will reach the expected values $N^{(n)}/N$. 

The time scale of evaporation and mixing in this model is set by the parameter $\sigma$ and the number of Hilbert subspaces $\cH^{(n)}$. We cannot pick $\sigma$ to be different for different $H_{\tsc{hop}}^{(n,n+1)}$, to tune the ratio between mixing and evaporation time scales, as this affects the ergodicity of the Hamiltonian. However, the number of Hilbert spaces that a state has to go through before it has evaporated will also impact the evaporation time. The hopping Hamiltonian only connects $\cH^{(n)}$ with $\cH^{(n+1)}$ in our model, so a black hole initial state has to pass through all of these subspaces to fully evaporate.

It is important to preserve the semiclassical nature of the state. At each moment in time, the subspace probabilities should only be nonzero in a narrow range, $p_n,\dots p_{n+m}$ with $m = O(1)$; this keeps the variances of energies in the black hole and radiation subsystems small. This reflects the fact that for an observer, up to $O(1)$ corrections, it is possible to measure how much of the energy is in the black hole and how much is in the radiation. Our expectation is that in the large $N$ limit, where the number of Hilbert subspaces $\cH^{(n)}$ is large, the many blocks in the nearly-block-diagonal Hamiltonian will ensure that the density matrix has only $O(1)$ nonzero neighbouring subspace probabilities $p_n$.

\subsection{Numerics}
\label{sec: numerics}

We will numerically simulate the dynamics in our toy model after specifying its parameters. 
Let us set $E=2$ and the Hilbert space dimensions to
\begin{equation}
\{N^{(0)}_{\tsc{BH}},N^{(1)}_{\tsc{BH}},N^{(2)}_{\tsc{BH}}\} = \{1,2,8\}, \quad   \{N^{(0)}_{\tsc{rad}},N^{(1)}_{\tsc{rad}},N^{(2)}_{\tsc{rad}}\} = \{1,16,64\}.
    \label{eq: H sizes}
\end{equation}
This sets $\{N^{(0)},N^{(1)},N^{(2)}\} = \{8,32,64\}$ and $N = 104$. We also choose $\sigma^2 = 50$.

We will compare the dynamics of the coarse-grained state for two initial pure states:
\begin{equation}
\ket{\psi_{\tsc{BH}}(0)} \in \cH^{(0)}, \quad \ket{\psi_{\tsc{rad}}(0)}\in \cH^{(2)}.
    \label{eq: initial states}
\end{equation}
The $\ket{\psi_{\tsc{BH}}(0)}$ initial state represents a small black hole at the start of its evaporation, and $\ket{\psi_{\tsc{rad}}(0)}$ is a state with radiation and no black hole.
For $\ket{\psi_{\tsc{BH}}(0)}$, it does not matter which state in $\cH^{(0)}$ we choose, because of the unitary invariance of the ensemble mixing Hamiltonian; for our plots, we chose an eigenstate of a single draw $H_0$ from the Hamiltonian ensemble. For $\ket{\psi_{\tsc{rad}}(0)}$, it is important that it is not an energy eigenstate of $H_0$, since that means it will be an energy eigenstate of all $H_{\tsc{mix}}$ and is therefore not a generic state. Instead, we take a Haar random state in $\cH^{(2)}$.

\begin{figure}[h!]
    \centering
\includegraphics[width=0.8\linewidth]{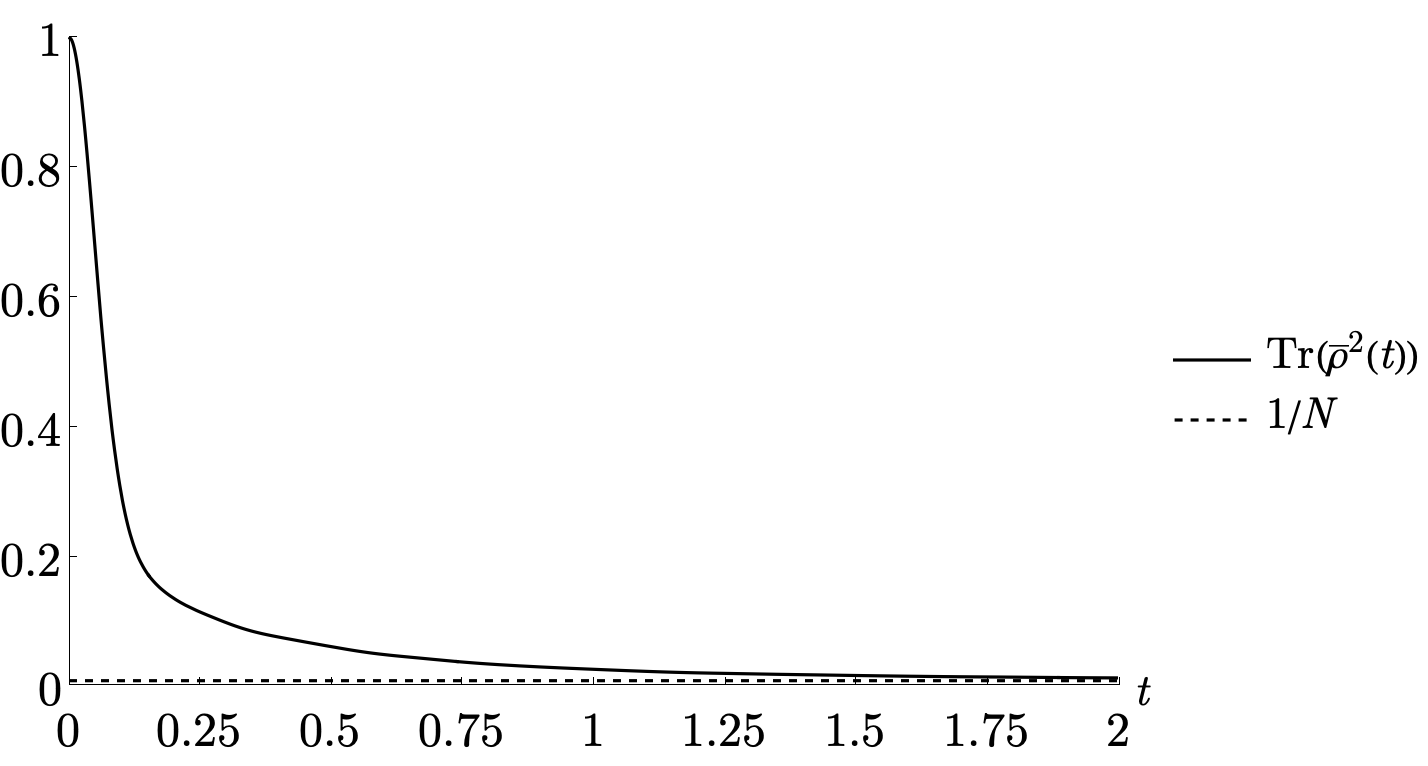}
    \caption{A plot of the purity of the averaged density matrix $\overline{\rho_{\tsc{BH}}(t)}$ from \eqref{eq: initial states}, where the initial state is a black hole. In this plot and all others in the appendix, we have set the sizes of the Hilbert space as in \eqref{eq: H sizes}, and  $\sigma^2 = 50$. We average over 1000 instances of $H_{\tsc{mix,BH}}^{(E-n)}$, and draw one instance of $H_{\tsc{hop}}$ and $H_{\tsc{mix,rad}}^{(n)}$. }
    \label{fig:purityBH}
\end{figure}

\begin{figure}[h!]
    \centering
\includegraphics[width=0.8\linewidth]{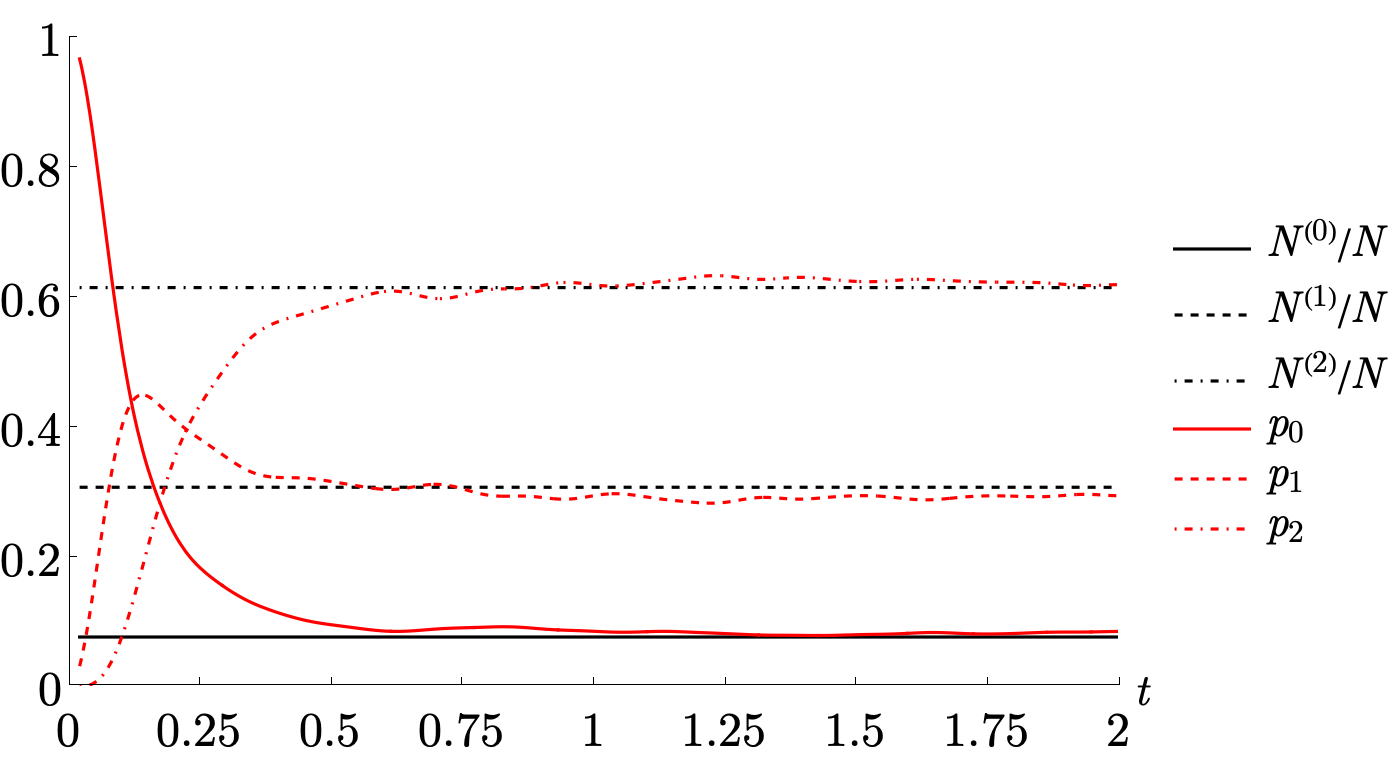}
    \caption{
    This plot shows in red the subspace probabilities $p_n = \Tr(P_n \rho P_n)$, where $P_n$ is the projection operator onto $\cH^{(n)}$ and $\rho(0) = \rho(0)_{\tsc{BH}}$ as in \eqref{eq: initial states}. In black are the expected values $N^{(n)}/N$. }
    \label{fig:subprobBH}
\end{figure}

\begin{figure}[h!]
    \centering
\includegraphics[width=0.8\linewidth]{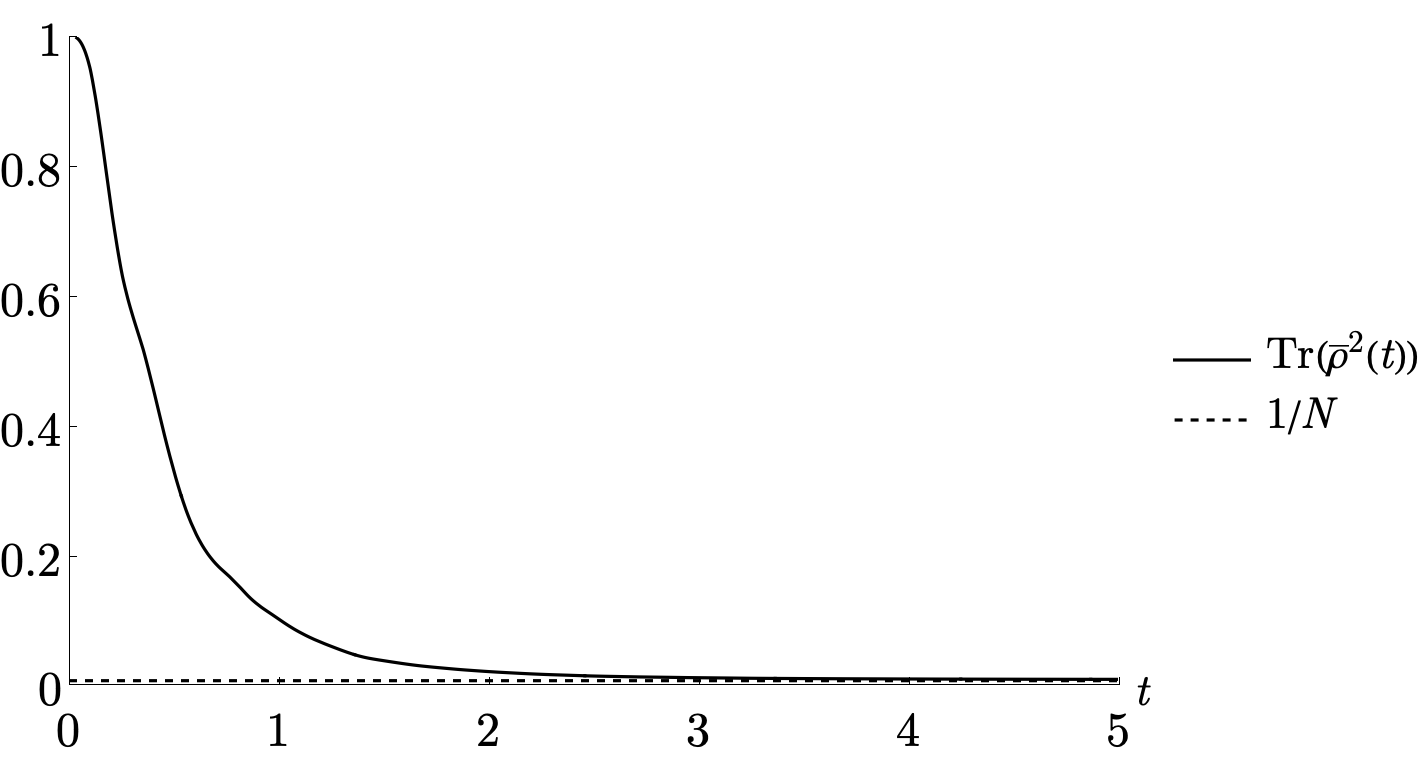}
    \caption{A plot of the purity of the averaged density matrix $\overline{\rho_{\tsc{rad}}(t)}$ from \eqref{eq: initial states}, where the initial state is radiation.}
    \label{fig:purityrad}
\end{figure}

\begin{figure}[h!]
    \centering
\includegraphics[width=0.8\linewidth]{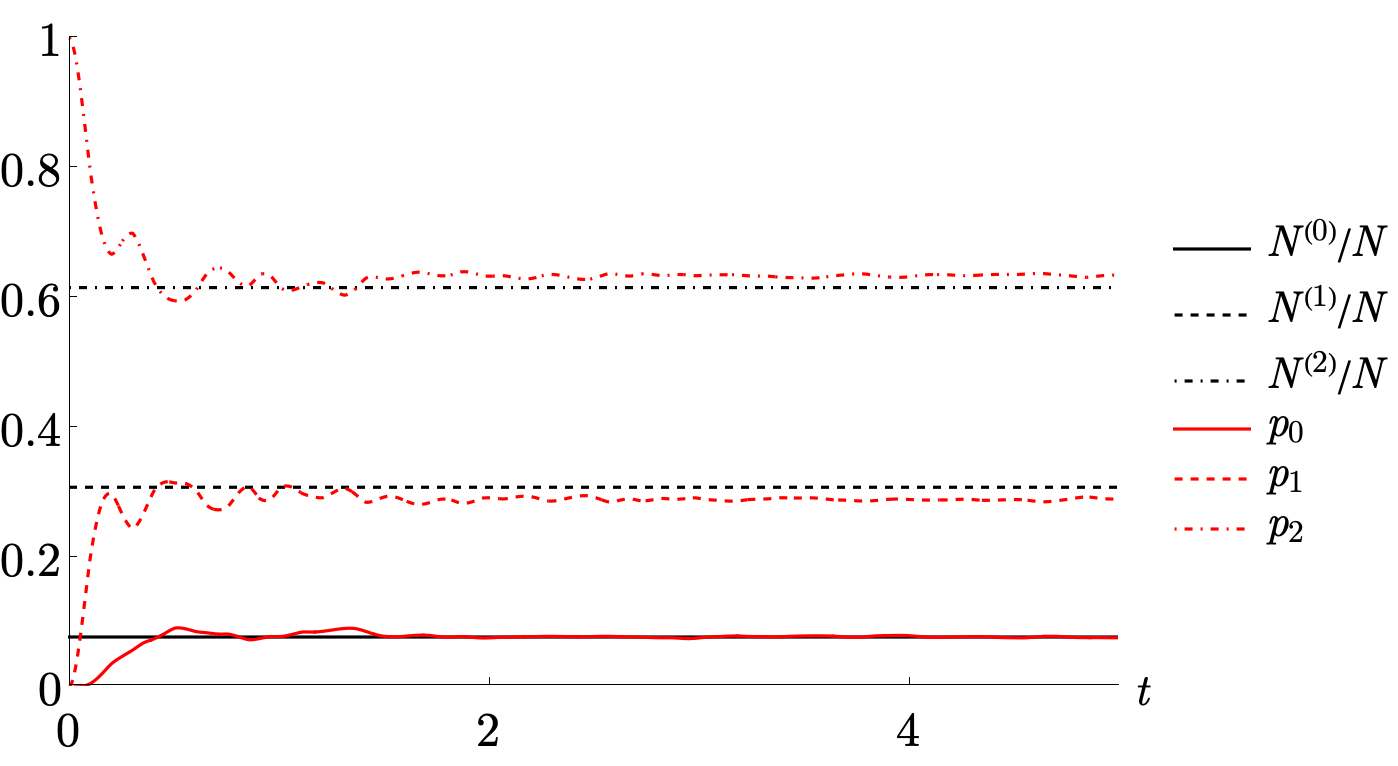}
    \caption{This plot shows in red the subspace probabilities $p_n = \Tr(P_n \rho P_n)$, where $P_n$ is the projection operator onto $\cH^{(n)}$ and $\rho(0) = \rho(0)_{\tsc{rad}}$ as in \eqref{eq: initial states}. In black are the expected values $N^{(n)}/N$. }
    \label{fig:subprobrad}
\end{figure}
We plot the purity of the coarse-grained time evolution of the pair of states~\eqref{eq: initial states} in Fig. \ref{fig:purityBH} and Fig. \ref{fig:purityrad}. We also verify that the subspace probabilities $p_n$ reach the expected values $N^{(n)}/N$ in Fig. \ref{fig:subprobBH} and Fig. \ref{fig:subprobrad}. In all the plots, we average over 1000 instances of $H_{\tsc{mix,BH}}^{(E-n)}$, and draw one instance of $H_{\tsc{hop}}$ and $H_{\tsc{mix,rad}}^{(n)}$.
It can be seen that starting with a radiation state gives rise to slower mixing. This is as expected: part of the radiation first has to collapse into a black hole before it will mix and reduce the purity. 

\begin{figure}
    \centering
\includegraphics[width=\linewidth]{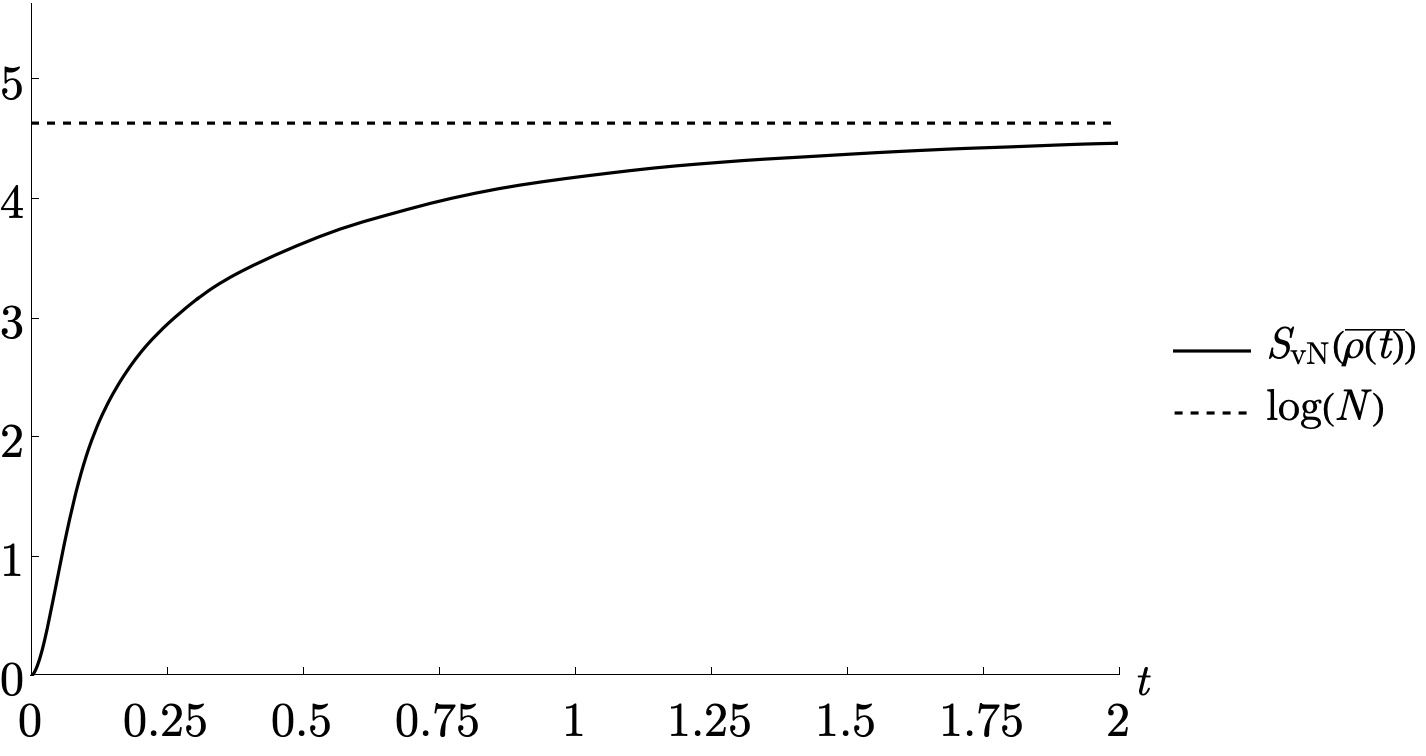}
    \caption{A plot of the non-unitary entropy curve that is obtained by computing the von Neumann entropy of the averaged density matrix $\overline{\rho_{\tsc{BH}}(t)}$ from \eqref{eq: initial states}.}
    \label{fig:SBH}
\end{figure}

For each draw of the Hamiltonian, the dynamics are unitary. So, if the initial state is pure, the averaged von Neumann entropy is trivially zero. However, one obtains a non-unitary entropy curve from the von Neumann entropy of the averaged density matrix, as shown in Fig. \ref{fig:SBH}.

\bibliographystyle{JHEP}
\FloatBarrier
\bibliography{references.bib}

\end{document}